\documentclass[twocolumn,showpacs,preprintnumbers,amsmath,amssymb]{revtex4-1}

\usepackage{graphicx,color}
\usepackage{soul}
\usepackage{dcolumn}
\usepackage{bm}
\usepackage{epstopdf}

\begin{document}

\preprint{APS/123-QED}

\title{Electric-field control of the exchange interactions}

\author{S.~Mankovsky$^1$,  E.~Simon$^2$, S.~Polesya$^1$,
  A.~Marmodoro$^3$, and H.~Ebert$^1$}
\affiliation{%
$^1$Department of Chemistry/Phys. Chemistry, LMU Munich,
Butenandtstr. 11, D-81377 Munich, Germany \\
$^2$Materials Center Leoben Forschung GmbH, Roseggerstr. 12, A-8700
Leoben, Austria\\
$^3$Institute of Physics, Czech Academy of Sciences,Cukrovarnická 10, 162 00 Praha 6, Czech Republic
 \\
}%

\date{\today}

\begin{abstract}
 The impact of an applied electric field on the exchange coupling
  parameters has been investigated based on  
first-principles electronic structure calculations by means of the KKR
Green function method. The calculations have been performed for a Fe
film, free-standing and deposited on two different substrates, having 1
monolayer (ML) thickness to minimize the effect of screening of the
electric field typical for metallic systems. By comparing the results
for the free-standing Fe ML with those for Fe on the various substrates, we could
analyze the origin of the field-induced change of the exchange
interactions. Compared to thefree-standing Fe ML, in particular rather
pronounced changes have been found 
for the Fe/Pt(111) system due to the localized electronic
states at the Fe/Pt interface, which are strongly affected by the
electric field and which play an important role
for the Fe-Fe exchange interactions.
\end{abstract}

\maketitle

\section{Introduction}

The control of  magnetic properties by applying an
electric field is discussed in the literature since many years
\cite{SK80,JSM+06,MTO15,KBN07,MSN+09,IYS+18, NSM+12,DVS+08,Rad61,YSP+18a,YUF+11,OHH+15}. 
Apart from the well known
example of the magneto-crystalline
anisotropy (MCA) 
influenced by an electric field
\cite{MSN+09,NSF+09,WFM+07,LMV+09,CNMO10}, various types of the
magneto-electric (ME) effects have been discussed.
Accordingly, quite a number of investigations have been devoted to antiferromagnetic
(AFM)  \cite{DKT61,Rad61,RF62,LSM+06}, non-collinear magnetic
 \cite{SK80, KNB05} or ferromagnetic (FM)
 \cite{OCM+00,CYMO03,MTO15,OW09,DVS+08,SCO+12} systems. 
In the case of FM materials the investigations have been focused in
particular on the dependence of the exchange interactions
on applied electric field
aiming to manipulate that way by the
ferromagnetic-to-paramagnetic transition. 
The features of the ME effect depend in turn on the  dominating exchange
mechanism in the material \cite{MTO15}. In the diluted magnetic
semiconductor (DMS) (In,Mn)As \cite{OCM+00,CYMO03}, for instance,
used as a prototype system within such investigations, 
the dominating Ruderman-Kittel-Kasuya-Yosida (RKKY) exchange is
mediated by holes and strongly depends on the hole concentration which
may be efficiently controlled by the applied electric field.
In the case of metallic materials, the situation is more
complicated\cite{OW09} as their magnetic properties are governed by the
exchange interactions having quite a different origin and as a
consequence a different behavior under an applied electric field.
It is worth noting in addition that despite a short screening length 
in metals \cite{MTO15} a rather pronounced magneto-electric effect 
was demonstrated within DFT calculations performed for thin metallic
Fe(001), Ni(001), and Co(0001) films\cite{DVS+08}.
Experimentally, a significant field induced change of the Curie
temperature $T_C$ was observed for Co ultrathin films embedded into 
different layered structures  \cite{AYK+18,IYS+18}. The dependence of
$T_C$ on the electric field  strength was attributed to the corresponding modification
of the interatomic exchange interactions $J_{ij}$. As it was mentioned
above, the origin of these changes in metallic films is different
compared to DMS materials, and  
explicit first-principles calculations of the exchange coupling
parameters would be very desirable to find out the relationship between
the field induced modification of $J_{ij}$ and the electronic structure
in the system, as it has been done for instance for free-standing
Fe(001) and Co/Pt(111) \cite{ONA+15} FM films.

Because of the central role of the Dzyaloshinskii-Moriya interaction (DMI) in
two-dimensional layered systems for the
formation of magnetic skyrmions, control of the DMI
by an applied electric field is of great interest as it gives an
  access to manipulate the stability of skyrmions. 
In fact, a strong variation of the DMI with electric field was found
experimentally for the Ta/FeCoB/TaO \cite{SSJ+18} and 
MgO/Fe/Pt \cite{ZZZ+18} trilayer systems.
An impact of the electric field on
the size of the magnetic domain wall, investigated experimentally for Pt/Co
\cite{AKK+16} and  Pt/Co/AlO$_x$ \cite{SRB+21} thin films as well as an
impact on the domain wall motion in
Pt/Co/Pd films \cite{KNIC18},  was also associated with the field-induced
change of the Dzyaloshinskii-Moriya interaction. 
Furthermore, there is also great interest in the ME effect in bulk non-collinear magnetic
materials \cite{SK80,KNB05}, which is also associated with the electric
field induced DMI.

Despite significant attention  devoted in the literature to the
magneto-electric effect, so far no systematic investigations have been
performed on a first principles level. To our
knowledge there are only few corresponding reports in the literature. This is
the above mentioned report by Oba et al. \cite{ONA+15} on the
field-dependent $J_{ij}$ for free-standing Fe(001) and Co/Pt(111).
Yang et al. \cite{YBC+18} have studied the electric field control of the
DMI for the NM/Co/Pt trilayers with different non-magnetic NM layers.
Recently Paul and Heinze \cite{PH21} reported on the stability of skyrmions
controlled by the electric field, where $J_{ij}$, $\vec{D}_{ij}$,
biquadratic interactions and MCA have been calculated on the same
footing on ab-initio level.


In this work we focus on three prototype systems, the free-standing Fe
monolayer, 1ML Fe deposited on Pt(111) and on 1H-WS$_2$ substrates.
The first principles calculations of the electronic structure and the
exchange interactions and DMI for these systems have been performed
without external electric field as well as in the presence of the
electric field, to reveal the relation between the field induced changes of
the electronic structure and the exchange parameters.

\section{Computational details}
\label{SEC:Computational-scheme}

Within the present work,
 exchange coupling parameters were calculated
using the spin-polarized relativistic KKR (SPR-KKR) Green
function method  \cite{SPR-KKR7.7,EKM11}.
The fully-relativistic mode was used throughout except for those cases,
for which a scaling of the spin-orbit interaction was applied. 
All calculations have been performed within the framework of the local 
spin density approximation (LSDA) to spin density 
functional theory (SDFT), using a parametrization for the exchange and
correlation potential as given by Vosko et al.\ \cite{VWN80}.
The charge and spin  densities as well as the
potentials were treated on the level of the atomic sphere approximation
(ASA). A cutoff $l_{max} = 3$ was used for the angular momentum 
expansion of the Green function.
The $\mathbf{k}$-space integration over the two dimensional (2D) Brillouin zone 
(BZ) was done using a $109 \times 109$ 2D k-mesh.  

The calculations for 1ML of Fe deposited on the  Pt(111) surface 
have been performed for a geometry consisting
of semi-infinite Pt and vacuum
subspaces to the left and to the right, respectively, of
the so-called interaction zone consisting of 3 atomic
layers of Pt, one layer of Fe and 5 layers of
empty spheres (vacuum).
The calculations for a free-standing Fe monolayer and 1ML Fe on WS$_2$
have been performed in slab geometry.
In the latter case there are two possibilities for the arrangement of the Fe
atoms with respect to the position of W in 1H-WS$_2$;
i.e.\ being either above the 
W atoms  or above vacancies within the W layer. As the former occupation
is energetically more preferable, all calculations here have been done
for Fe occupying positions above the W atoms, with identical Fe-S
and W-S distances. Finally,
it should be noted that for all calculations
the structure relaxation of the surface
layers has not been taken into account.

\medskip

Within the present work, we investigate
the  impact of an electric field on the magnetic
properties of  metallic FM films restricting to the situation
when the field  is applied along the
normal to the surface.
This implies that the electric field will lead to some charge
rearrangement but not to a steady state electric current. 
We focus here on  ultrathin films considering one Fe monolayer, as in this
case one can expect pronounced  effects while these will be
reduced in metallic bulk  materials   
because of  screening  (see, e.g.\ Ref.\  \onlinecite{SMME21}). 
Accordingly, we consider here as representative examples
a 1ML Fe film deposited on different substrates.
In order to clarify the role of the substrate, we
represent also  results for an  unsupported Fe monolayer.

Dealing with an Fe monolayer, one has to make a remark concerning its
magnetic ordering.
It is well known that the magnetic order in an ideal 2D system  
should be broken at $T > 0$K due to spin-wave (SW) excitations, as it  is
to be expected on the basis of the  Mermin-Wagner theorem\cite{MW66}.
This, however, does not hold in the presence of magnetic anisotropy
that causes for the SW spectrum the opening of an energy gap at the $\Gamma$ point of the 2D BZ
and blocks that way the low-energy SW excitations\cite{Bru91}.
As a consequence, this mechanism leads to a dependence of the 
Curie temperature, $T_C$, on the MCA\cite{TKDB06}.
Accordingly, one may expect  a possible impact of the
electric field on $T_C$ due to the
field-dependent changes of the MCA discussed in the literature.

In the present work we assume a finite MCA to ensure
a unique FM order in the  Fe film.
However, we do not discuss the impact of the electric field on the MCA
and this way on the Curie temperature. Instead, we focus on the behavior
of the exchange parameters $J_{ij}$ and  $\vec{D}_{ij}$
that represent the isotropic and anisotropic
Dzyaloshinskii-Moriya interaction, respectively.
Their field dependence will be monitored
using 
the Curie temperature $T^{MFA}_C$ evaluated via mean-field theory. 

For the electronic structure calculations,
the effect of a homogeneous external electric field was modeled by a
periodic array of point charges in the vacuum region
that behave essentially like a charged capacitor plate. 
This leads to a homogeneous electric field of strength,
\begin{equation}
 E = \frac{Q}{A \epsilon_{0}} \, ,
\label{eq:Q-E}
\end{equation}
where $Q$ is the charge of the capacitor in unit of the electron's charge,
$\epsilon_{0}$ is the permittivity of vacuum and $A$ is the area per
charged site in the capacitor plate.
As suggested by Eq.\ (\ref{eq:Q-E}), the orientation of the electric
field can be controlled via the sign of $Q$. A positive charge  $Q$ in
front of a surface gives rise to a field $\vec{E} = E_z \hat{z} $ 
pointing inwards to the bulk and anti-parallel to the surface normal
$\hat{z}$. As $E_z$ in this case, the electric field will be denoted
briefly 'negative', while 'positive' denotes an outward-oriented
electric field.

\section{Results \label{results}}

\subsection{Free-standing 1 ML Fe  \label{Fe}}

\begin{figure}[h]
\includegraphics[width=0.25\textwidth,angle=270,trim= 0 4.5cm 0
7.54cm,clip]{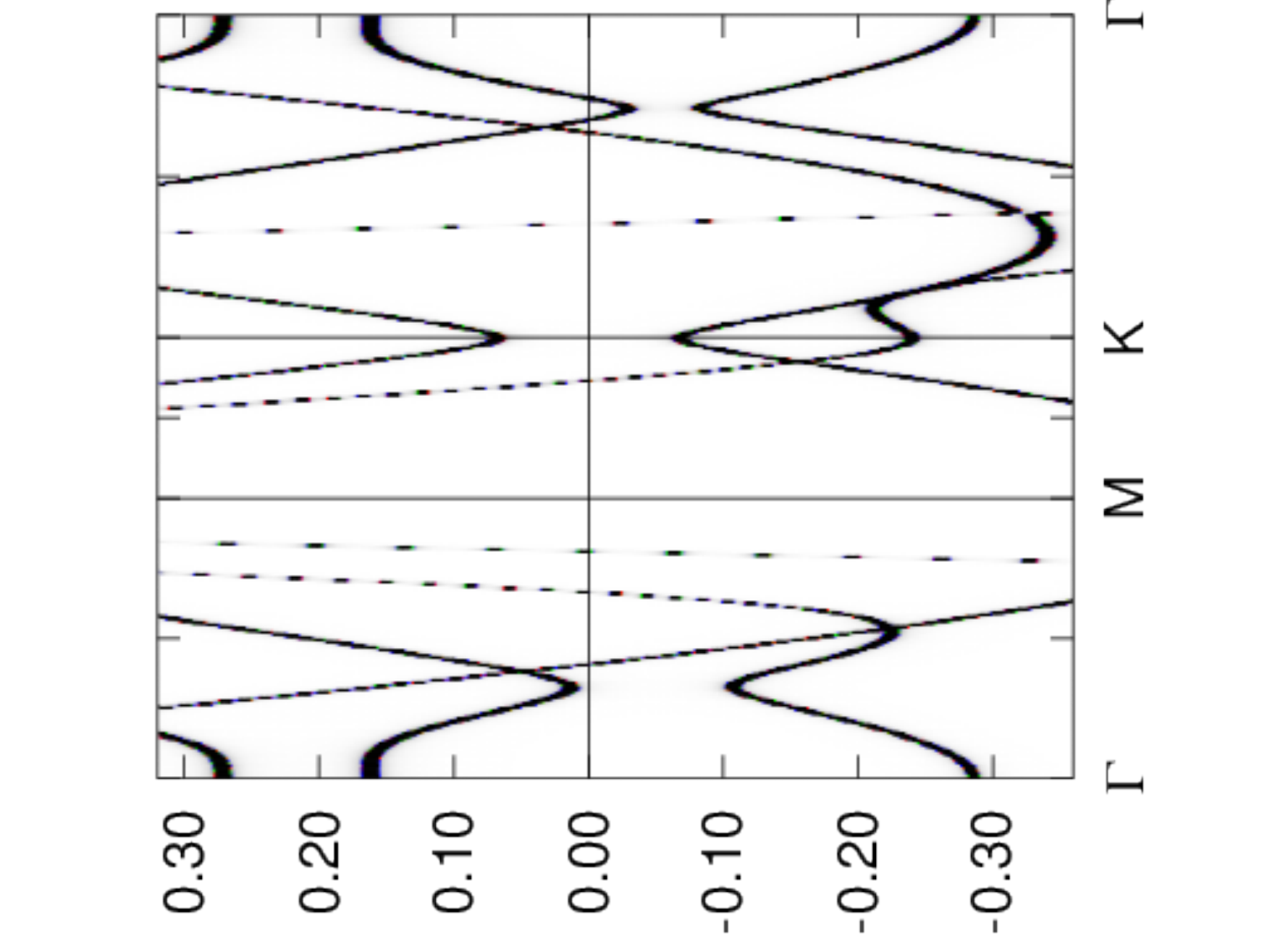}
\includegraphics[width=0.25\textwidth,angle=270,trim= 0 4.5cm 0
4.3cm,clip]{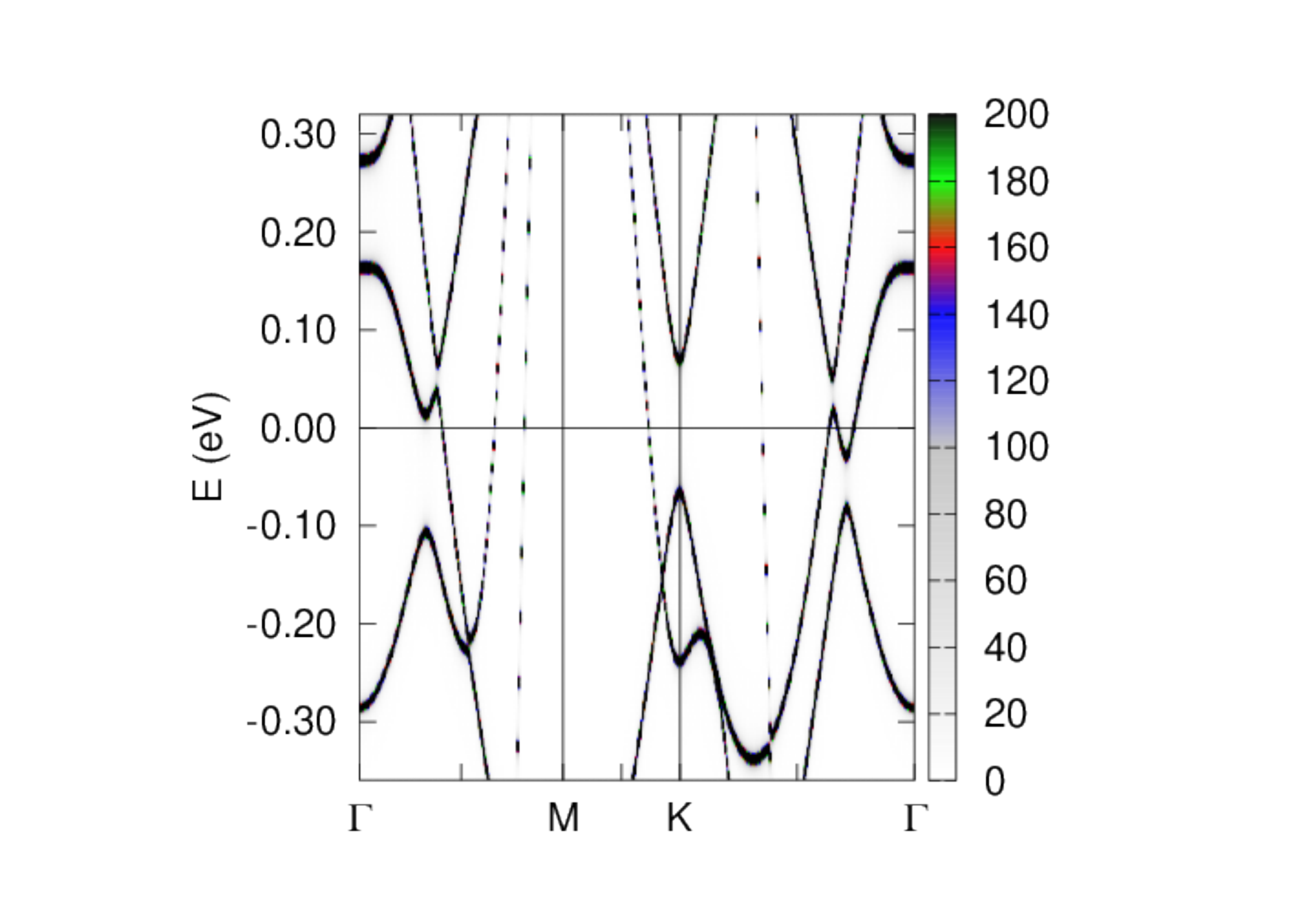}(a)
\includegraphics[width=0.25\textwidth,angle=270,trim= 0 4.5cm 0
7.52cm,clip]{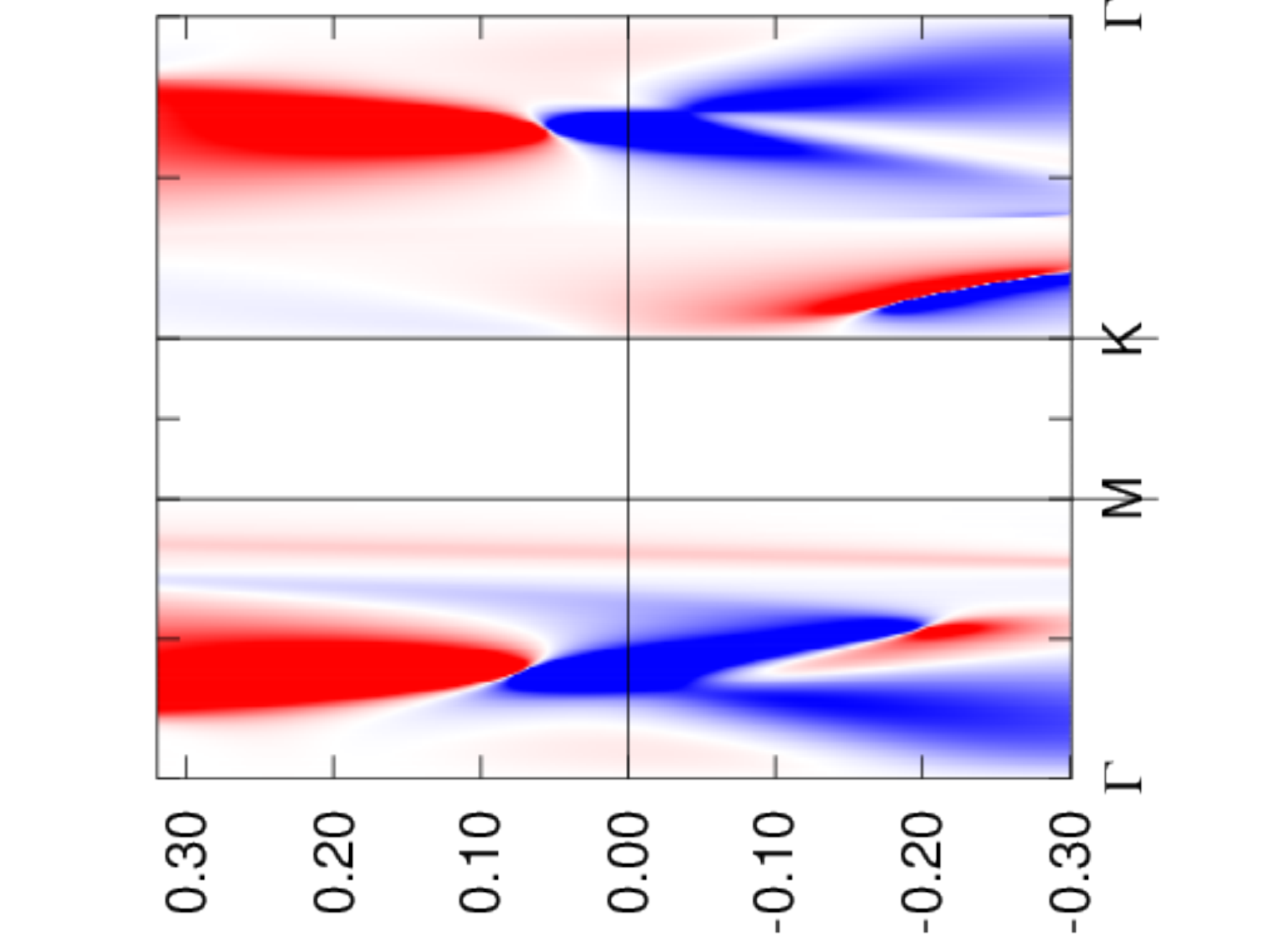}
\includegraphics[width=0.25\textwidth,angle=270,trim= 0 4.5cm 0
4.3cm,clip]{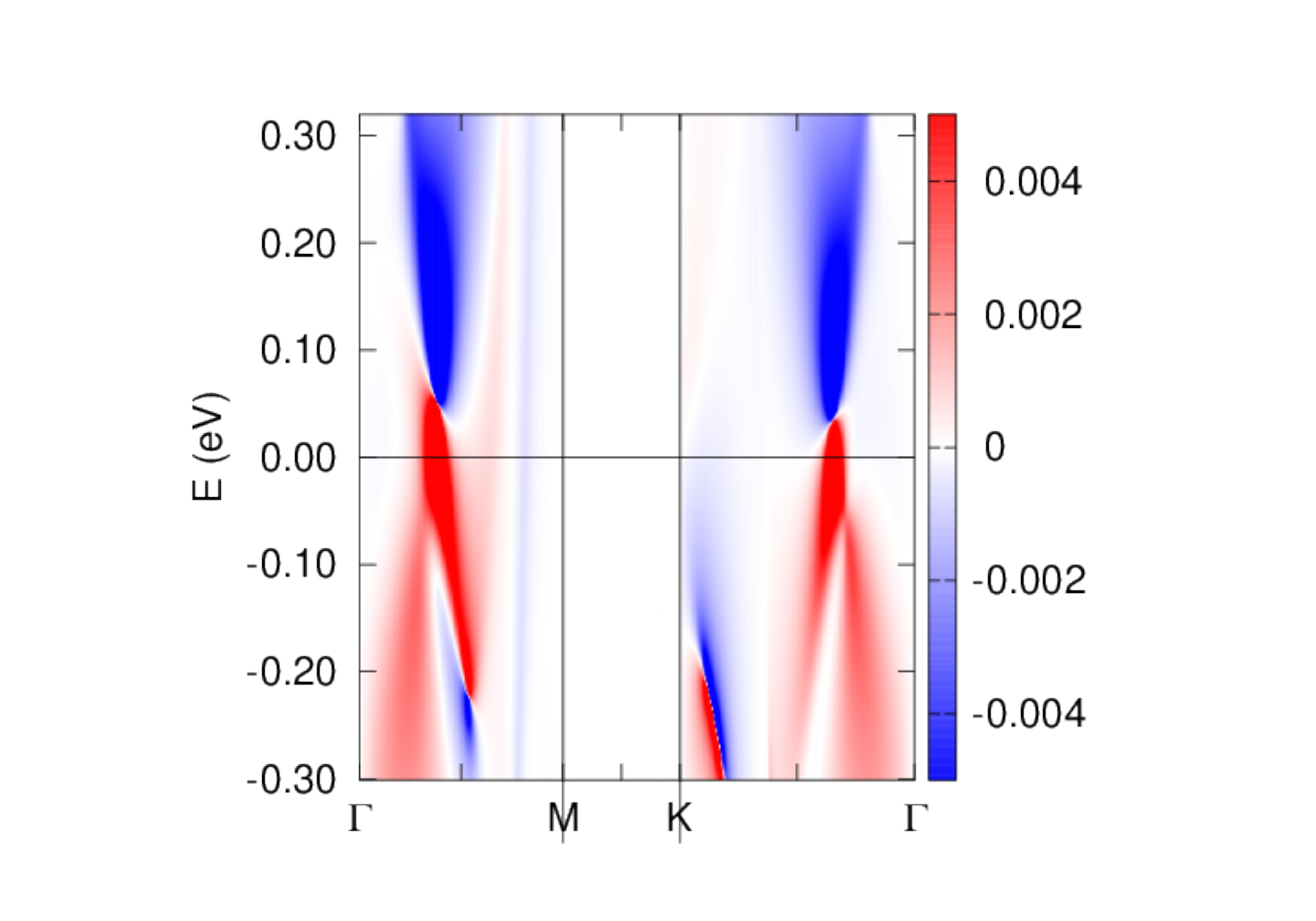}(b)
\caption{\label{fig:BSF_Fe_1MLFe} 
(a) Calculated Bloch spectral
  function $A({\cal E},\vec{k},0)$ (left panel) and $A({\cal
    E},\vec{k},E)$ with $E = 13.6\, \frac{V}{nm}$ (right panel) for
  the unsupported Fe monolayer. (b) Spin polarization along the $\hat{x}$
  direction of the electronic states in the presence of an electric field
  $E = -13.6$ (left) and $E = 13.6\, \frac{V}{nm}$ (right), }     
\end{figure}

Presenting our results, 
we start with the  free-standing Fe monolayer having a hexagonal structure with
its structure parameters corresponding to 1 ML Fe deposited on a Pt(111)
surface ($a_{lat} = 5.24$ a.u.).
The  spin magnetic moment of Fe
calculated for this 2D system is  $3.013 \mu_B$.
Considering a  monolayer film,
the system experiences accordingly
 the influence of the electric field without screening.
As it was
discussed by Nakamura et al.\ \cite{NSF+09} and Oba et al.\ \cite{ONA+15},
the external electric  field  $\vec{E} = E_z \hat{z} $ along 
the film normal $\hat{z}$, 
introduces a perturbation to the system
according to the Hamiltonian
 ${\cal H}^{(1)} = - \sqrt{\frac{4\pi}{3}} e E_z Y_{1,0}$ 
creating a coupling of the $l$- and $l \pm 1$-type  orbitals with
equal magnetic quantum number $m$, i.e.\ for examples $d_{z^2}$ and $p_z$ as well as
$d_{xz}, d_{yz}$ and $p_x, p_y$ orbitals.
This can be seen in the Bloch spectral function (BSF) plotted in 
Fig.\ \ref{fig:BSF_Fe_1MLFe}(a), 
right panel, in comparison with the BSF for
the non-distorted system shown in the left panel as a reference. 
%
 In the former case the avoided crossings can be seen for the
  energy bands in the middle of the $\Gamma-M$ and $\Gamma-M$
  directions, being a consequence of such a $p-d$ hybridization.
The change in the charge density $\rho^{(1)} \sim -\frac{1}{\pi} \mbox{Im \;
  Tr} \int d{\cal E}\; Y_{1,0}\; G({\cal E})\; {\cal H}^{(1)} \;G({\cal E})$ arising as a
response to the external electric field creates in turn an 
induced electric polarization breaking the inversion symmetry of the film.

\begin{figure}
\includegraphics[width=0.18\textwidth,angle=0,clip]{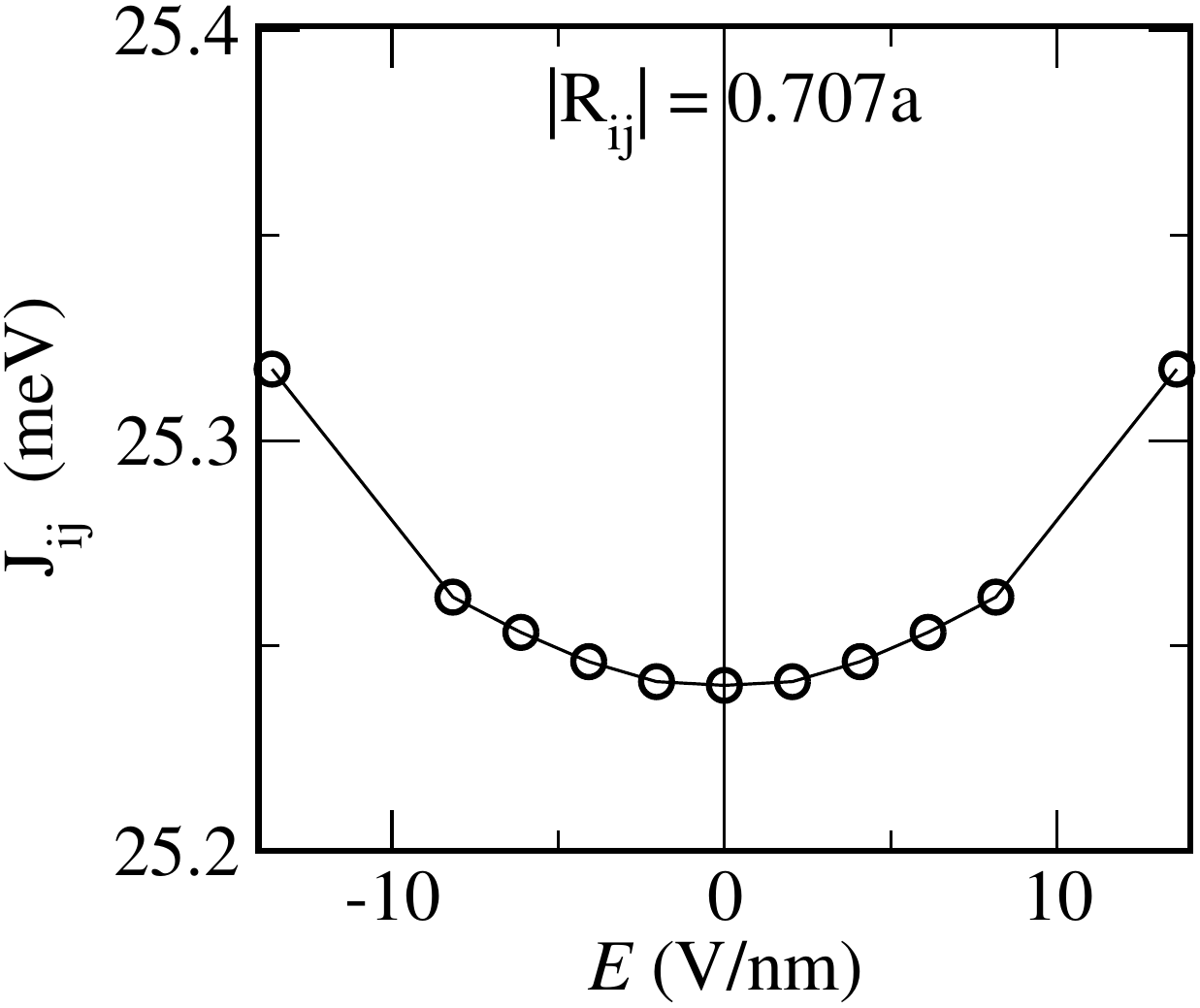}
\includegraphics[width=0.18\textwidth,angle=0,clip]{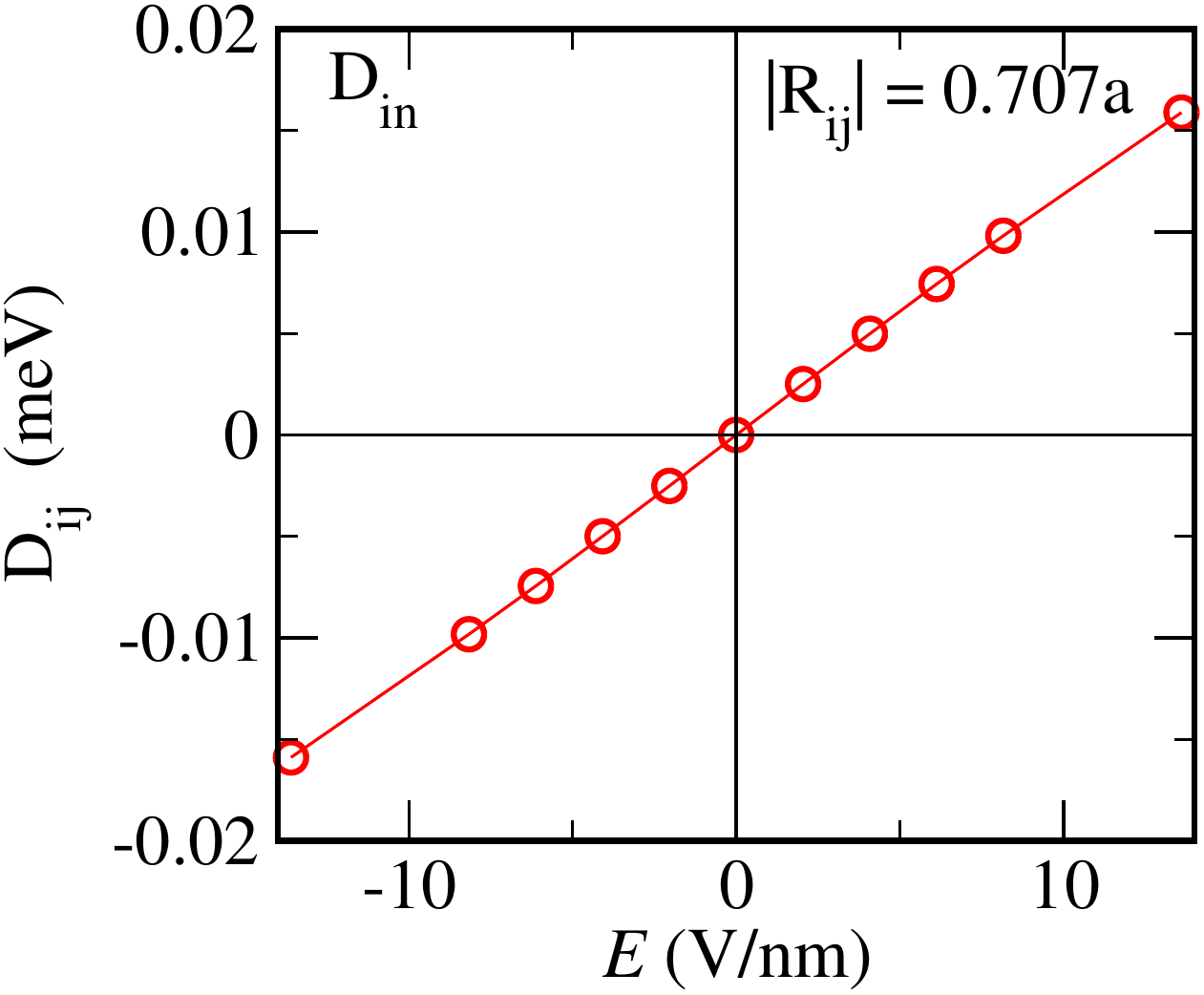}(a)
\includegraphics[width=0.18\textwidth,angle=0,clip]{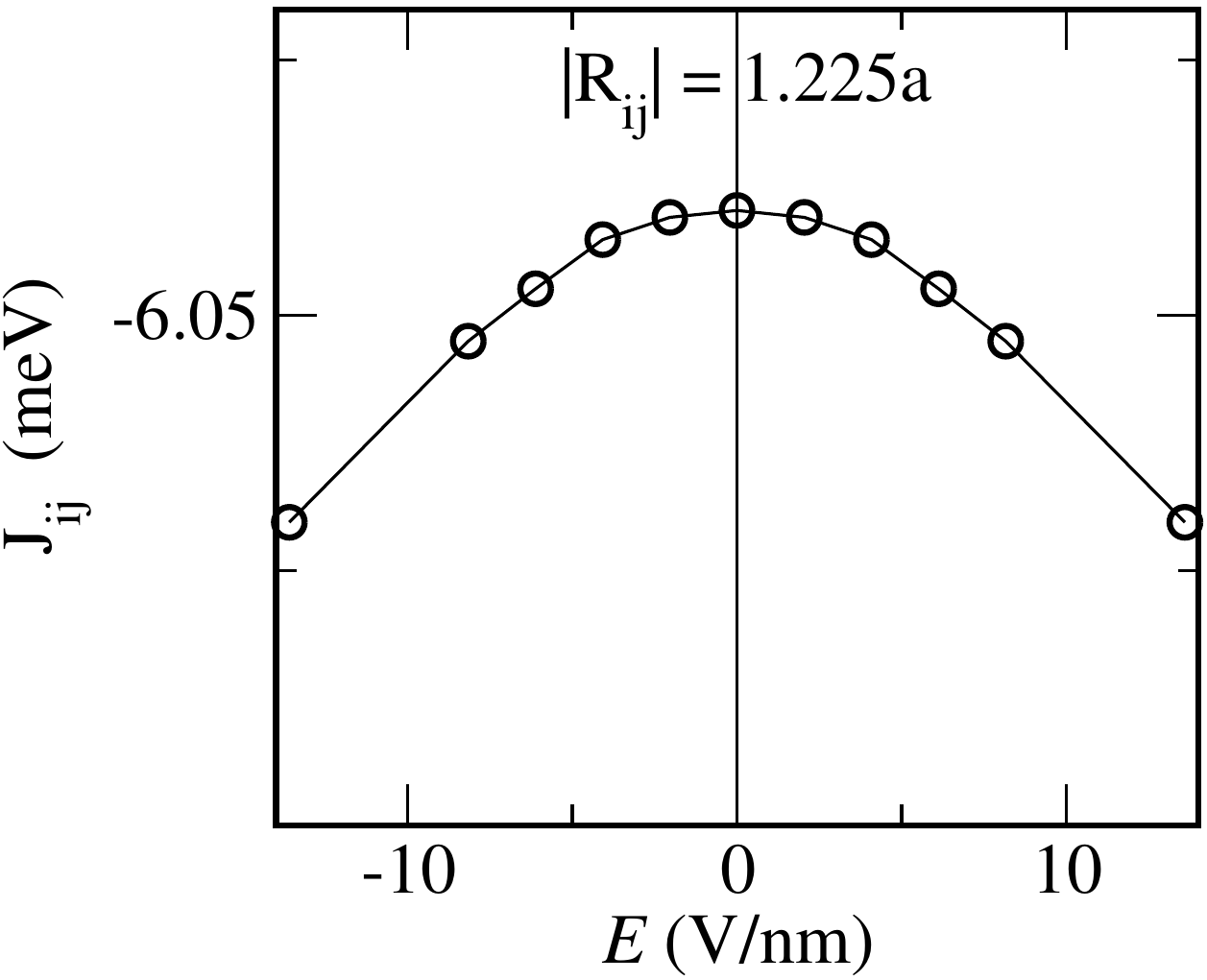}
\includegraphics[width=0.18\textwidth,angle=0,clip]{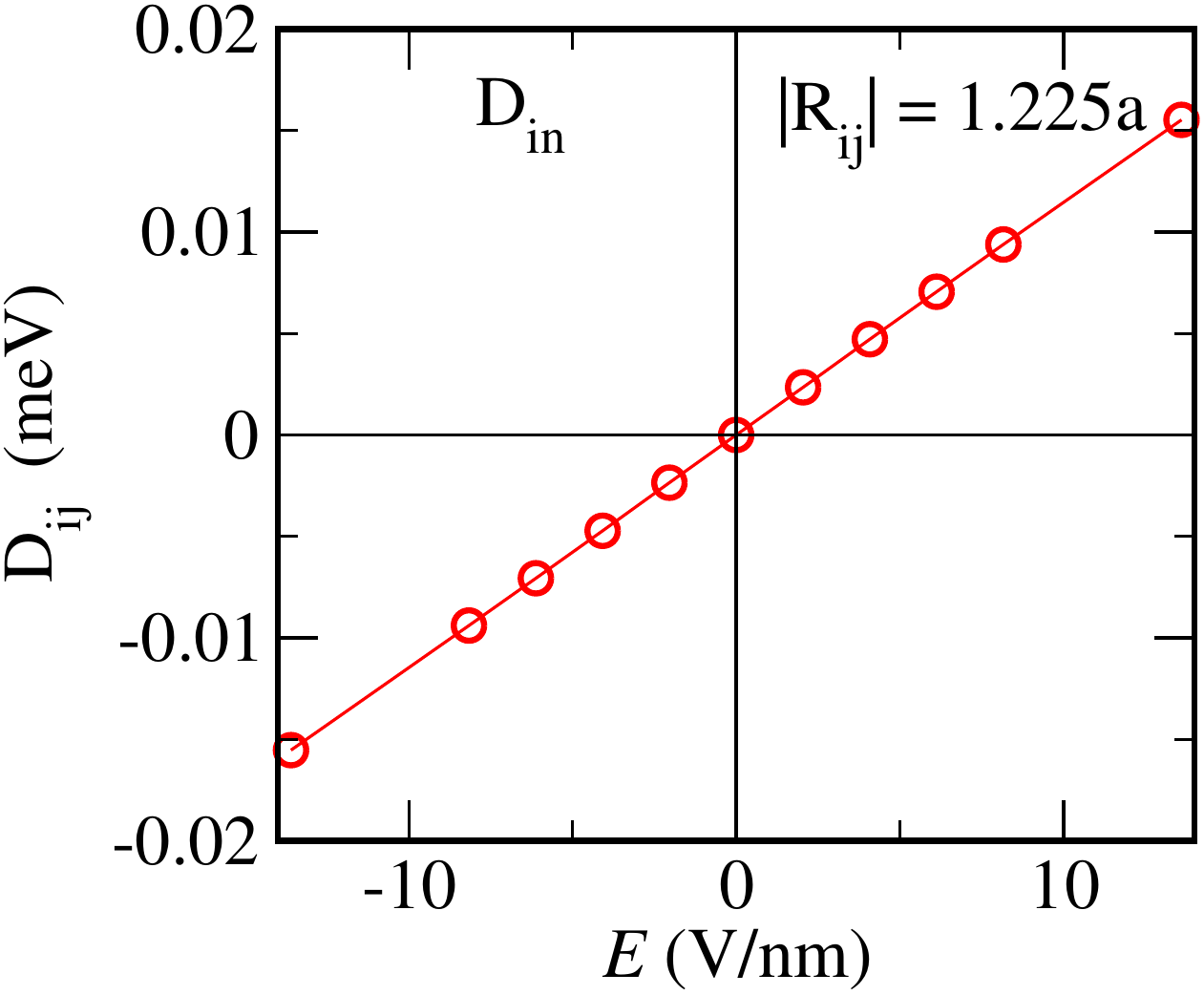}(b)
\includegraphics[width=0.18\textwidth,angle=0,clip]{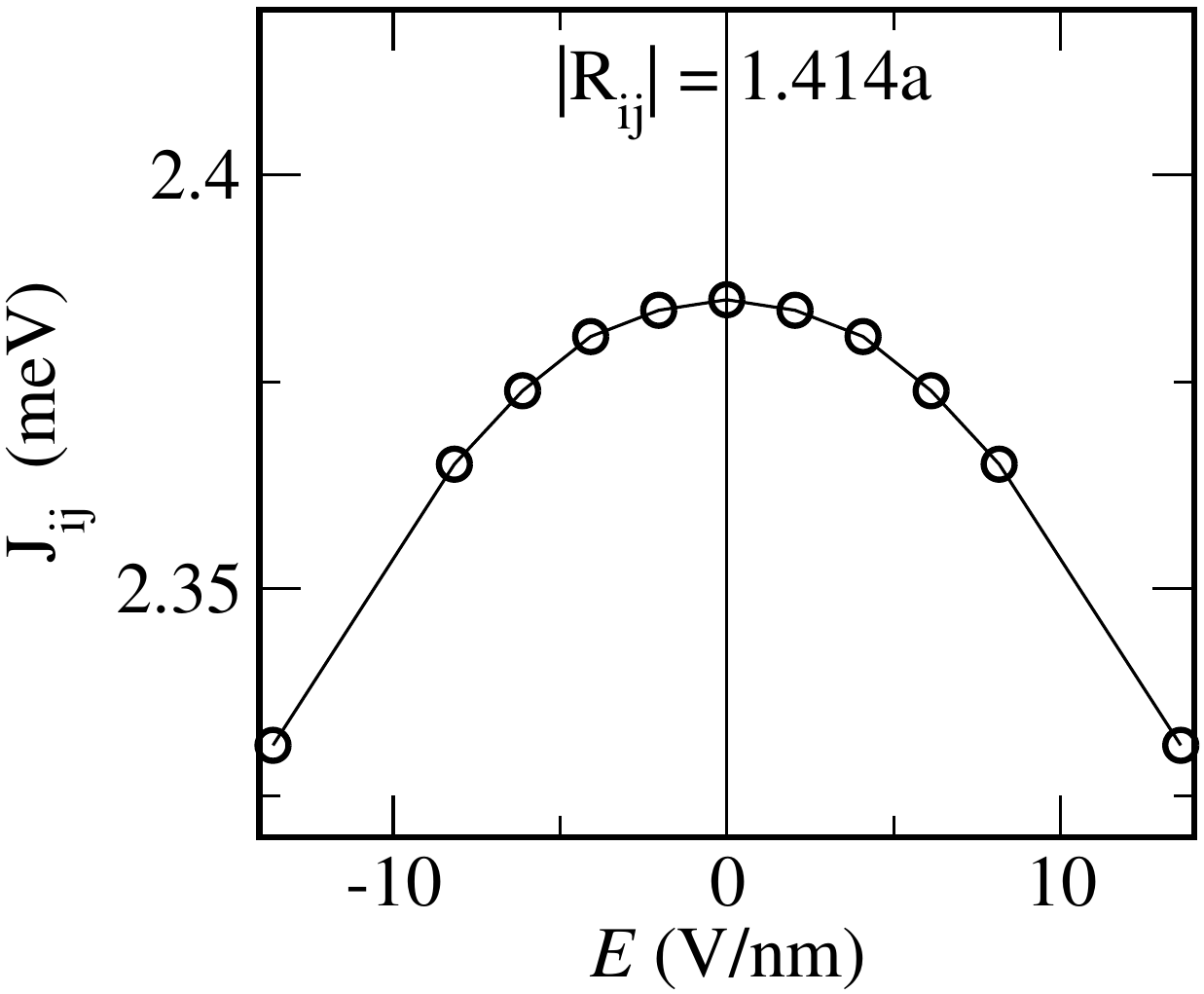}
\includegraphics[width=0.18\textwidth,angle=0,clip]{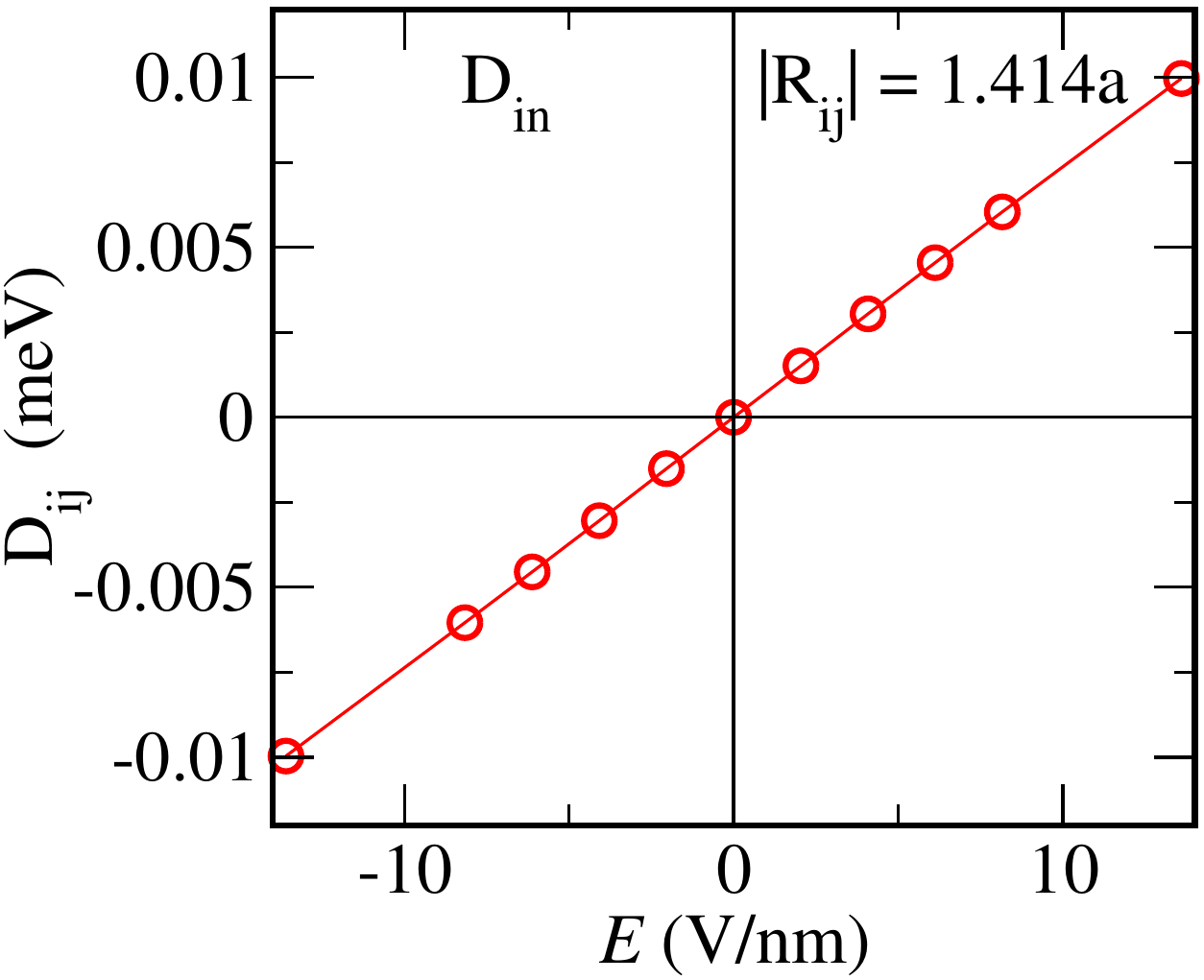}(c)
\caption{\label{fig:JXC_1MLFe} Isotropic Fe-Fe exchange coupling parameter
  $\vec{J}_{ij}$ (left panel) and the maximal in-plane component of
  DMI, $\vec{D}^{in}_{ij}$, characterizing the interactions with the
  first-neighbor at $R_{01} = 0.707a$ (a),  second-neighbor at $R_{02} =
  1.225a$ (b), and third-neighbors at $R_{03} = 1.414a$ (c), for
a  free-standing Fe monolayer.    }   
\end{figure}

As it follows from the calculations, the Fe spin magnetic moment is
practically unchanged by the applied electric field. 
Nevertheless, there is a noteworthy change for the exchange
coupling parameters, that are shown up to third atomic shells 
in Fig.\ \ref{fig:JXC_1MLFe} (left panel) 
as a function of the  electric field strength.
%
As one can see, the  weak modification of $J_{ij}$ 
due to the electric field has an almost parabolic dependency.
It can be ascribed to the shift of the Fermi level and to the 
change of hybridization of the  
localized $d$ and delocalized  $p$ electrons mediating interatomic exchange
interactions.
Considering the electric field as a perturbation, one can expand the
exchange coupling parameter in powers of the electric field. This expansion
has contributions only from even powers with respect to $\vec{E}$ to ensure a
scalar character for $J_{ij}$. This leads to a parabola-like dependence
of this quantity on the electric field. As will be shown below, the field-dependent
variation of $J_{ij}$ parameters corresponding to deposited Fe monolayers
is more complicated due to others field-induced effects having an impact on
the exchange interactions.

\begin{figure}[h]
\includegraphics[height=0.15\textwidth,angle=0,trim= 1.cm 1.cm 5.9cm
1.0cm,clip]{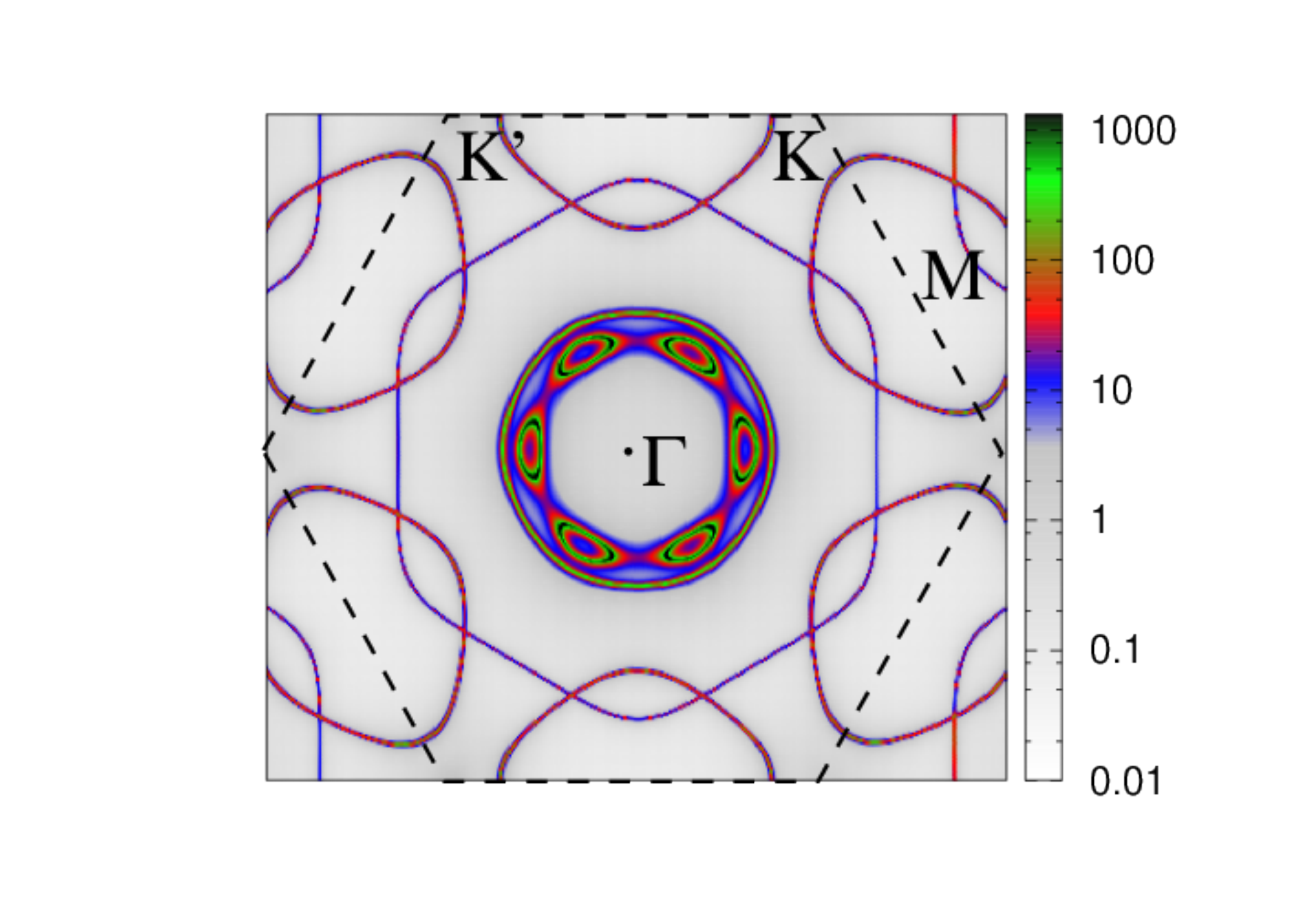}
\includegraphics[height=0.15\textwidth,angle=0,trim= 1.cm 1.cm 2cm
1.0cm,clip]{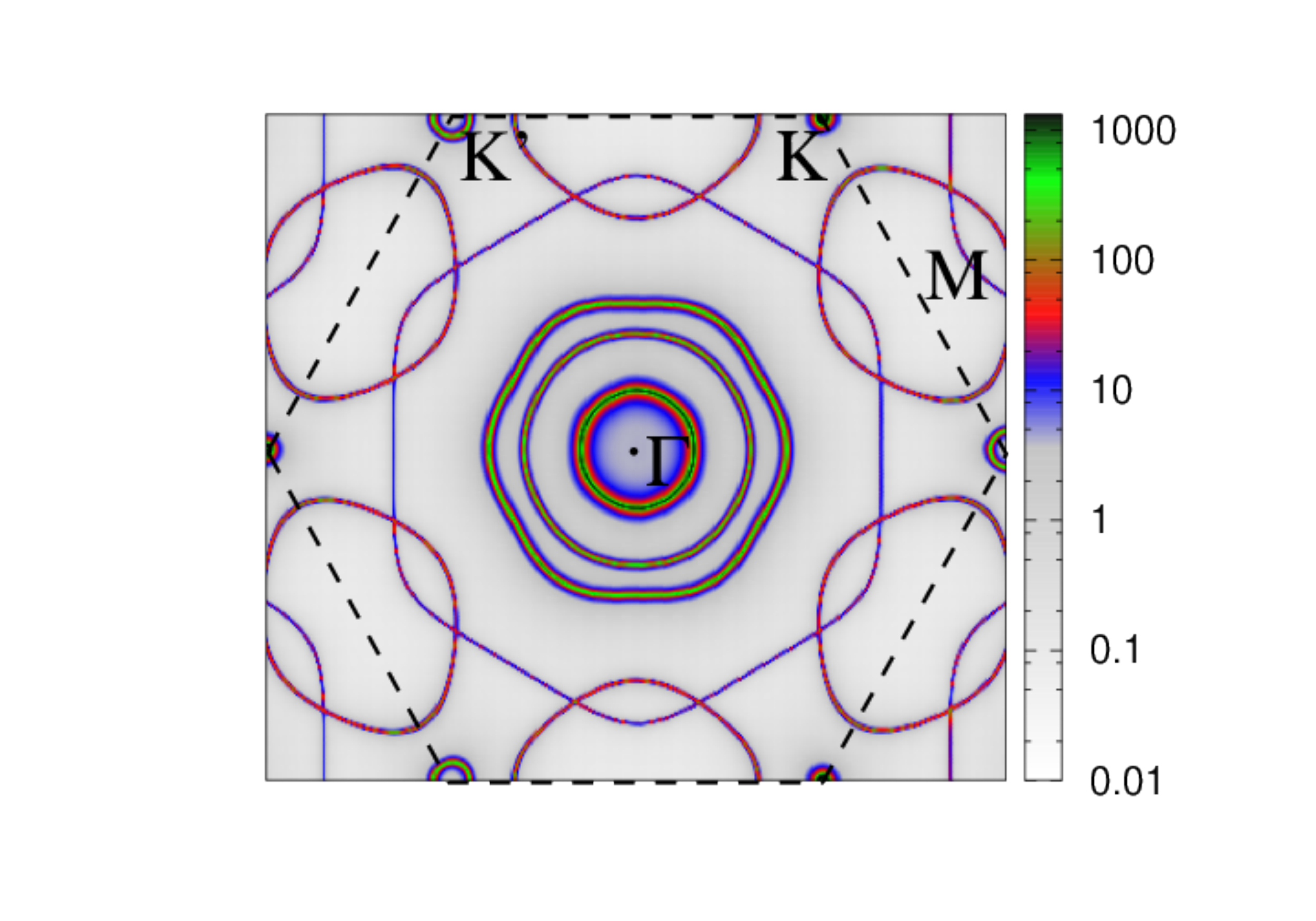}(a)
\includegraphics[height=0.15\textwidth,angle=0,trim= 1.cm 1.cm 5.9cm
1.0cm,clip]{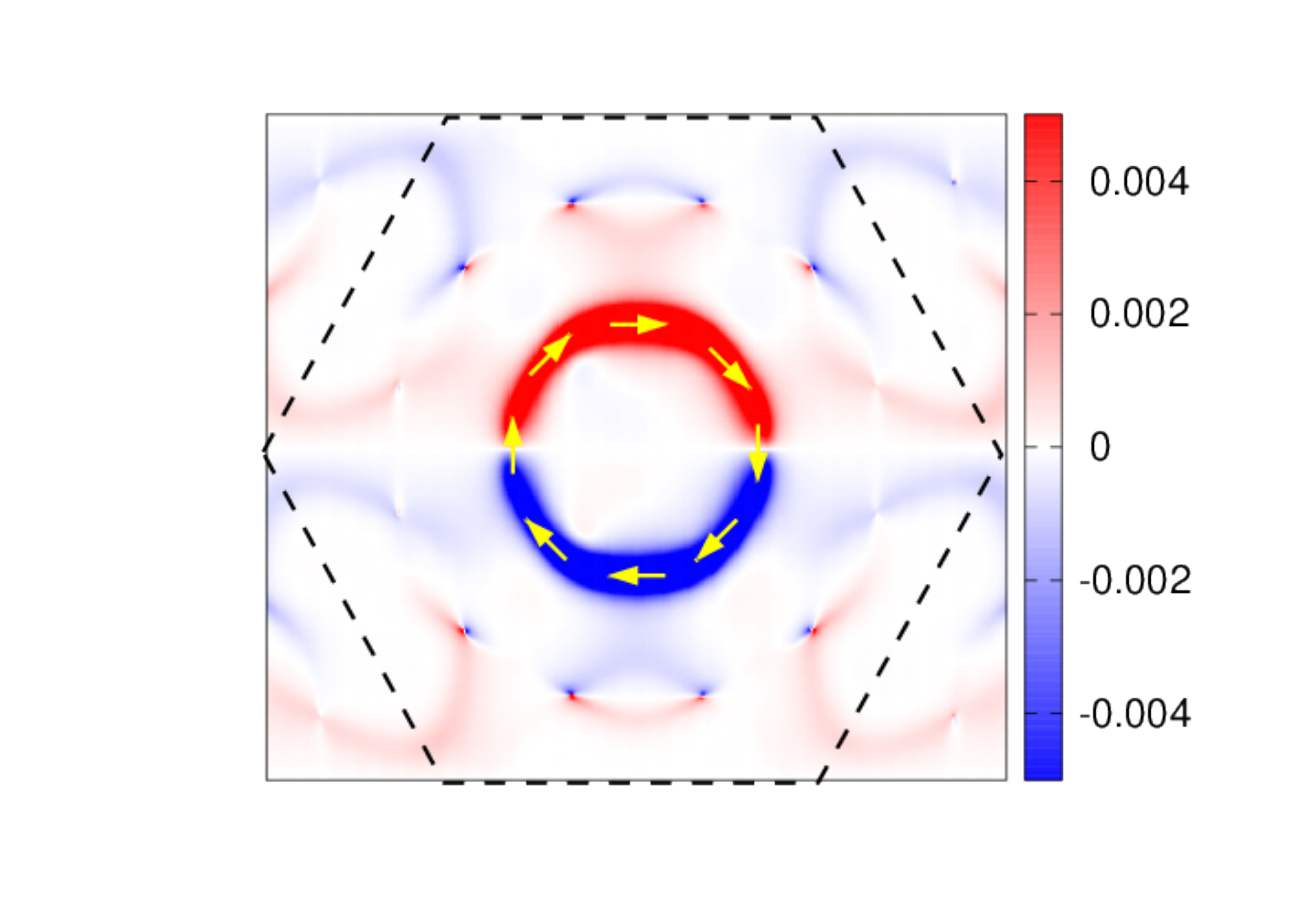}
\includegraphics[height=0.15\textwidth,angle=0,trim= 1.cm 1.cm 2cm
1.0cm,clip]{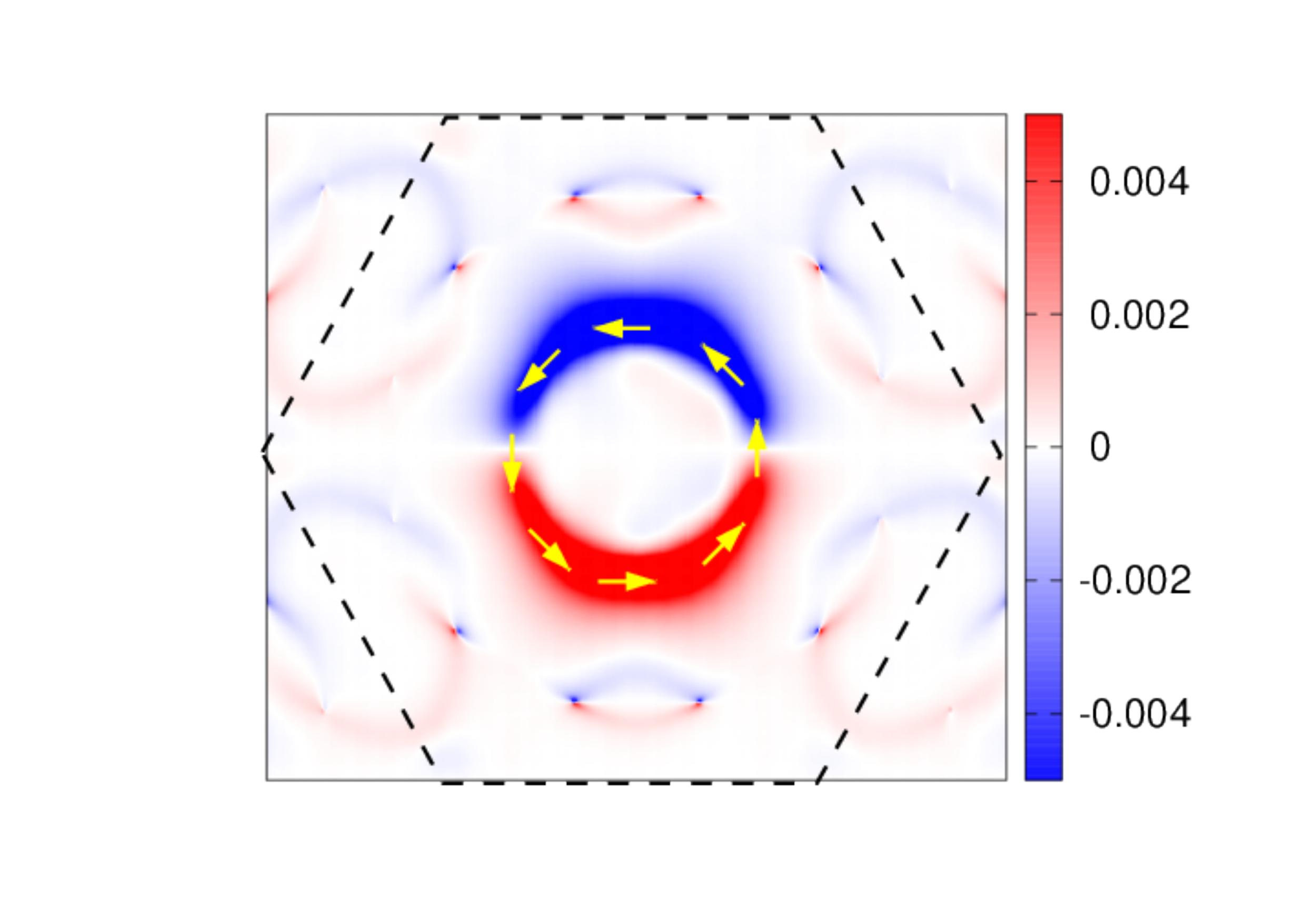}(b)
\caption{\label{fig:BSF_kk_Fe_1MLFe} 
(a) Calculated Bloch spectral
  function $A({\cal E}_F,\vec{k}_{||},E)$ (left panel) and  $A({\cal
    E}_F + \delta,\vec{k}_{||},E)$ (right panel)
   with $\delta = 0.2$ eV and $E = 13.6\, \frac{V}{nm}$, for 
  the unsupported Fe monolayer and (b) corresponding $\hat{x}$-spin
  polarization of these electronic states. }     
\end{figure}

The broken inversion symmetry due to  the presence of an external electric field
$\vec {E}$ creates a non-zero DMI, that is shown in the right panel of Fig.\
\ref{fig:JXC_1MLFe}. As one can see, the 
DMI increases linearly  with the field strength and accordingly 
changes sign when the field changes its direction. 
This behavior is associated with the field-induced Rashba SOC 
that introduces a perturbation 
to the delocalized electrons mediating the Fe-Fe exchange interaction
according to the expression \cite{BR84}
%
\begin{eqnarray}
\cal{H}_R &=& \frac{\alpha_R}{\hbar} (\hat{z} \times \vec{p}) \cdot \vec{\sigma}
\label{eq:Rashba} \; ,
\end{eqnarray}
%
 with the Rashba parameter $\alpha_R \sim E_{eff}$, where the effective
  elecric field $E_{eff}$ is created by the external and 
  induced, $E_{ind} = - \nabla_z \delta \phi[\rho^{(1)}]$, electric
  fields (where $\delta \phi[\rho^{(1)}]$ is the field-induced potential).   
   To demonstrate the impact of the Rashba-type SOC on the electronic
   structure, we plot in Fig.\ \ref{fig:BSF_Fe_1MLFe}(b) the in-plane
   spin polarization ($\hat{x}$-projection) of the electronic 
   states, which does not show up in the field-free system as well as in
   the case of SOC strength artificially scaled down to zero.  
   Moreover, this spin polarization having different directions for different 
   energy bands, changes sign to the opposite together with the electric field
   direction (similar behavior to that observed for the DMI), that can be seen by
   comparing left and right panels in  Fig.\ \ref{fig:BSF_Fe_1MLFe}(b).
   To have a more complete picture, Fig.\ \ref{fig:BSF_kk_Fe_1MLFe} (a)
   shows the calculated Bloch spectral functions $A({\cal
     E}_F,\vec{k}_{||},E)$ (left panel) and  $A({\cal 
     E}_F + \delta,\vec{k}_{||},E)$ (right panel), with $\delta = 0.2$ eV
   and $E = 13.6\, \frac{V}{nm}$, 
   representing the cut of the
   energy bands by the energy planes  ${\cal E} = {\cal E}_F$ and
   ${\cal E} = {\cal E}_F + 0.2$ eV. The corresponding
   $\hat{x}$-projected spin polarization for these energy bands is
   displayed in Fig.\ \ref{fig:BSF_kk_Fe_1MLFe} (b). It has different
   sign for the  states with ${\cal E} = {\cal E}_F$ and  ${\cal E} = {\cal E}_F + 0.2$ eV, located
   around $\Gamma$ point. Taking into account also the
   $\hat{y}$-projected spin polarization, one can reproduce the
   orientation of the  in-plane spin polarization shown in  Fig.\
   \ref{fig:BSF_kk_Fe_1MLFe} (b) by yellow arrows.   
  It would be worth noting about the  Rashba SOC induced modification of
  the electronic structure seen in  \ref{fig:BSF_kk_Fe_1MLFe} (a), right
  panel, with small electronic pockets in the $K$  and $K'$ symmetric
  points of 2D BZ have having different size in the presence of the
  electric field, but getting identical in the case of SOC switched off.


\begin{figure}
\includegraphics[width=0.12\textwidth,angle=0,clip]{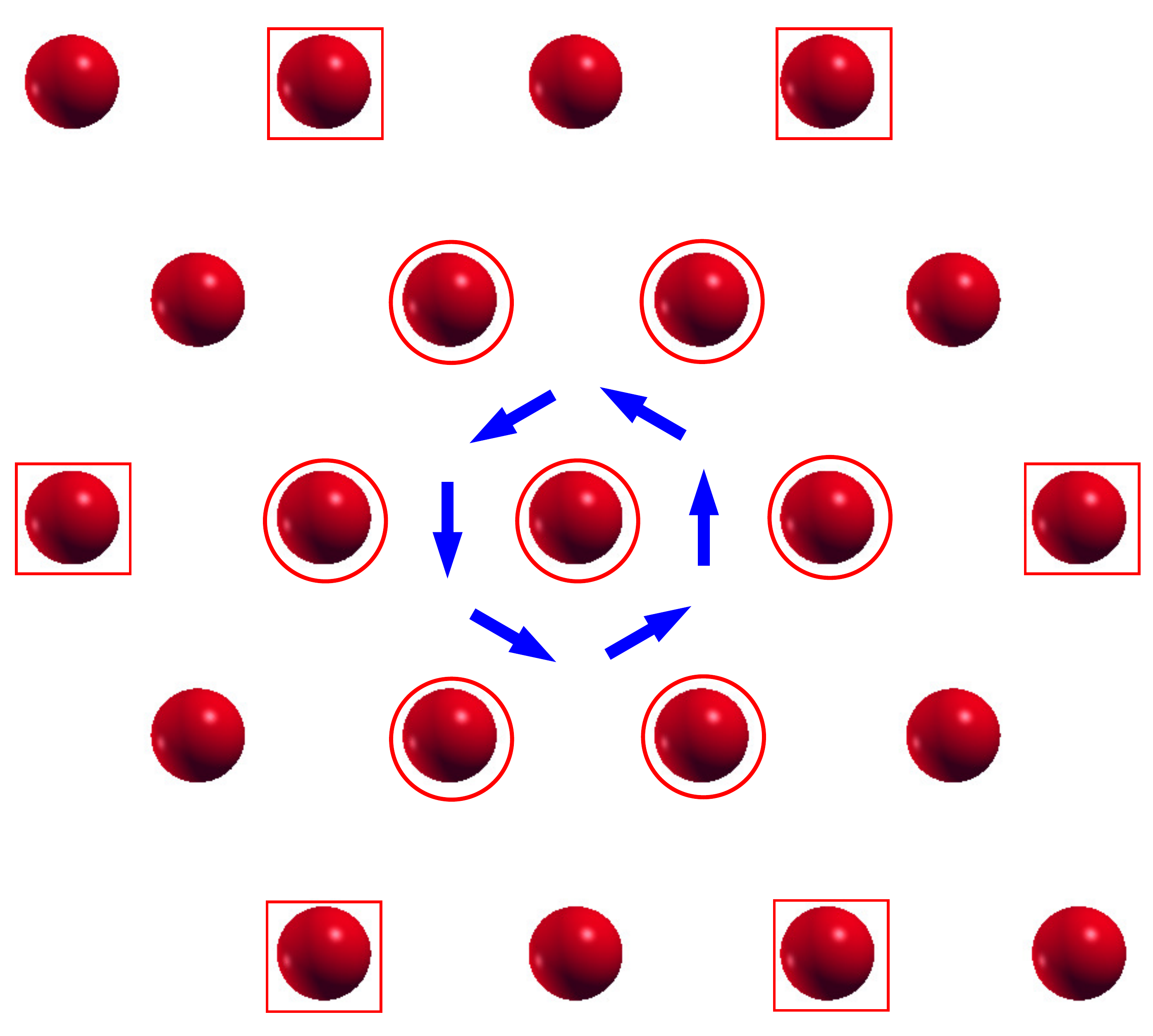}\;(a)
\includegraphics[width=0.12\textwidth,angle=0,clip]{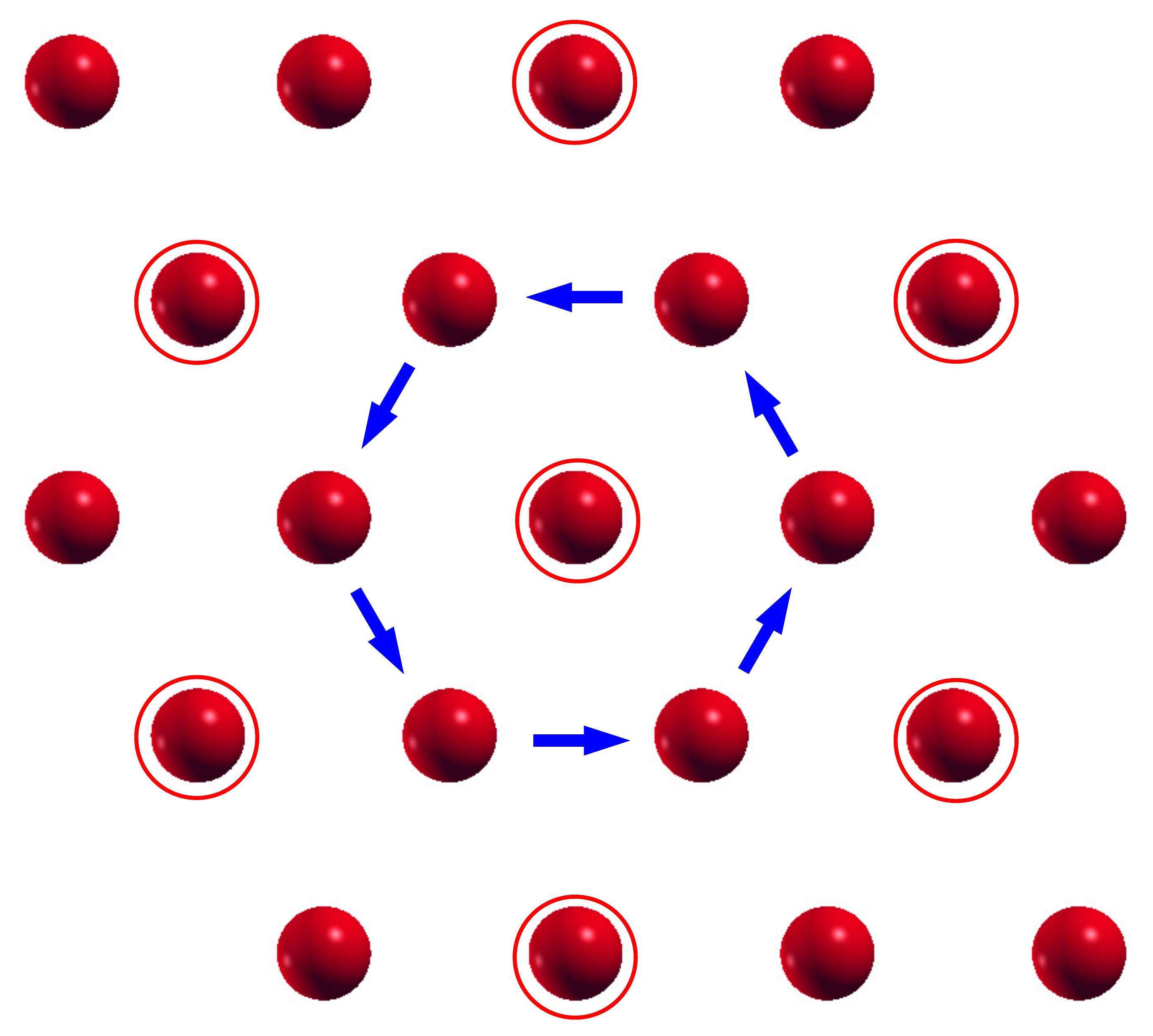}\;(b)
\includegraphics[width=0.12\textwidth,angle=0,clip]{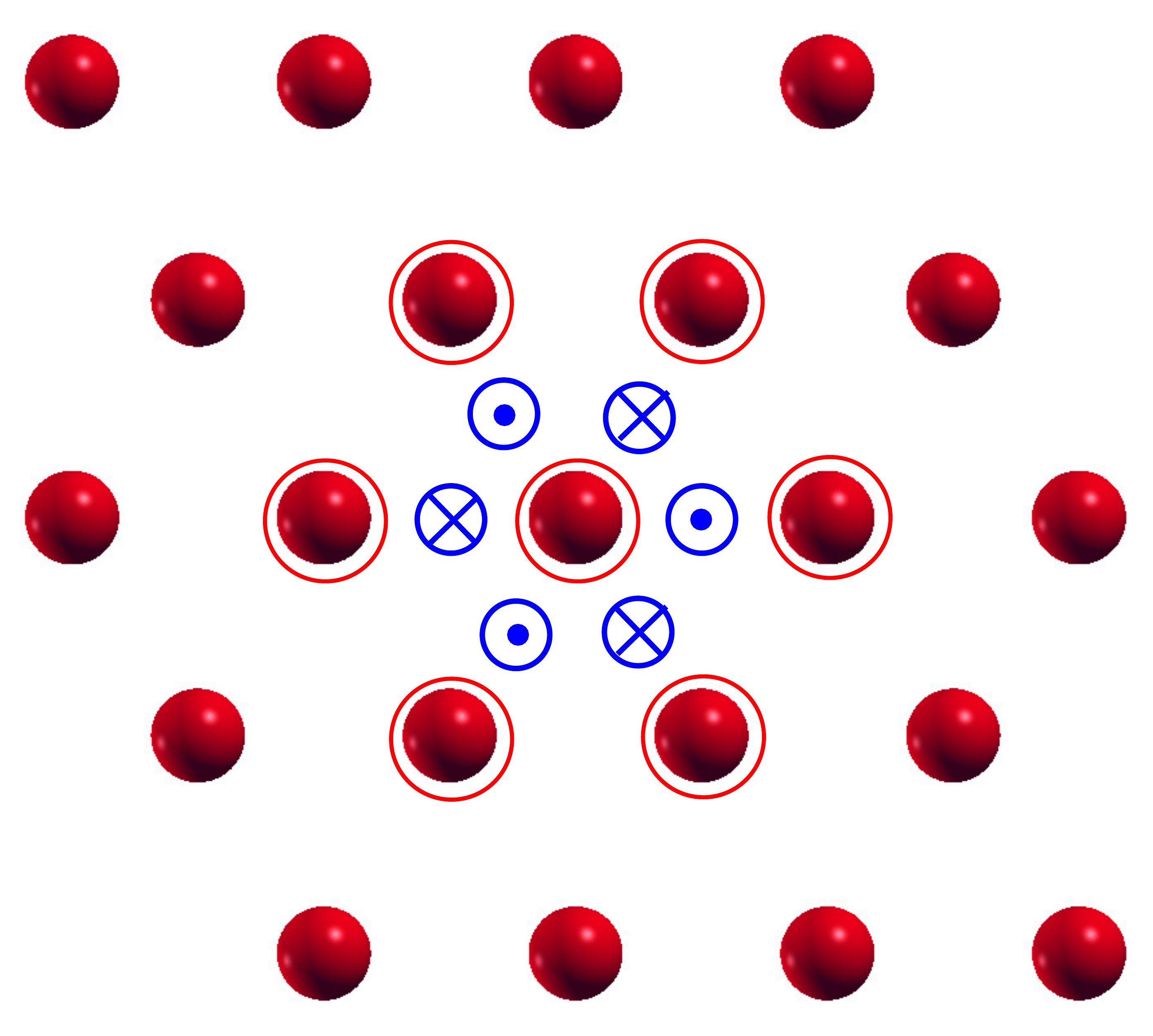}\;(c)
\caption{\label{fig:GEOM_DMI} The structure of the Fe monolayer and the
  positions of $1^{st}$- ((a), shown by cycles) $2^{nd}$- (b) and
  $3^{rd}$-neighbor ((a), shown by squares) atoms. Positive directions
  of the in-plane components of DMI, $\vec{D}^{in}$, are shown by blue
  arrows. (c) represents the out-of-plane DMI component,
  $\vec{D}^{out}$, alternatively changing sign coming from atom to atom
  within the atomic shell.}    
\end{figure}

The DMI caused by the Rashba SOC was discussed already in the
literature e.g.\ by Kundu and Zhang \cite{KZ15}.
For this reason, we only point out that
in the particular case of a free-standing Fe monolayer, all three DMI
parameters presented in Fig.\ \ref{fig:JXC_1MLFe_Pt}(right panel)
have the same order of magnitude. Positive orientations of the DMI
vectors are shown in Fig. \ref{fig:GEOM_DMI}.
%

\subsection{1 ML Fe/ Pt(111) \label{Fe-Pt}}

\begin{figure}
 \includegraphics[width=0.4\textwidth,angle=0,clip]{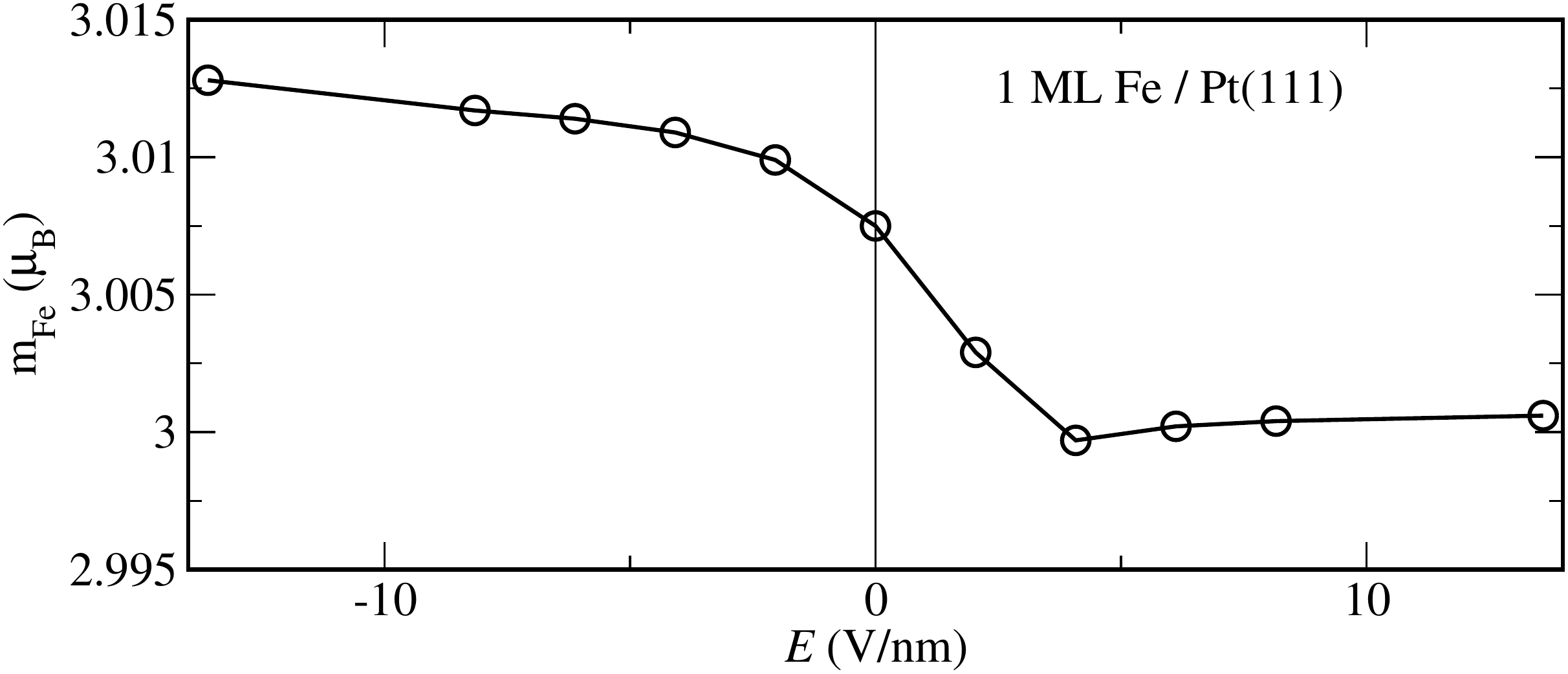}\;
 \caption{\label{fig:MMOM_1MLFe_Pt} Calculated spin magnetic moment of
   Fe, $ m_{Fe}(E)$, as a function of the external electric field $E$ for
   1 ML Fe on Pt(111).  }   
 \end{figure}

\begin{figure}[h]
\includegraphics[width=0.28\textwidth,angle=270,trim= 0 4.5cm 0 8.5cm,clip]{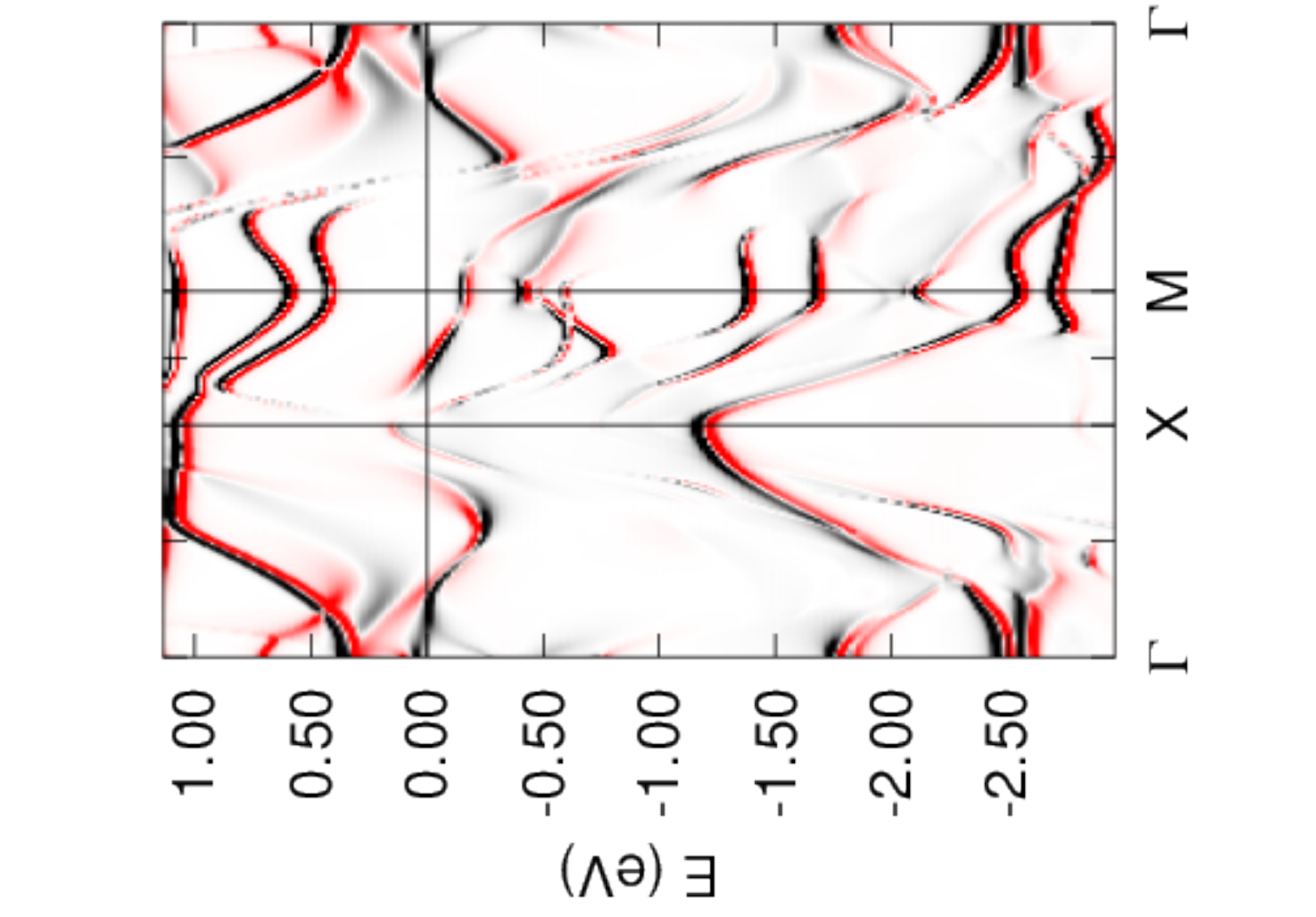}
\includegraphics[width=0.28\textwidth,angle=270,trim= 0 4.5cm 0 6cm,clip]{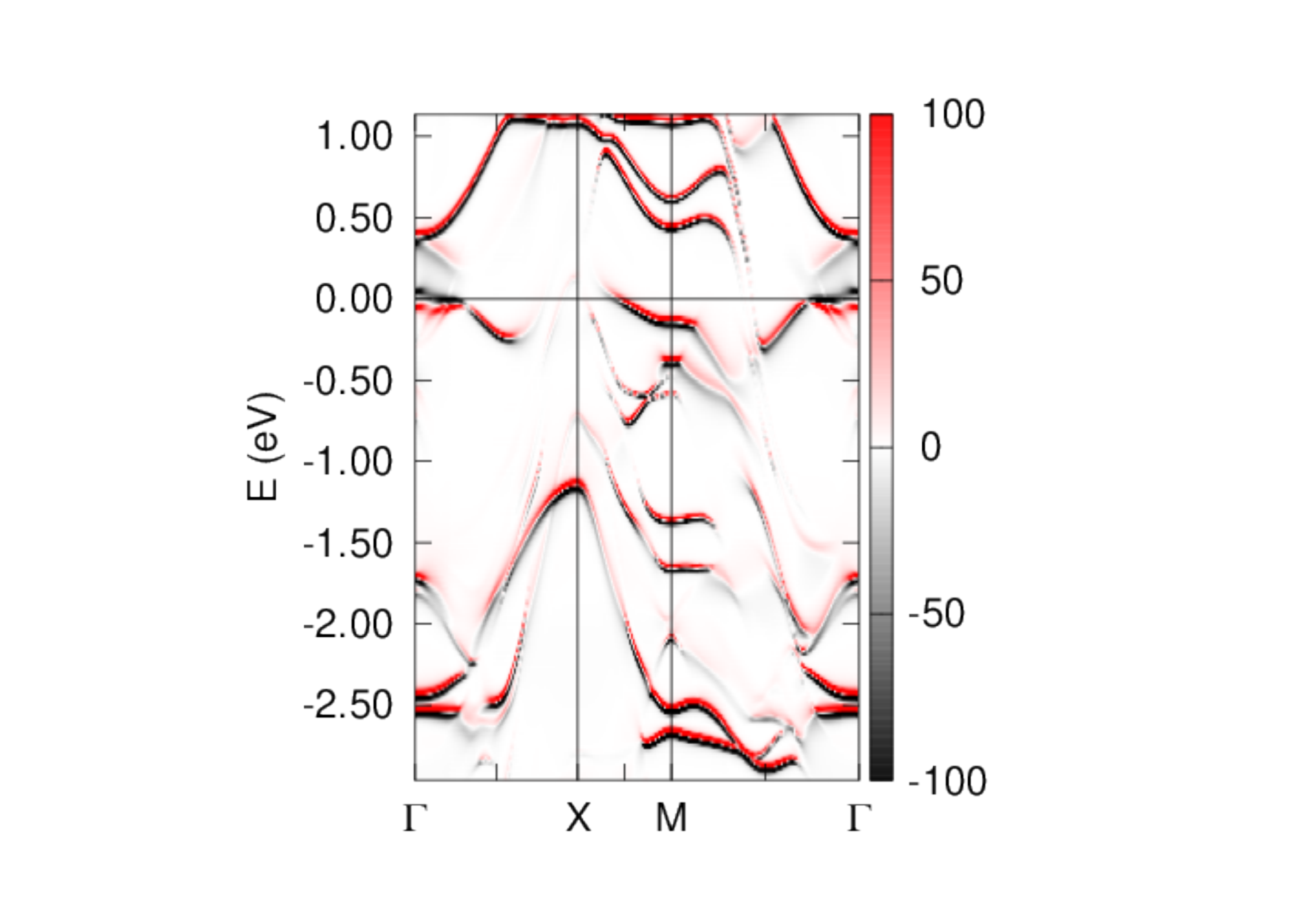}(a)
\includegraphics[width=0.28\textwidth,angle=270,trim= 0 4.5cm 0 8.5cm,clip]{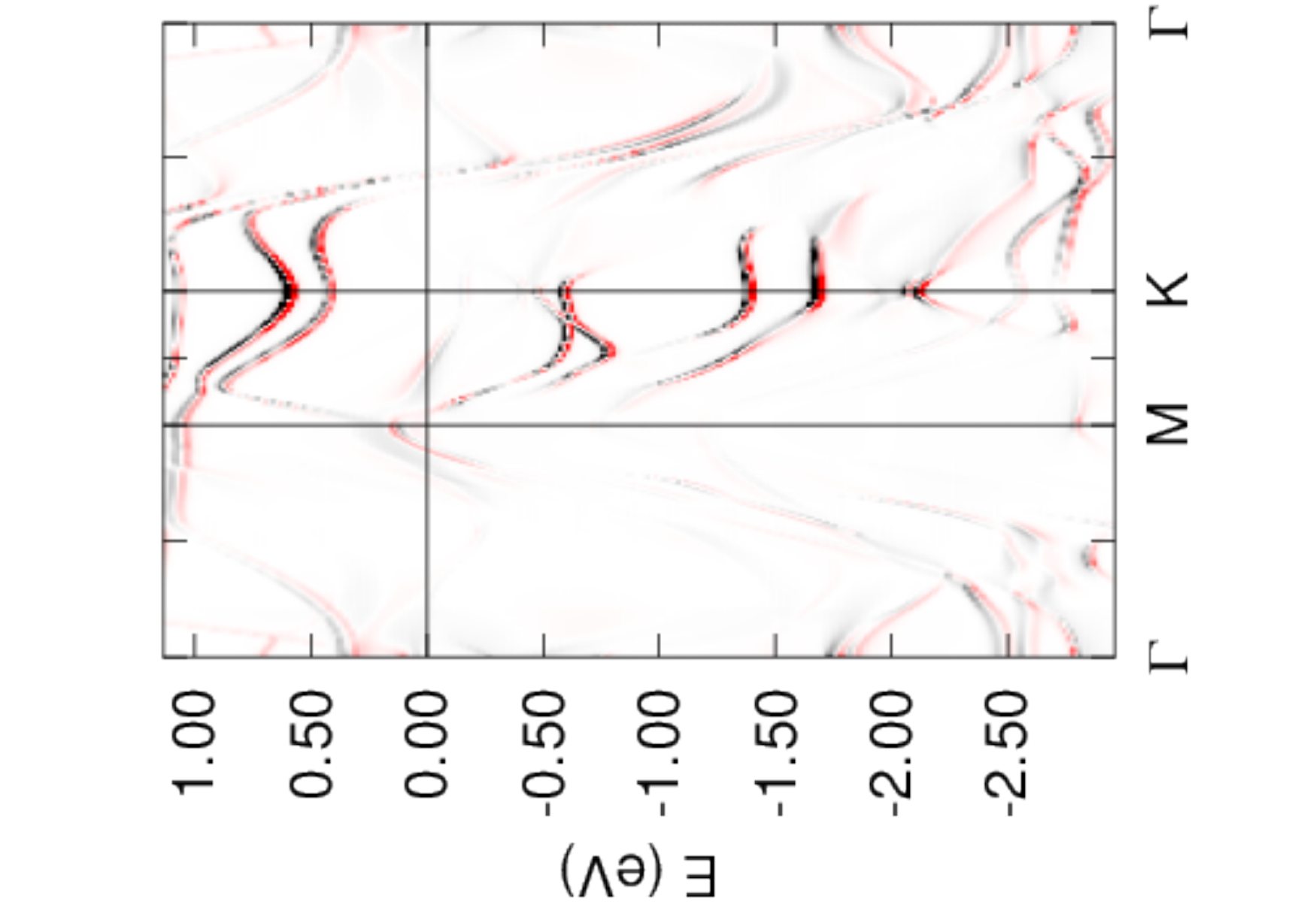}
\includegraphics[width=0.28\textwidth,angle=270,trim= 0 4.5cm 0 6cm,clip]{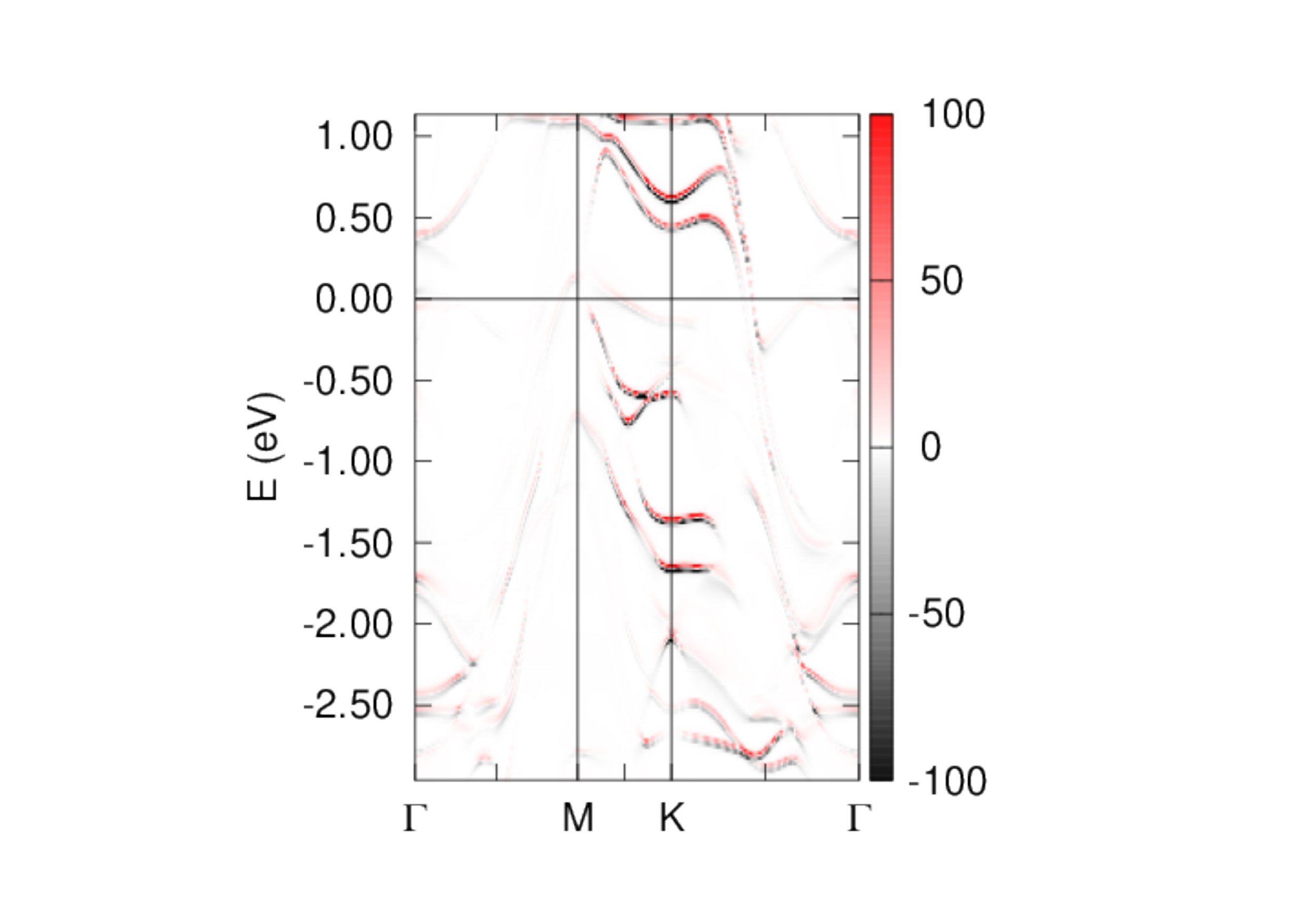}(b)
\caption{\label{fig:BSF_DIFF_Fe_1MLFe_Pt} Calculated difference in the Bloch spectral
  function $A({\cal E},\vec{k},E) - A({\cal E},\vec{k},0)$,
  demonstrating the field-induced changes of the Fe- (a) and
  Pt(I)-projected (b) electronic states (Pt(I) denotes the Pt layer at the
  Fe/Pt interface) in 1 ML Fe on Pt(111).  Red and black colors
  correspond to the modified and non-modified states, respectively, in
  the presence of negative, $E = -13.6$ (left) and positive, $E = +13.6
  \, \frac{V}{nm}$ (right).  }     
\end{figure}

Next, we consider a 1 ML Fe film on a Pt(111) substrate.
The change of the spin magnetic moment in the Fe monolayer, $m_{Fe}$, 
due to an applied electric field is shown in Fig.\ \ref{fig:MMOM_1MLFe_Pt}.
%
%
The most significant change occurs at a small strength of the
electric field, with an almost linear dependence of $m_{Fe}(E)$ on the
electric field strength. 
With further increasing field strength, $|\vec{E}| > 5 \frac{V}{nm}$, the
magnetic moment shows only a weak variation with  the field.
To demonstrate  the  impact of the electric field on the electronic structure, 
we plot in 
Fig.\  \ref{fig:BSF_DIFF_Fe_1MLFe_Pt}
 the difference in the Bloch spectral
function $A({\cal E},\vec{k},E) - A({\cal E},\vec{k},0)$,  representing the
changes of the (a)  Fe- and (b) Pt(I)-projected electronic states due to
a negative ($E = -13.6 \, \frac{V}{nm}$; left) and positive
($E = 13.6 \, \frac{V}{nm}$; right) electric field with 
Pt(I) denoting  the Pt atoms at Fe/Pt interface. 
%
As one can see, the Fe layer experiences the strongest influence of the
electric field. 
The modification of the band structure due to the
electric field for 1 ML Fe on Pt(111) is obviously
 more complex than for the free-standing Fe monolayer. 
As it was already discussed previously
\cite{SMME21}, one can distinguish several mechanisms for
 the observed  field-induced
changes in the electronic structure. First of all,
one notes  a shift of the electronic
states, which depends on their distance to  the
surface. This shift is the strongest for the states well localized in
the surface region with an unscreened or weakly screened electric field.
In Fig.\ \ref{fig:BSF_DIFF_Fe_1MLFe_Pt}(a) one can see 
in particular a pronounced
field-induced shift for the Fe $d$-states, while the shift
for  the Pt states at the Pt/Fe interface is weaker.
This is a result of a partial screening of the electric field as well as 
a weaker localization of the electron Pt states.

As it was pointed out above,
the applied electric field changes
in addition
 the hybridization\cite{ONA+15} of
the $p$- and $d$- states. This effect however is
hard to see in Fig.\ \ref{fig:BSF_DIFF_Fe_1MLFe_Pt}
because of the strong modification of the electronic states of deposited Fe
when compared to a single Fe monolayer. 

Finally, one notes that the bulk-like Pt states are almost unmodified. 
As a
result, the field induced shifts of the Fe $d$-states having
  essentially 2D character are accompanied by
a corresponding 
broadening, as it is seen in Fig.\
\ref{fig:BSF_DIFF_Fe_1MLFe_Pt},
which is a consequence of their modified hybridization with the
  bulk-like Pt energy bands.

\begin{figure}
\includegraphics[width=0.19\textwidth,angle=0,clip]{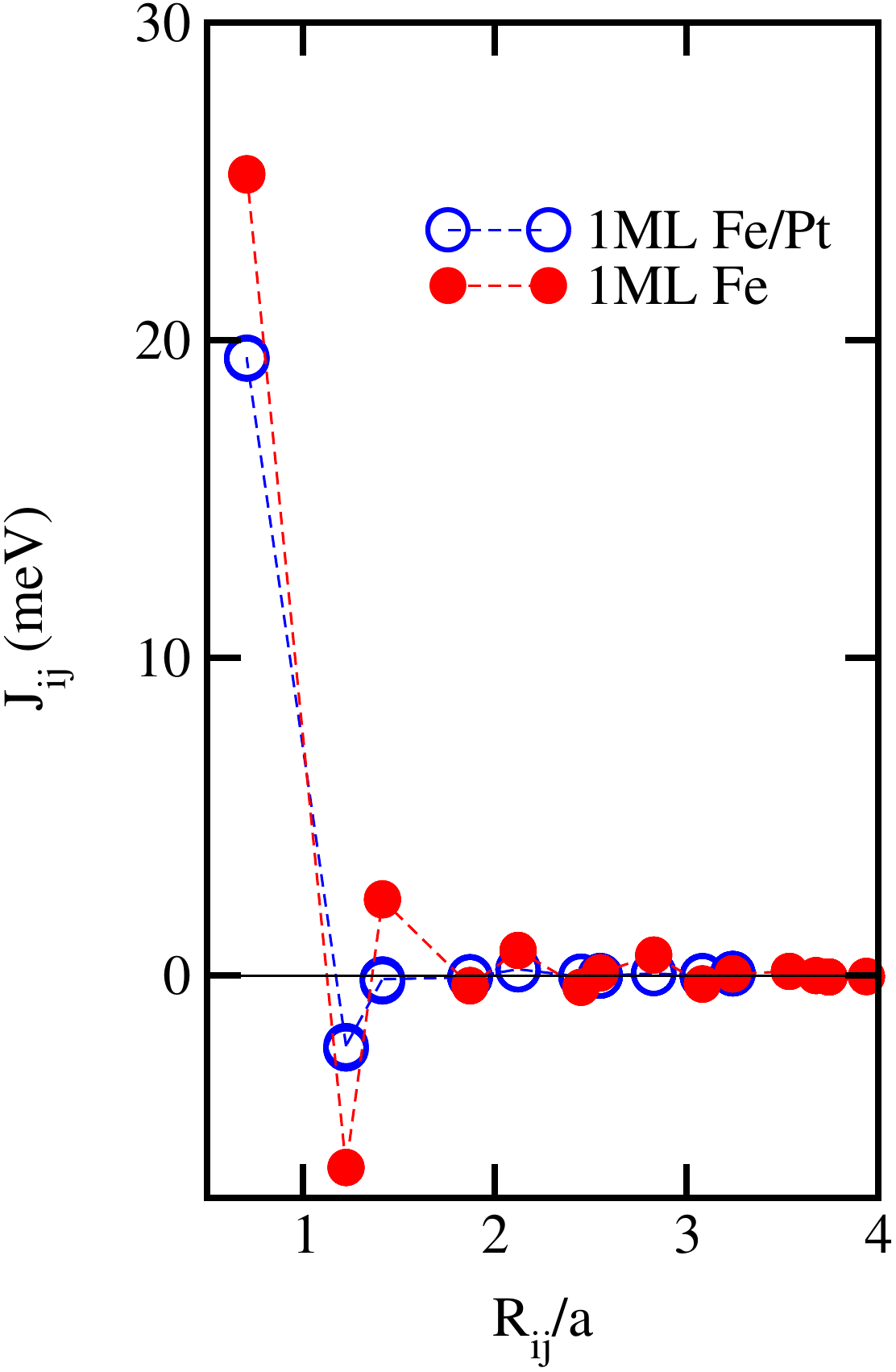}\;
\includegraphics[width=0.19\textwidth,angle=0,clip]{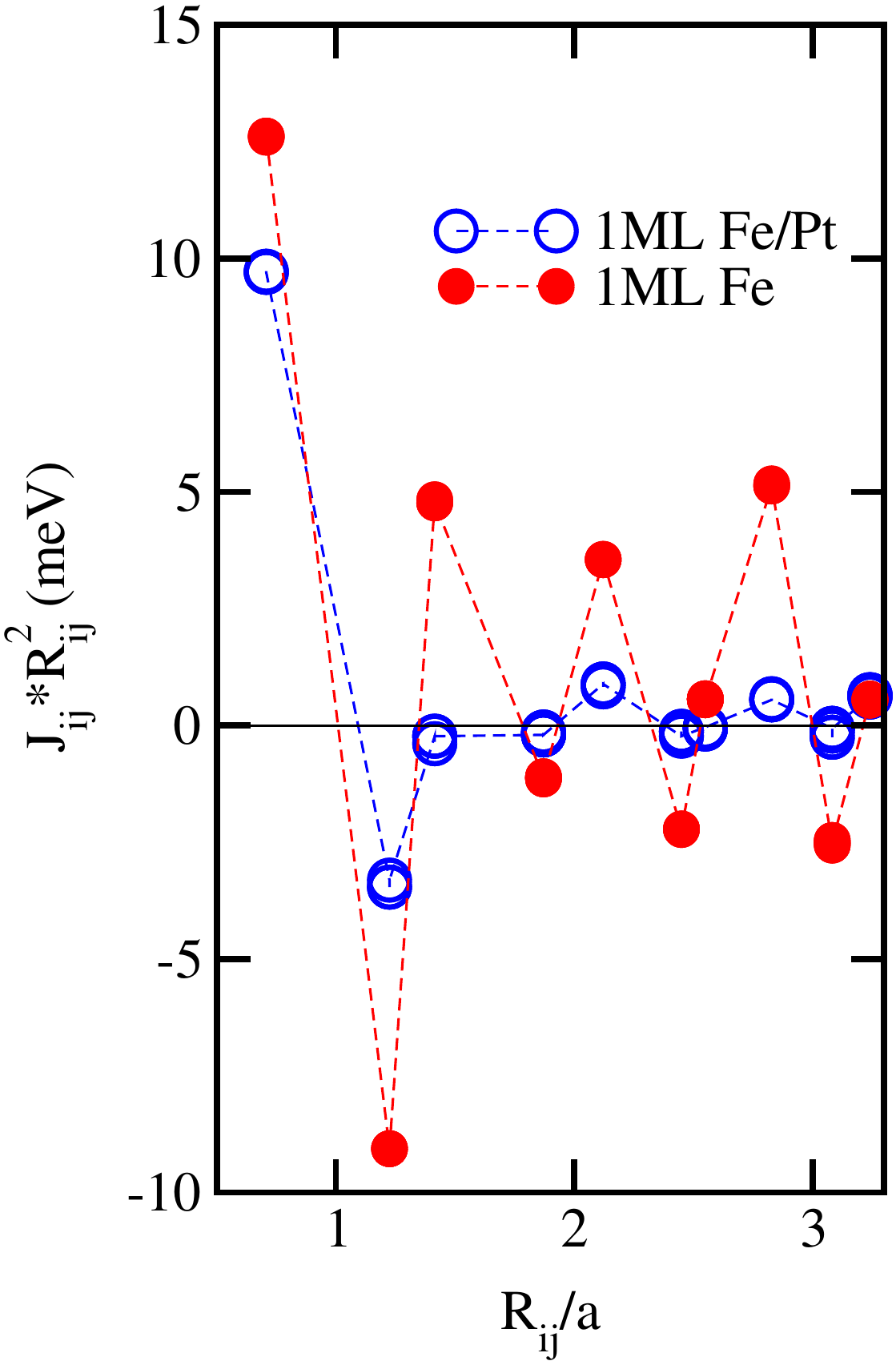}\;
\caption{\label{fig:JXC_Fe-Pt} Isotropic Fe-Fe exchange coupling parameter
  $J_{ij}$ as a function of the interatomic distance   $R_{ij}$ 
for a   non-distorted
  1 ML Fe  on Pt(111) (open symbols) and for a free-standing Fe monolayer
  (full symbols) (left panel).
The right panel represents the product 
  $J_{ij}\, R^2_{ij}$ as a function of the squared  distance $R^2_{ij}$.  }    
\end{figure}

\begin{figure}
\includegraphics[width=0.2\textwidth,angle=0,clip]{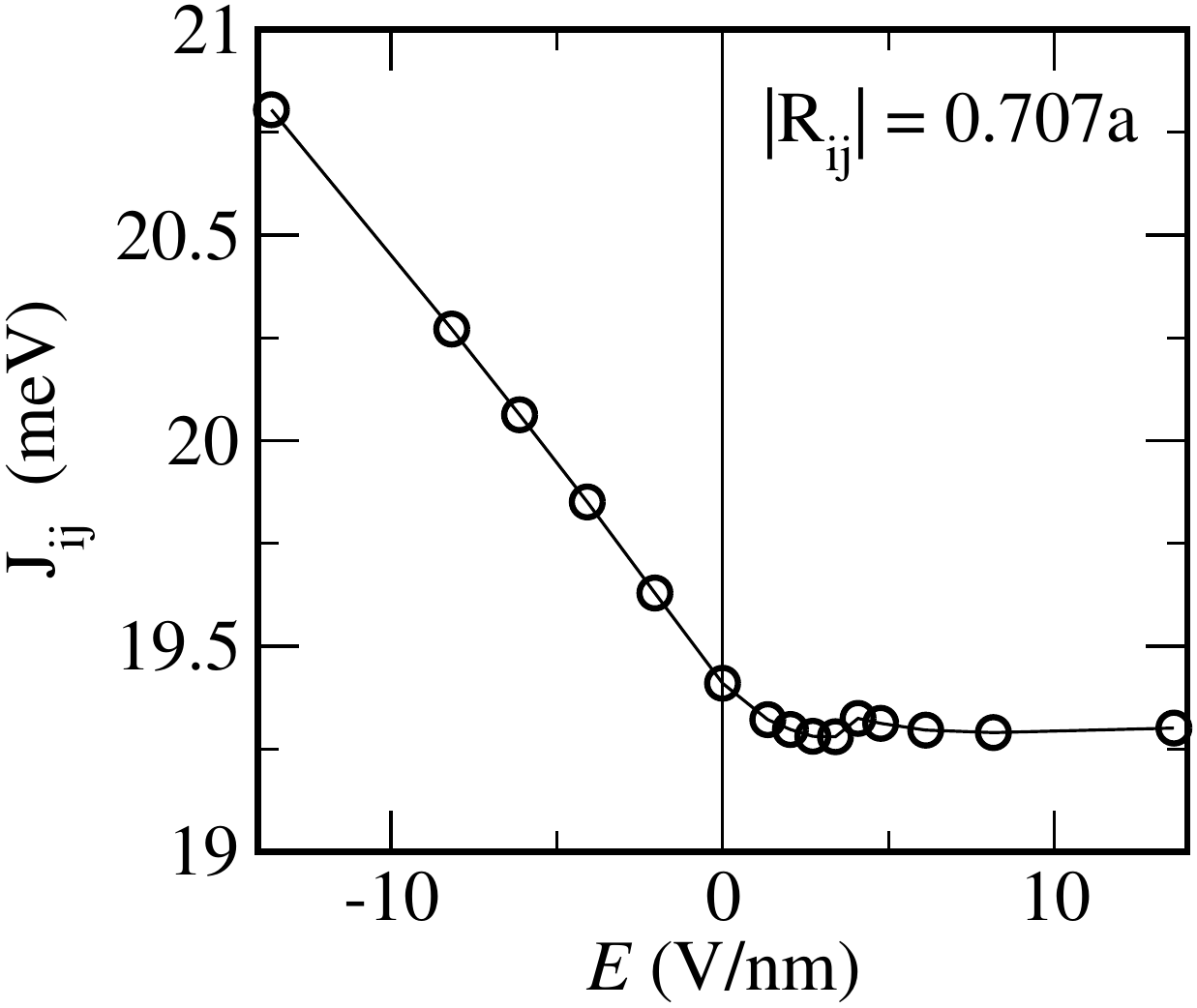}\;
\includegraphics[width=0.2\textwidth,angle=0,clip]{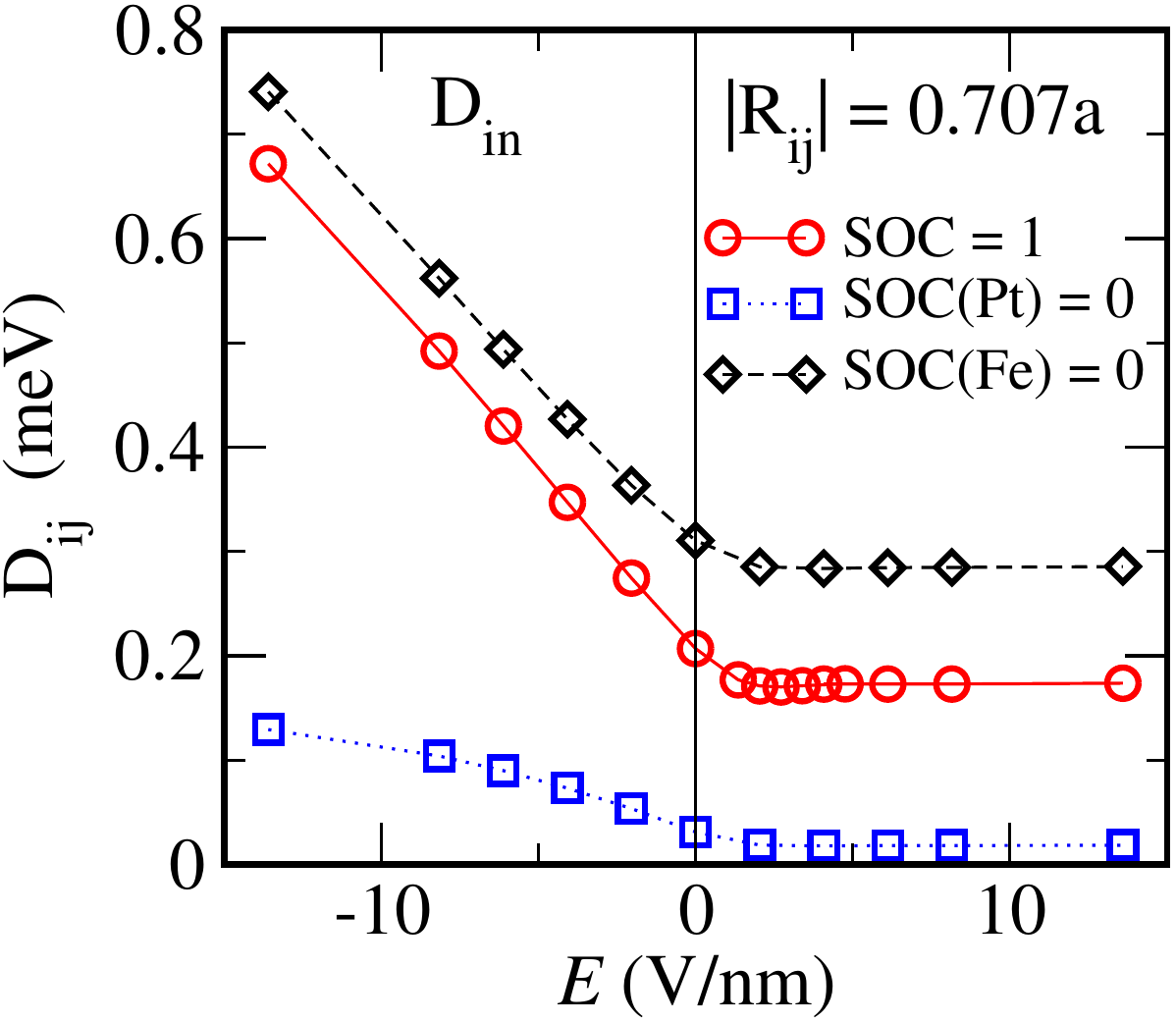}\;(a)
\includegraphics[width=0.2\textwidth,angle=0,clip]{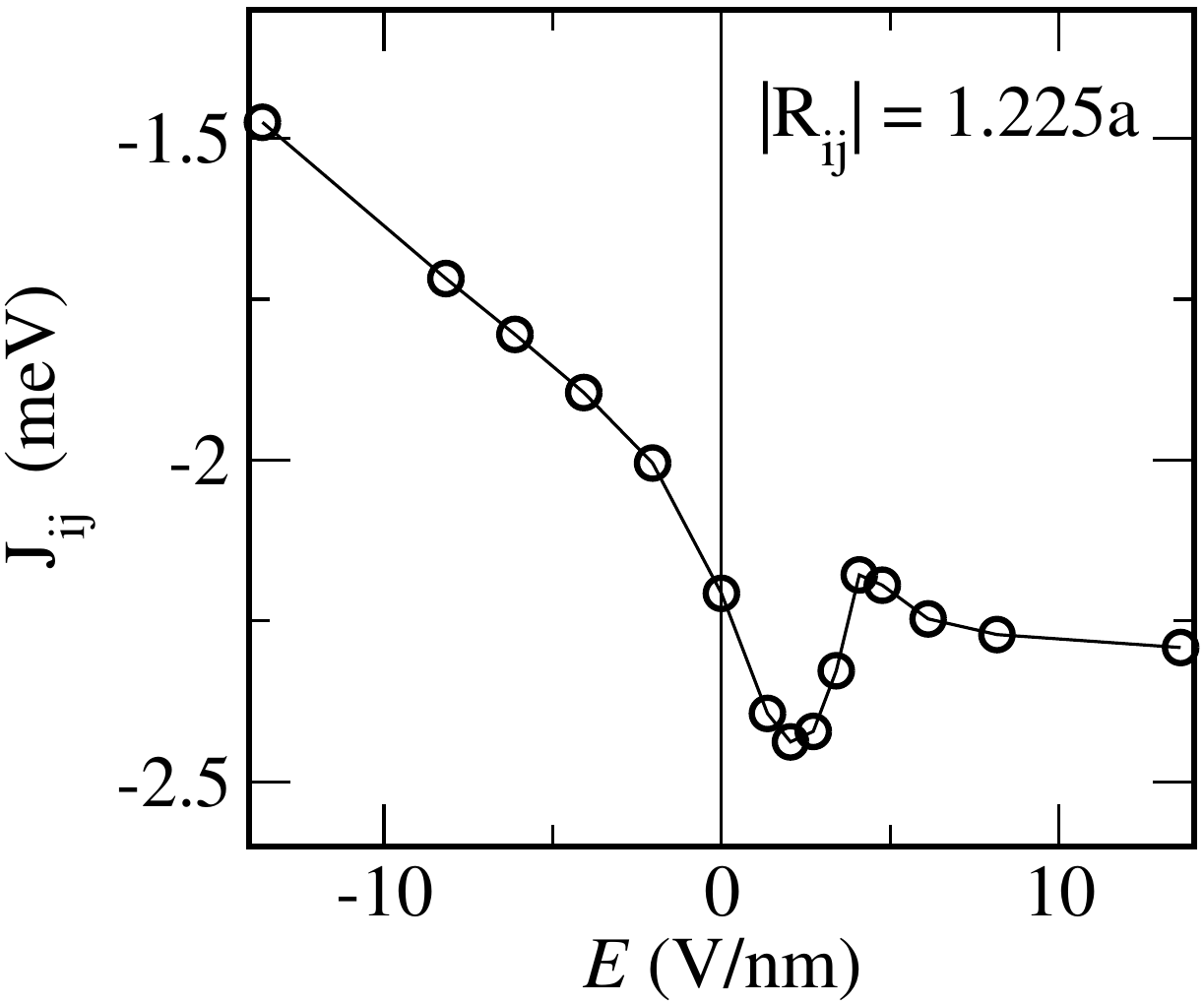}\;
\includegraphics[width=0.2\textwidth,angle=0,clip]{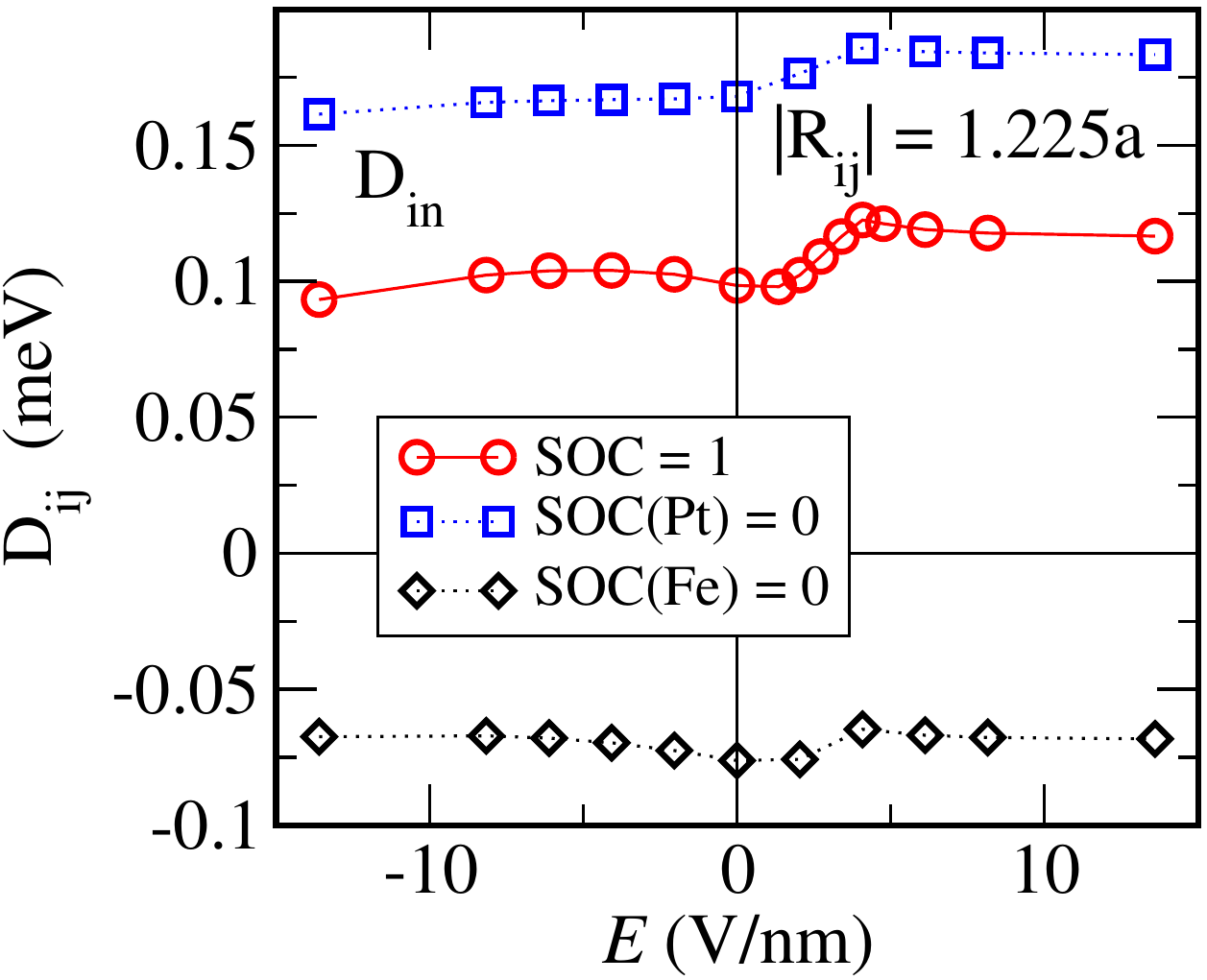}\;(b)
\includegraphics[width=0.2\textwidth,angle=0,clip]{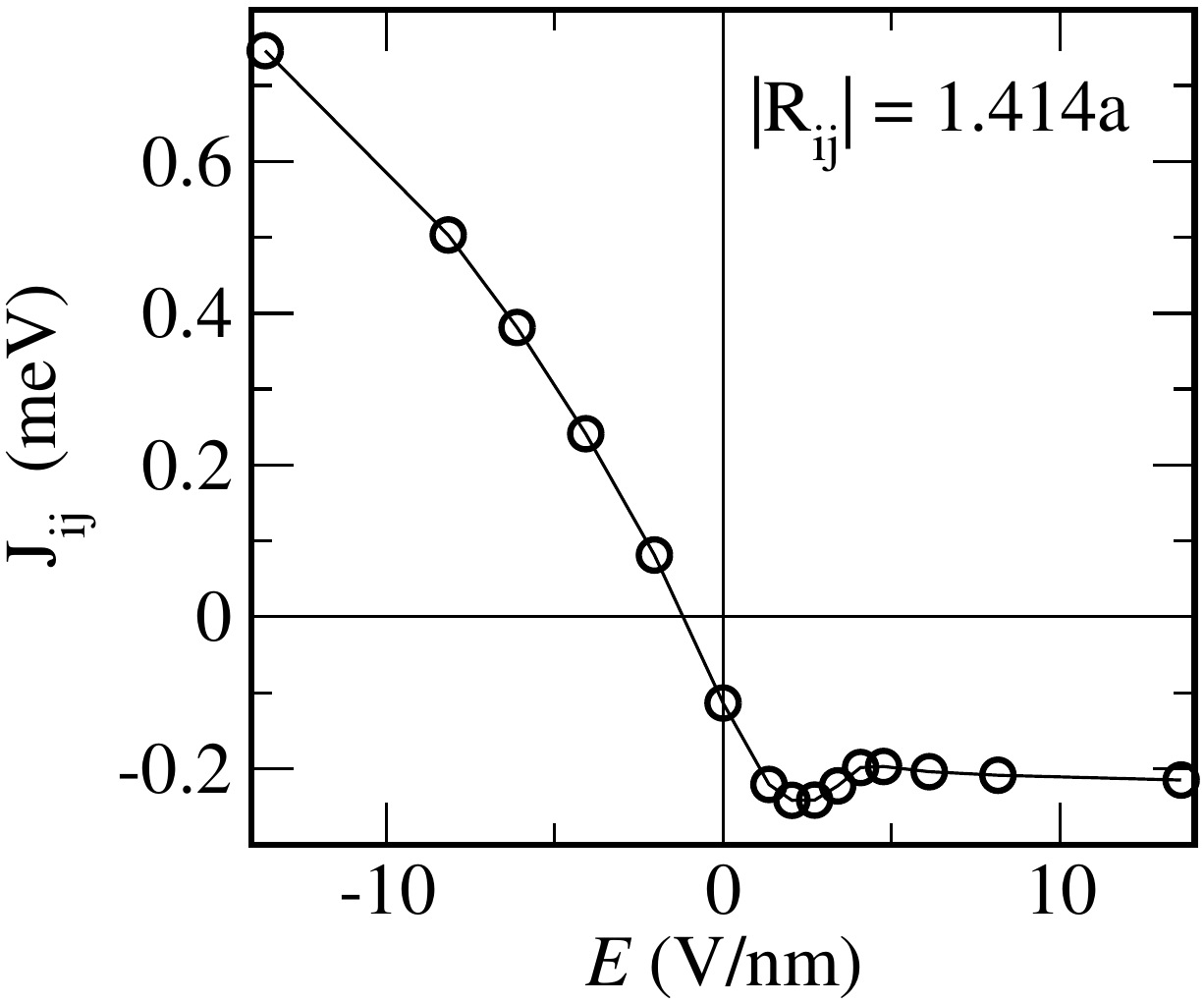}\;
\includegraphics[width=0.2\textwidth,angle=0,clip]{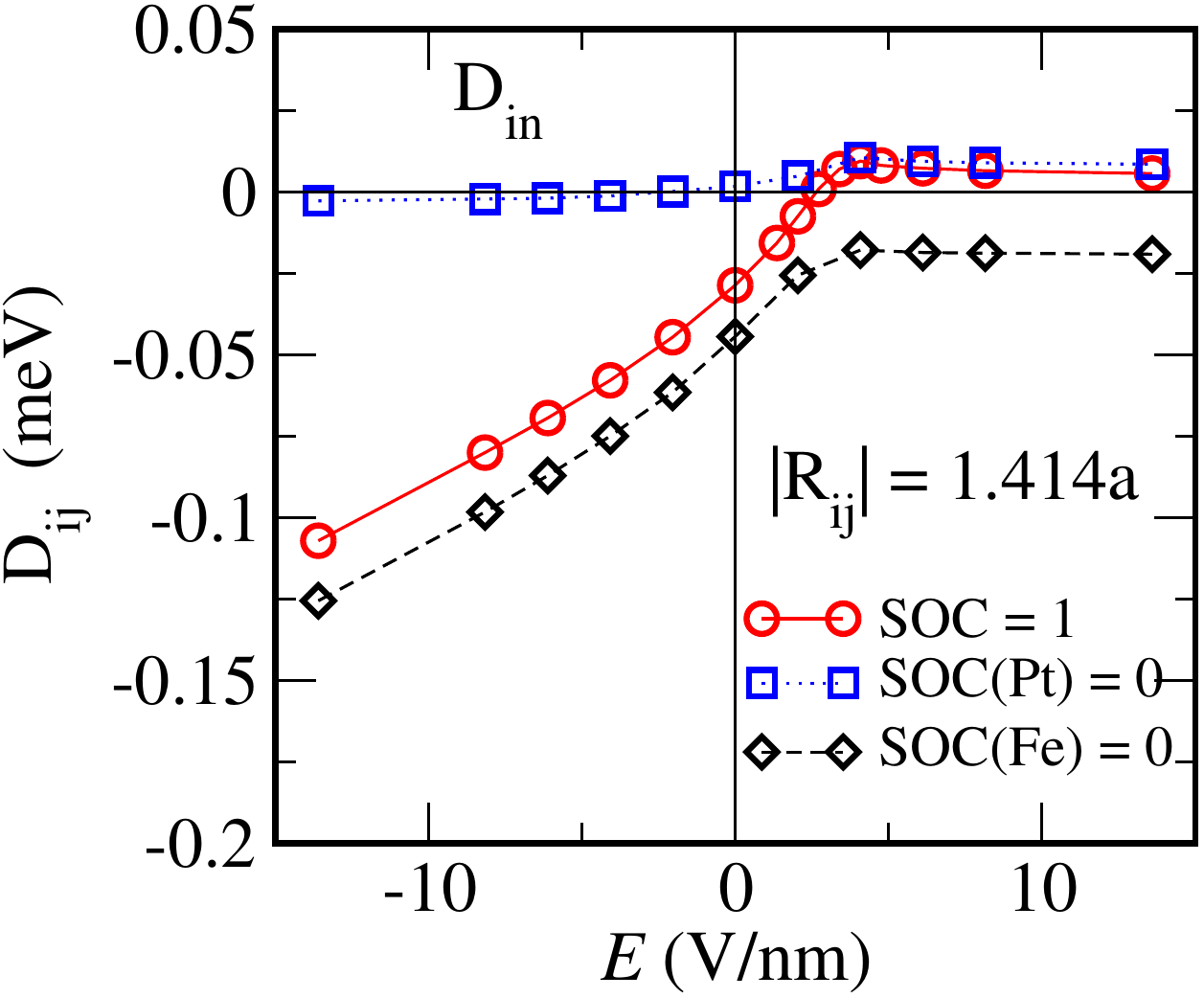}\;(c)
\includegraphics[width=0.4\textwidth,angle=0,clip]{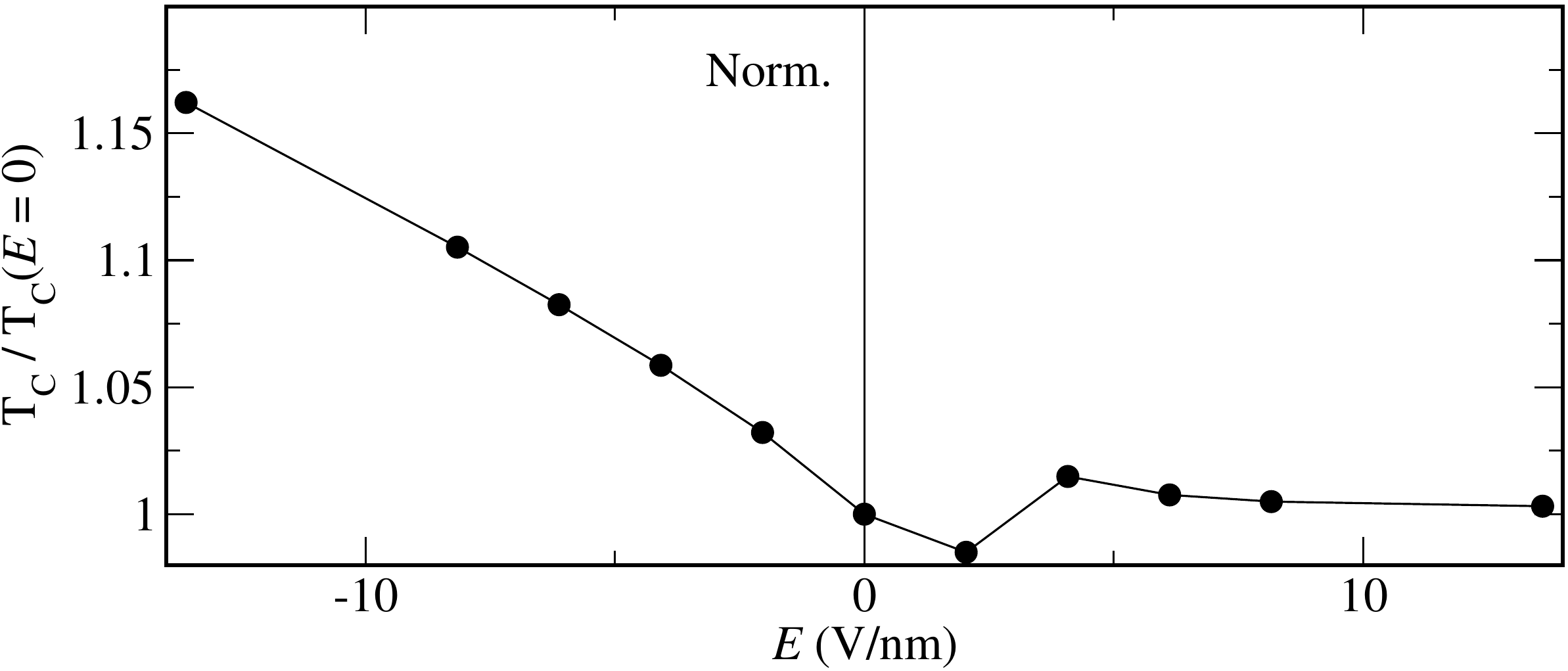}\;(d)
\caption{\label{fig:JXC_1MLFe_Pt} 
Isotropic Fe-Fe exchange coupling parameter
  $J_{ij}$ (left panel) and the maximal in-plane components of
  DMI, $\vec{D}^{in}_{ij}$, (right panel)
characterizing the interaction with the
  first-neighbors at  $R_{01} = 0.707a$ (a), the second-neighbors at $R_{02} =
  1.225a$ (b), and the third-neighbors at  $R_{03} = 1.414a$ (c), for 1 ML Fe
  on Pt(111). The parameters are plotted as a function of the applied
  electric field. (d): reduced Curie temperature $T_C/T_C(E=0)$  as a
  function of the  electric field, with $T_C(E=0) = 801$ K, determined
  on the basis of mean field theory. }   
\end{figure}

Concerning the field-dependence of the 
isotropic exchange coupling parameters, Fig.\ \ref{fig:JXC_Fe-Pt} 
shows these as a function of interatomic distance 
$R_{ij}$  for the field-free
case as a reference (left panel, open symbols). 
%
As can be concluded 
from the oscillating behavior of the product $J_{ij} \, R^2_{ij}$ 
 given in
the right panel in Fig.\ \ref{fig:JXC_Fe-Pt},
they show a well defined RKKY-like  characteristics for large distances. 
These interactions are
compared with those calculated for the free-standing Fe monolayer, shown by
full symbols. For all distances the Fe-Fe interactions are stronger for
the free-standing Fe monolayer, a finding that can be associated with a narrower
$d$-band and higher DOS at the Fermi energy, leading to a larger energy change
under the perturbation caused by spin tiltings. 
Although the oscillations in both cases have a different amplitude
 the parameter   $J_{ij}$
has a  similar RKKY-like dependency on the  distance   $R_{ij}$.
Fig. \ref{fig:JXC_1MLFe_Pt} (left panel) represents the exchange coupling
parameters $J_{01}$ (for $R_{01} = 0.707a$), $J_{02}$ (for $R_{02} =
  1.225a$) and $J_{03}$ (for $R_{03} = 1.414a$) as a function of the
  electric field.
As one can see, the relative change of the first-neighbor  
parameter $J_{01}$ is rather small, while the change is more pronounced
for the two other parameters that are negative in sign for the
field-free case.  
However, both 
parameters, $J_{02}$ and $J_{03}$, are  substantially smaller when
compared to $J_{01}$, implying that  they play a much weaker
role for the magnetic properties of the system, which is expected to be
ferromagnetic (FM) due to dominating FM interactions $J_{01}$.
As one can see in Fig. \ref{fig:JXC_1MLFe_Pt}, all parameters increase
almost linearly together with increasing 'negative' electric field,
leading for  $J_{03}$ to a change in sign. 
An increasing 'positive' field, on the other hand,
  leads to a saturation already at $E > 5  \, \frac{V}{nm}$. 
To get an impression for the influence of the electric field on the
Curie temperature  $T_C$, we plot in Fig.  \ref{fig:JXC_1MLFe_Pt} (d)
$T^{MFA}_C(|\vec{E}|)$ calculated using the exchange coupling parameters
in  Fig. \ref{fig:JXC_1MLFe_Pt}(a)-(c) on the basis of the mean-field
theory. Obviously, $T^{MFA}_C(|\vec{E}|)$
follows the field-induced changes of $J_{ij}$ shown in  Fig. \ref{fig:JXC_1MLFe_Pt}
(a)-(c), exhibiting a strong field dependence for
'negative' field and only weak changes for 'positive' fields.

\begin{figure}[]
\centering
\includegraphics[width=0.7\columnwidth]{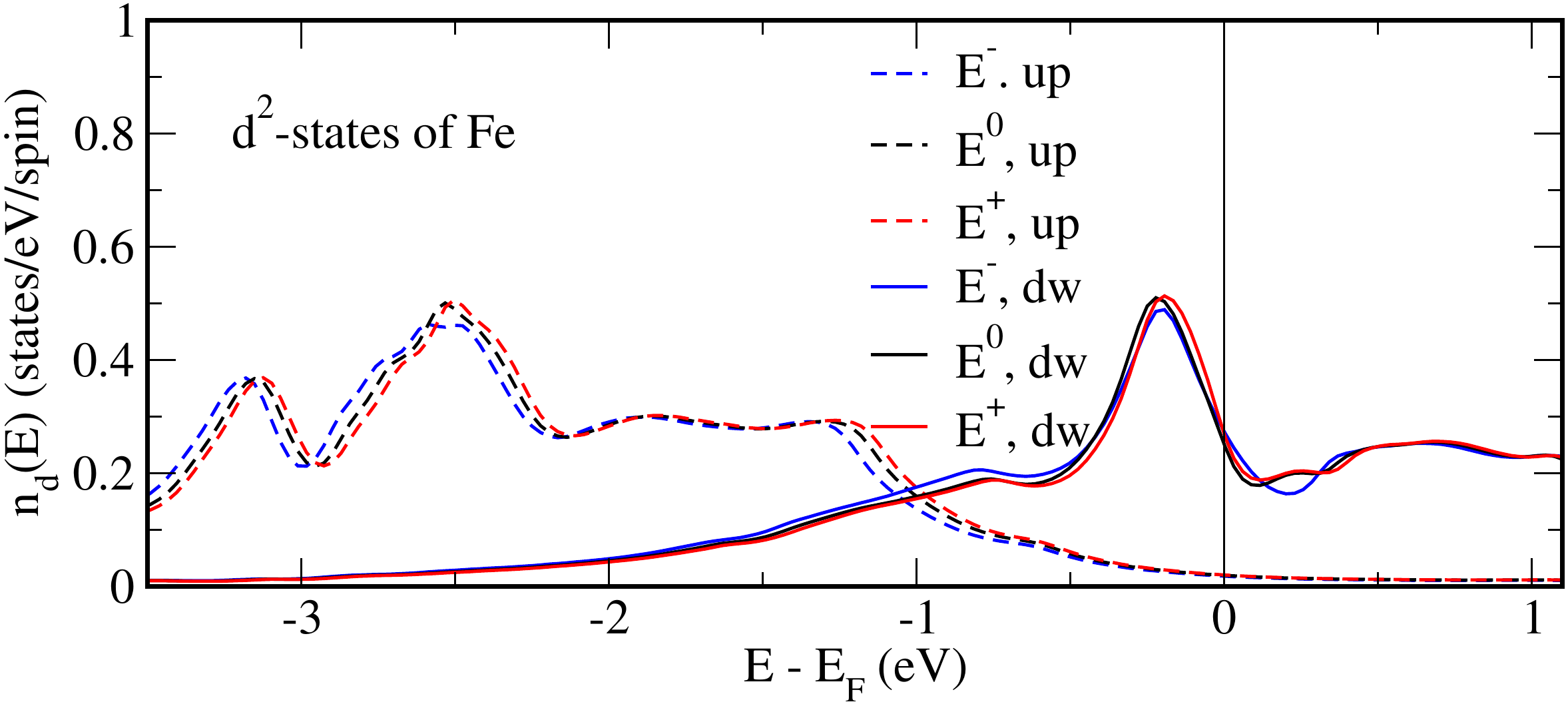}(a)
\includegraphics[width=0.7\columnwidth]{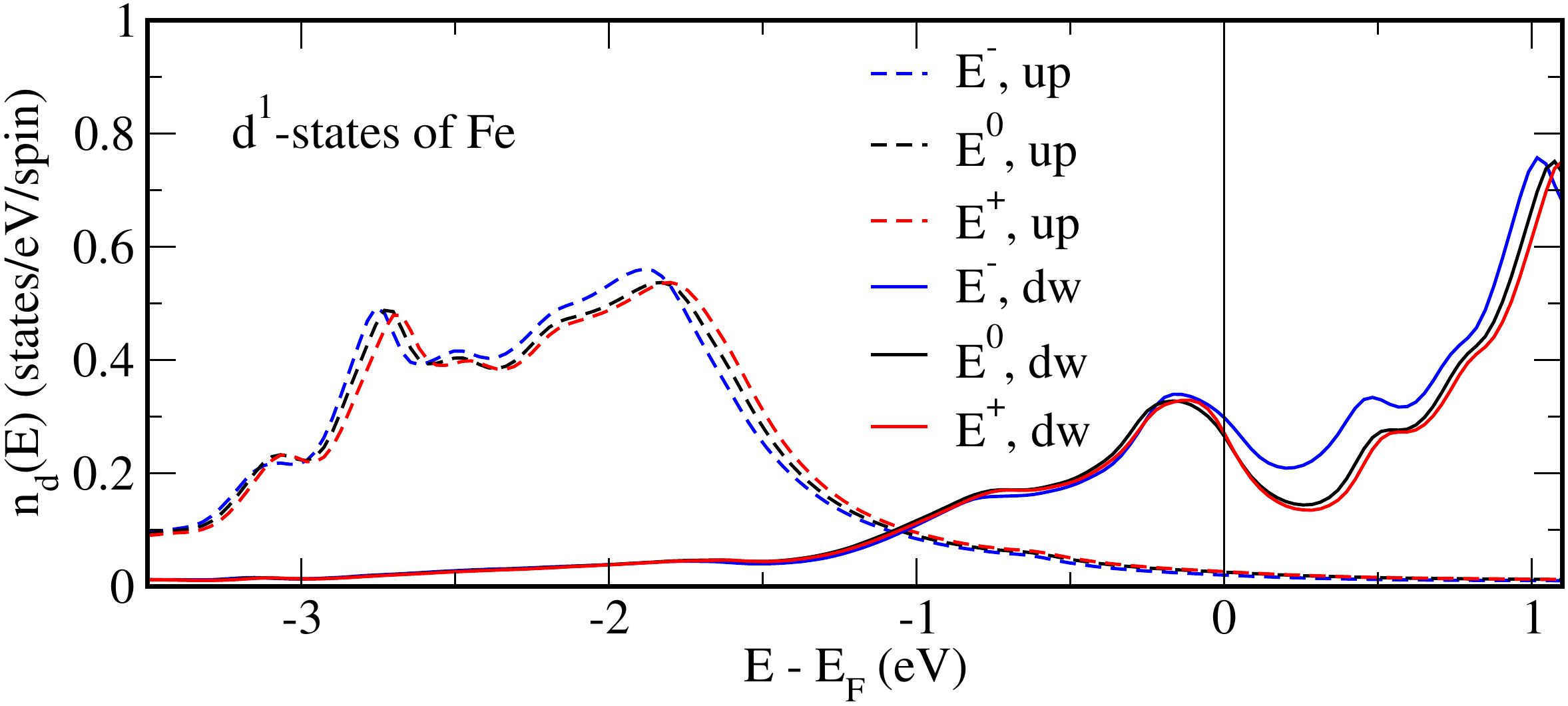}(b)
\includegraphics[width=0.7\columnwidth]{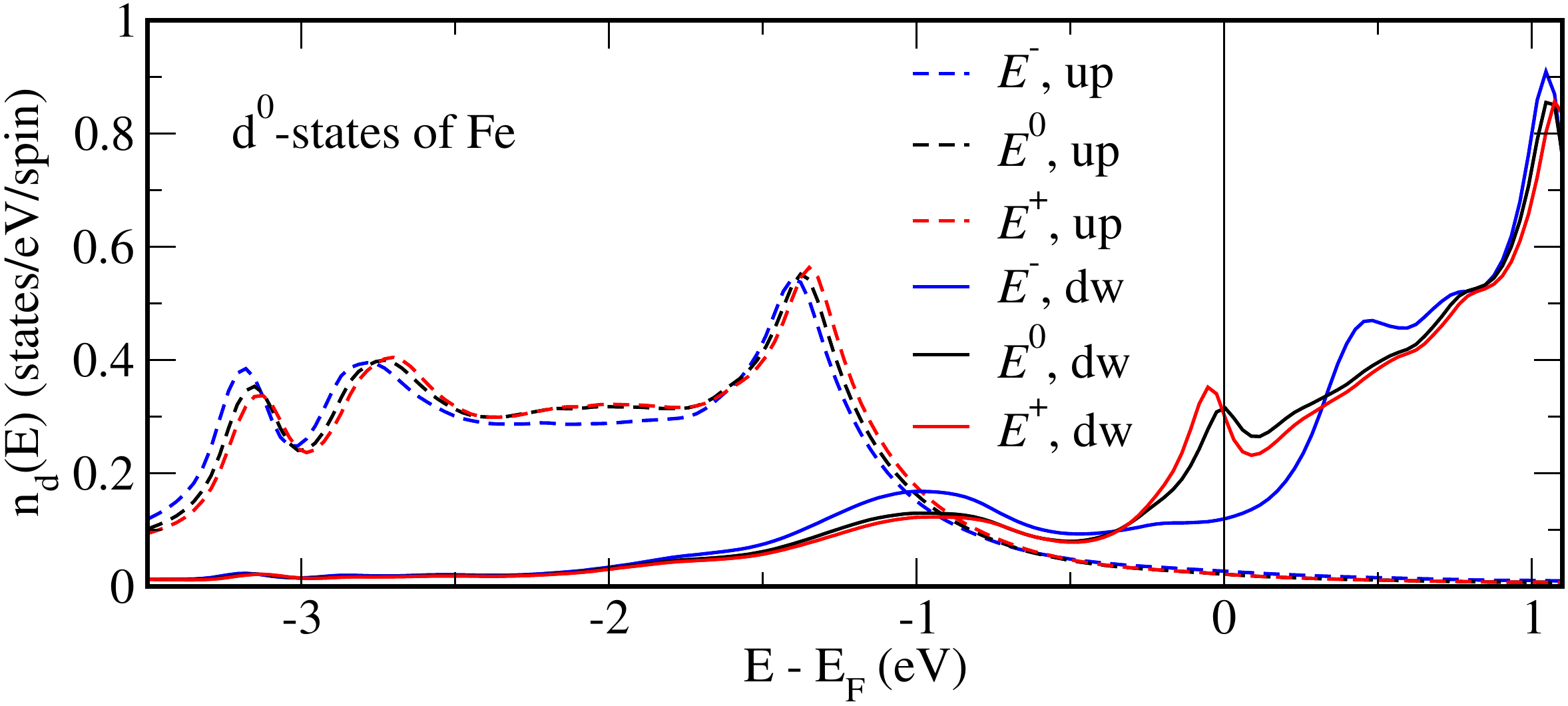}(c)
\caption{Electric-field-induced change of the $(l,m,s)$ resolved density
  of states in Fe layer on  Pt(111). The applied electric fied $E^0 =
  0.0$ V/nm,  $E^+ = +13.6$ V/nm and $E^- = -13.6$ V/nm. 
   }
\label{fig_DOS_RLM_Fe}
\end{figure}

\begin{figure}[h]
\includegraphics[width=0.28\textwidth,angle=270,trim= 0 4.5cm 0 8.5cm,clip]{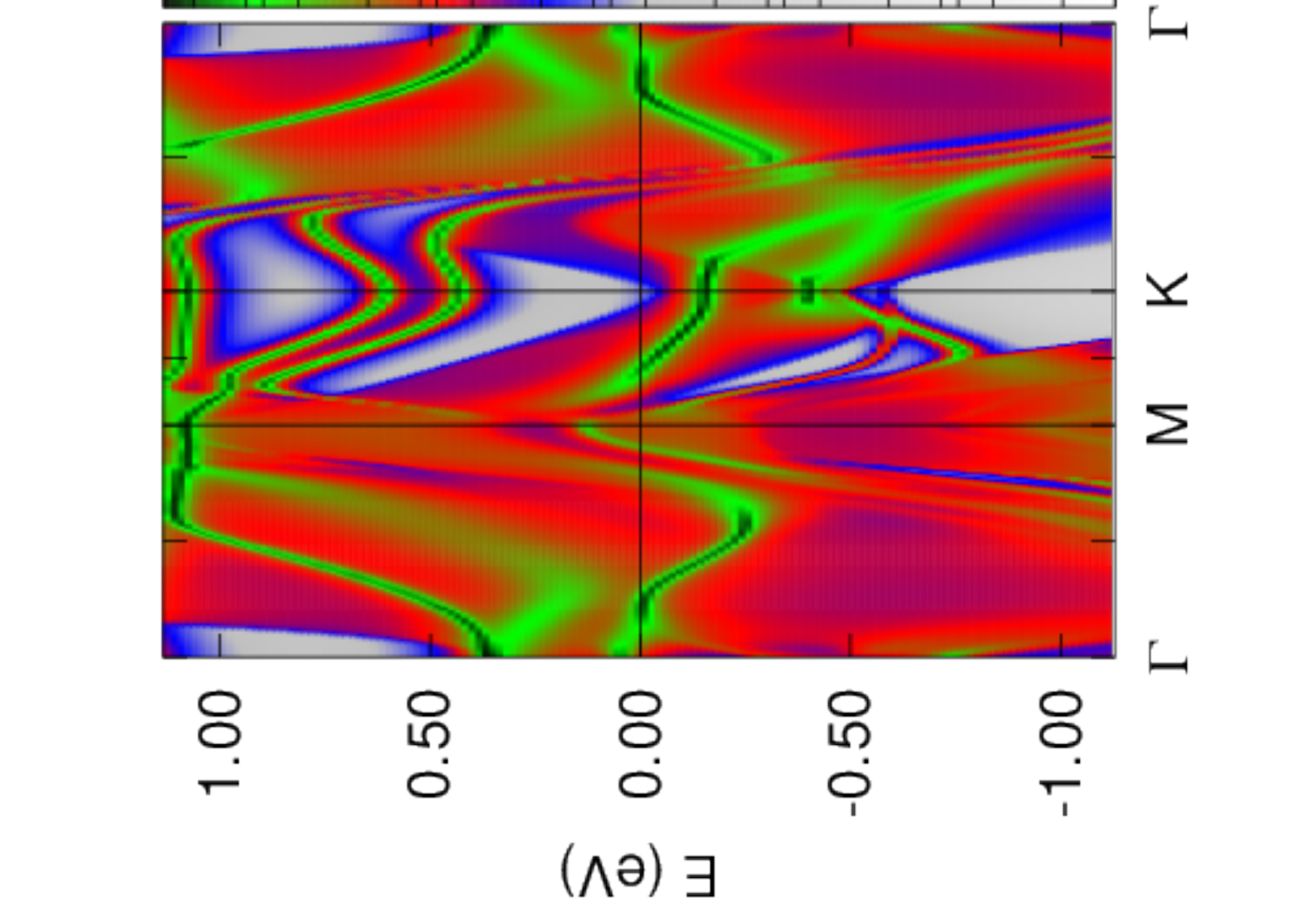}
\includegraphics[width=0.28\textwidth,angle=270,trim= 0 4.5cm 0
6cm,clip]{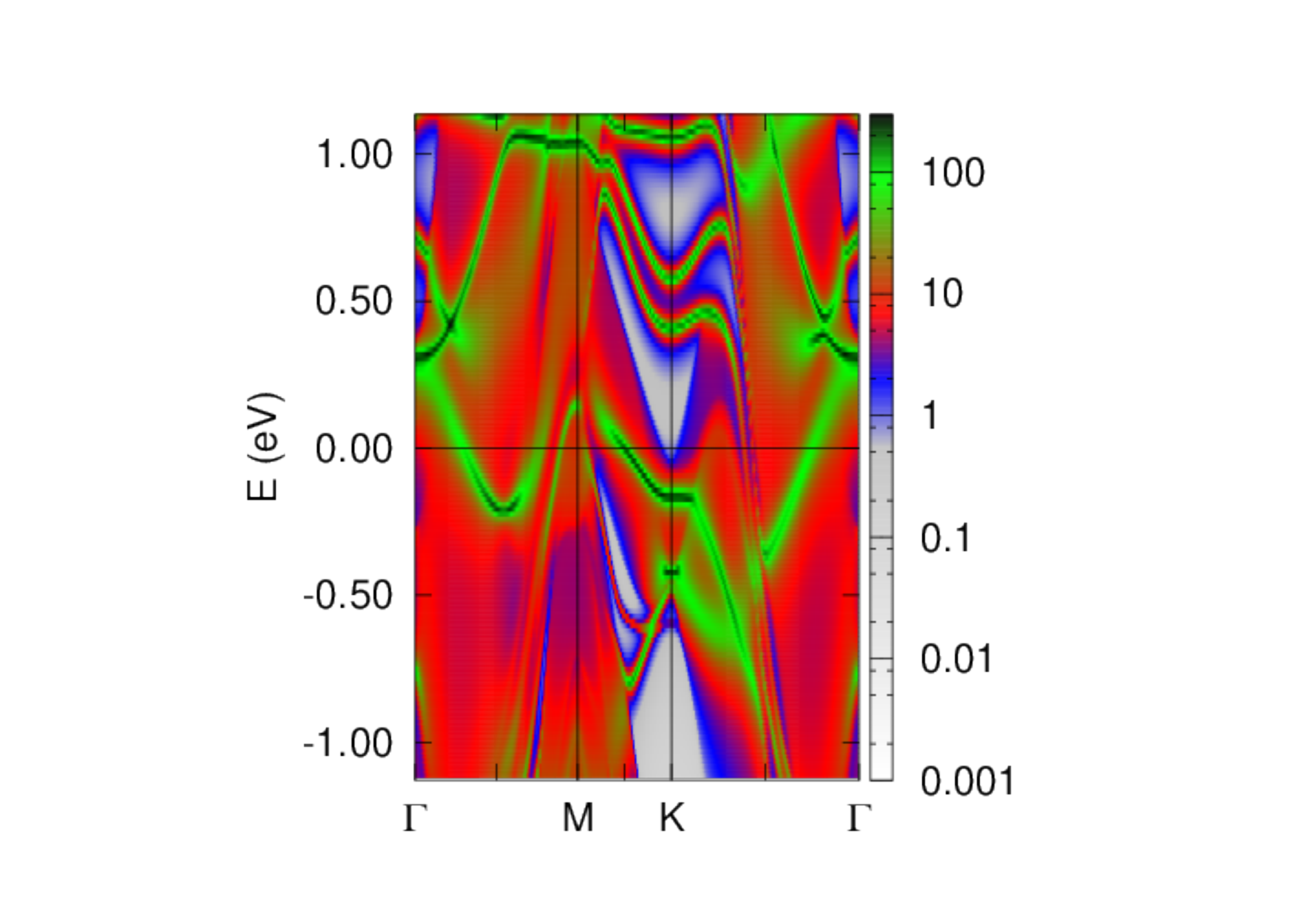}\,
\caption{\label{fig:Fe-BSF_1MLFe_Pt-plus} Calculated Bloch spectral
  function $A({\cal E},\vec{k},0)$ (left) and $A({\cal E},\vec{k},E)$,
  $E = -13.6 \frac{V}{nm}$ (right) representing the Fe-projected
  minority-spin states in 1 ML Fe on Pt(111). }     
\end{figure}

As one can see in Fig. \ref{fig:BSF_DIFF_Fe_1MLFe_Pt}, a 'positive'
electric field leads in general to an upwards shift for the $d$-states of Fe,
while 'negative' field leads to a shift of the states down in energy.
The same trend can be seen for the  $(l,m,s)$-resolved DOS  plotted in
Fig. \ref{fig_DOS_RLM_Fe}(a) and (b) showing the DOS for the $d_{x^2-y^2}$,
$d_{xy}$, $d_{xz}$ and $d_{yz}$ states.
However, pronounced field-induced shifts of the electronic states occur
around the $\Gamma$ point in the vicinity  of the  Fermi
energy for a 'negative' field . 
This shift is opposite in direction compared  to all others. As is shown in
Fig. \ref{fig_DOS_RLM_Fe}(c) a strong modification of the minority-spin
$d_{z^2}$ states occurs close to the Fermi energy for a 'negative' field.
%
This shift is also seen in Fig. \ref{fig:Fe-BSF_1MLFe_Pt-plus}
giving  the BSF for the minority spin states of Fe.
As one can see, the states at the  $\Gamma$ point move in the presence of
electric field from a position close to  the Fermi level upwards in energy
into  the energy gap of the bulk Pt
states. These states have to be seen as interface states strongly
affected by the weakly screened electric field due to their spatial position.
On the other hand, such a behaviour is  not seen in the case of a 'positive'
electric field.
From this one may conclude that the interface states are primarily
responsible for the strong modification of the exchange parameters in
case of 'negative' electric field.

Due to the broken inversion symmetry at the surface, the Fe-Fe DMI does not
vanish in the absence of the electric field in contrast to the unsupported Fe
monolayer.
As it was mentioned above, the out-of-plane component of the DMI,
$D^{out}_{ij}$, has an alternating orientation when going from one atom to
another within the same neighbour shell, as it is shown in   Fig.\ 
\ref{fig:GEOM_DMI}. This results in a mutual cancellation of their
influence on the magnetic structure. For that reason, we discuss here only the
in-plane components of DMI $D^{in}_{ij}$, that are plotted in Fig.\
\ref{fig:JXC_1MLFe_Pt} (right panel) 
as a function of the field strength and direction. One can see a pronounced
increase for the magnitude of the first- and third-neighbor parameters
$D^{in}_{01}$ and  $D^{in}_{03}$ in the case of a 'negative' electric
field. Note, however, that their sign is opposite.
In the case of a 'positive' electric field the parameters $D^{in}_{ij}$
have rather weak variation with the field strength. This behaviour is
rather similar to the behaviour of the 
isotropic exchange interactions and can be related to the
field-dependent changes of the interface electronic states.

To demonstrate the role of the substrate atoms for the DMI and its 
dependence on the electric field, additional calculations have been
performed with an artificial scaling of the SOC on the Fe and the interface Pt atoms.
When the SOC of Fe atom is taken to be zero, SOC(Fe) = 0, the parameters
$D^{in}_{01}$ and $D^{in}_{03}$ slightly increase, following the same
field dependence as in the case of an unscaled SOC. In the case of
$D^{in}_{02}$ the effect of SOC(Fe) scaling is much more pronounced,
leading even to a change of sign for $D^{in}_{02}$.
On the other hand, scaling SOC on the Pt atoms, SOC(Pt) = 0, leads to an
increase of the magnitude of $D^{in}_{02}$ (which is negative), having a
rather similar field dependence as in the case of the unscaled SOC.
This results obviously reflect a strong 
competition of the Pt and Fe SOC effects for the parameter
$D^{in}_{02}$, with a leading effect of SOC(Fe). In contrast, in the
case of SOC(Pt) = 0 the parameters $D^{in}_{01}$ and 
$D^{in}_{03}$ drop down significantly, that implies that their strength is
governed by the SOC of the Pt atoms.

\section{1ML Fe on 1H-WS$_2$ \label{TMDC}}

\begin{figure}
\includegraphics[width=0.18\textwidth,angle=0,clip]{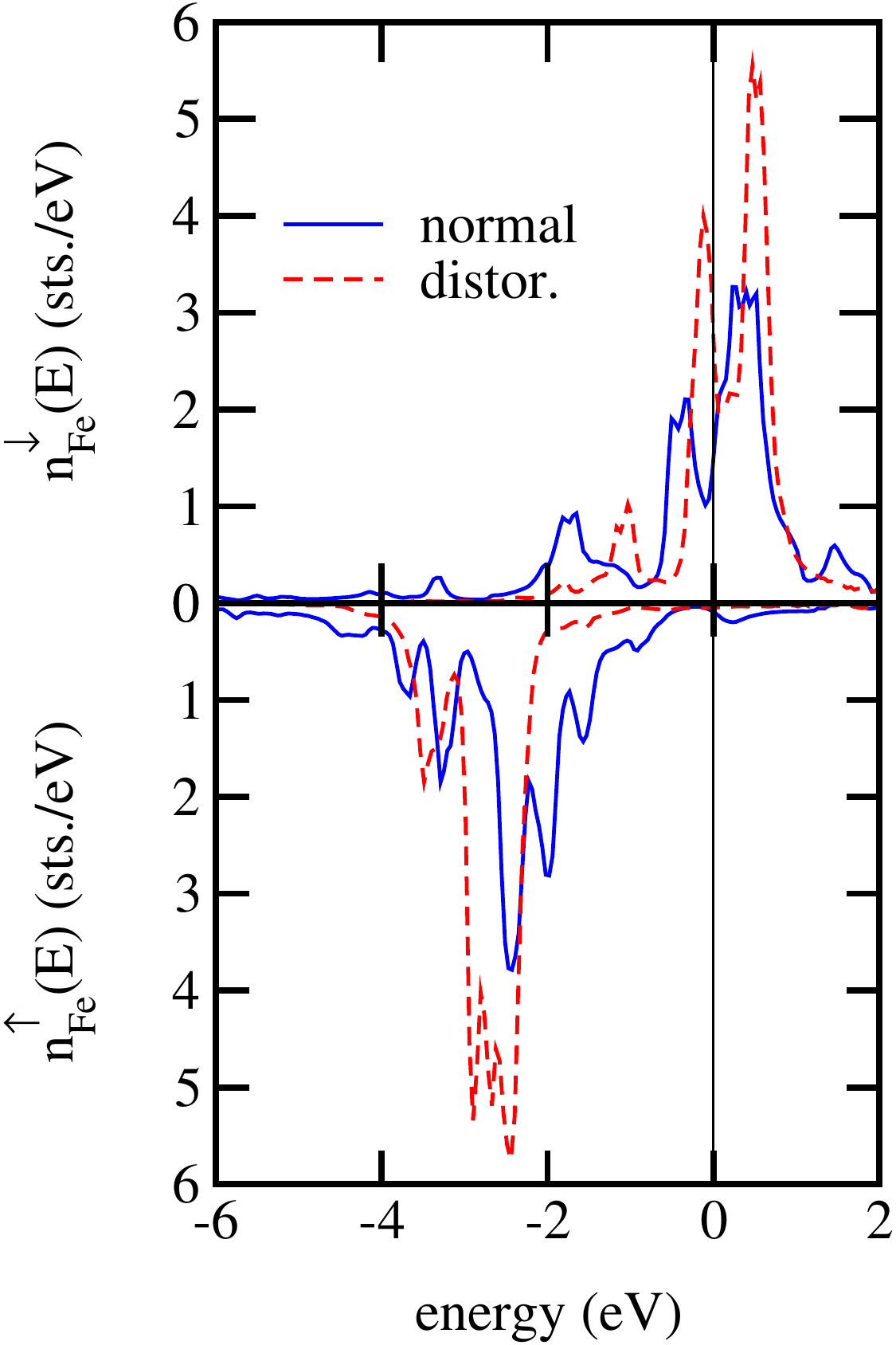}\;
\includegraphics[width=0.18\textwidth,angle=0,clip]{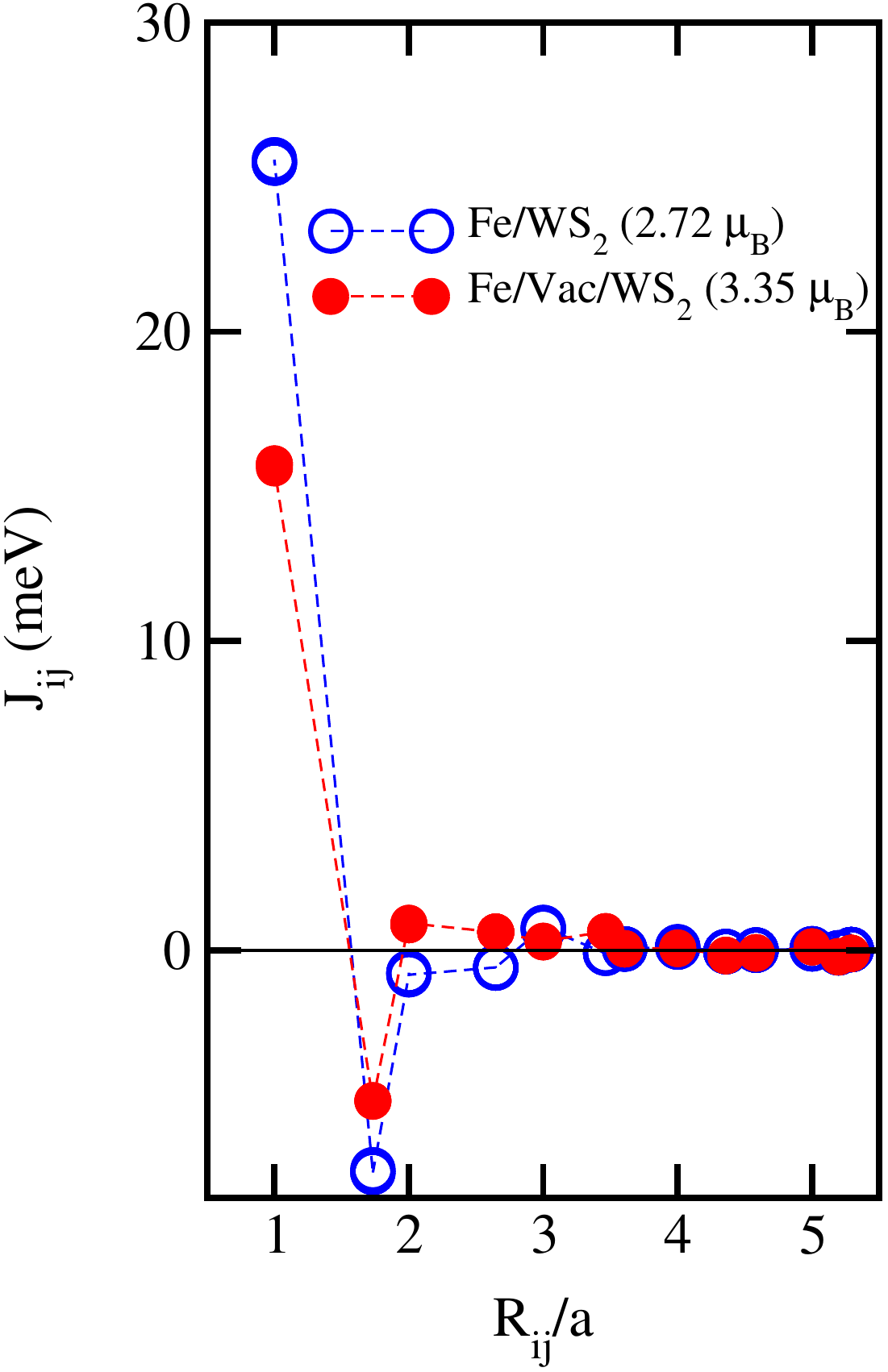}\;
\caption{\label{fig:CMP_JXC_WS2} Left panel: DOS for non-distorted system (solid
  line) and for distorted Fe/WS$_2$ system with empty monolayer inserted
  between Fe and substrate (dashed line). Right panel: isotropic Fe-Fe exchange
  parameters calculated for the non-distorted (open symbols) and
  distorted (full symbols) system.  }   
\end{figure}

In the case of 1ML Fe on 1H-WS$_2$, the substrate was chosen as an
example for an insulator in contrast to metallic Pt considered above. 
In this case one can expect a different impact of the substrate on the Fe-Fe
exchange interactions and in turn a different field-dependent
behavior. 
The spin magnetic moment of Fe  on 1H-WS$_2$ in the absence of an electric
field is $2.72 \mu_B$, that is essentially smaller when compared to
$3.01 \mu_B$ in the case of Fe/Pt(111), despite the larger Fe-Fe 
interatomic distance, $5.96$ a.u. for 1H-WS$_2$,  in comparison to
$5.24$ a.u. for Pt(111) as a substrate.
Moving Fe monolayer away from the surface of WS$_2$ by inserting
an empty layer in-between
leads to an increase of the Fe spin magnetic moment
to $3.35 \mu_B$, which is a consequence of the narrower $d$-bands of Fe leading
to more pronounced exchange splitting of the majority- and minority-spin
states, as can be seen in the DOS for the Fe $d$-states plotted in
Fig. \ref{fig:CMP_JXC_WS2} (left panel). 
This allows to conclude that the decreased magnetic moment of Fe  on 
1H-WS$_2$ is a result of the strong hybridization of the Fe $d$ states with
the $p$ states of S and $d$ states of W.

\begin{figure}[h]
 \includegraphics[width=0.4\textwidth,angle=0,clip]{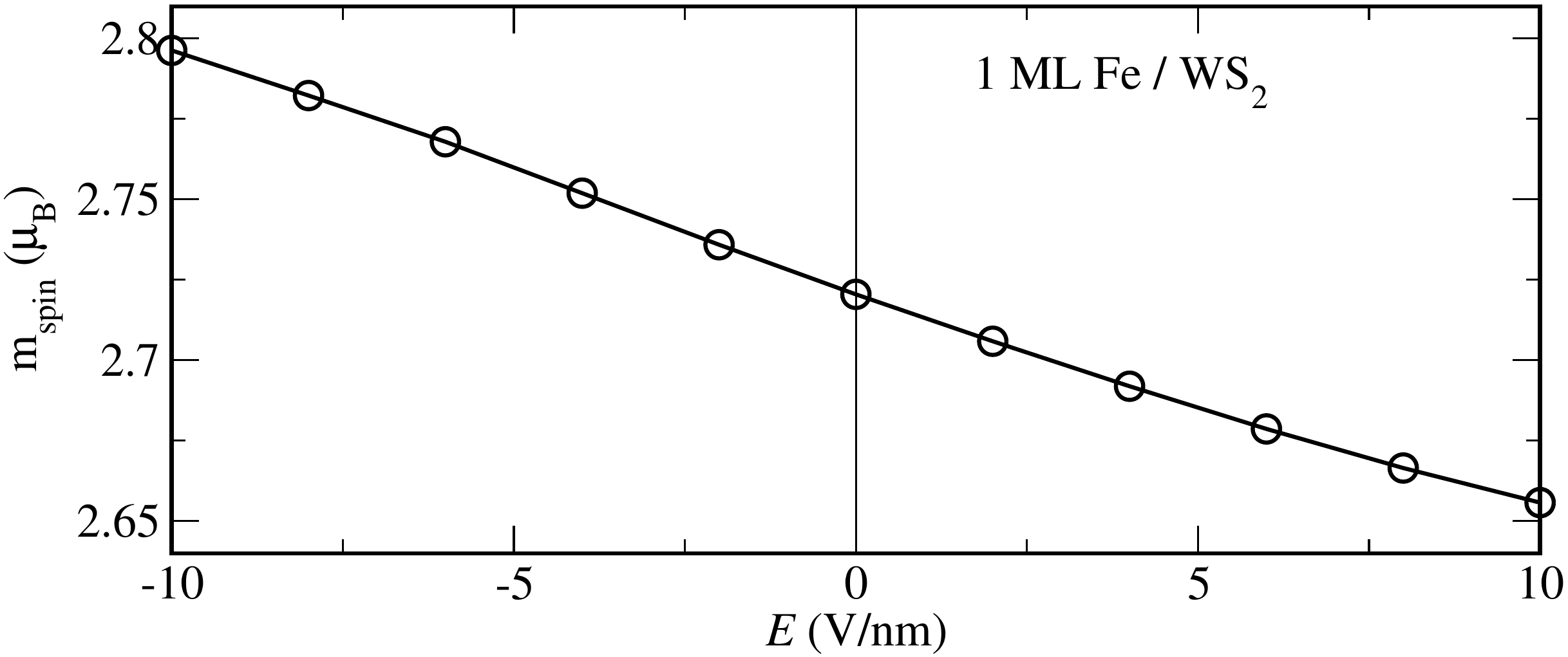}\;
 \caption{\label{fig:MMOM_1MLFe_WS2} Calculated spin magnetic moment of
   Fe,$ m_{Fe}(E)$, as a function of the external electric field $E$ for
   1 ML Fe on  1H-WS$_2$.  }   
 \end{figure}

The Fe-Fe exchange parameters calculated for the non-distorted Fe/WS$_2$
system (i.e., with identical Fe-S and W-S distances)
are also plotted in Fig. \ref{fig:CMP_JXC_WS2} (right panel, open symbols) in
comparison with those calculated for Fe monolayer spaced further away from the
substrate (full symbols). From this one can see that depositing a Fe ML on
WS$_2$ results in an increase of the exchange parameters despite the
decrease of the spin magnetic moment of Fe. This trend is opposite to that
found for 1ML Fe/Pt(111) and may indicate a crucial role of the
hybridization of the Fe $d$-states with the $p$-states of S responsible
for a Fe-Fe superexchange in this system.
%
As it is shown in Fig.\ \ref{fig:MMOM_1MLFe_WS2}, the Fe magnetic moment
$ m_{Fe}$ in Fe/1H-WS$_2$ has an almost linear dependence on the electric field.
The field-induced change of $m_{Fe}$ is by about order of magnitude
larger than in the case of Fe/Pt(111), as a result of the different
impact of the substrate on Fe in these two cases.

\begin{figure}[h]
\includegraphics[width=0.28\textwidth,angle=270,trim= 0 5.5cm 0 9.5cm,clip]{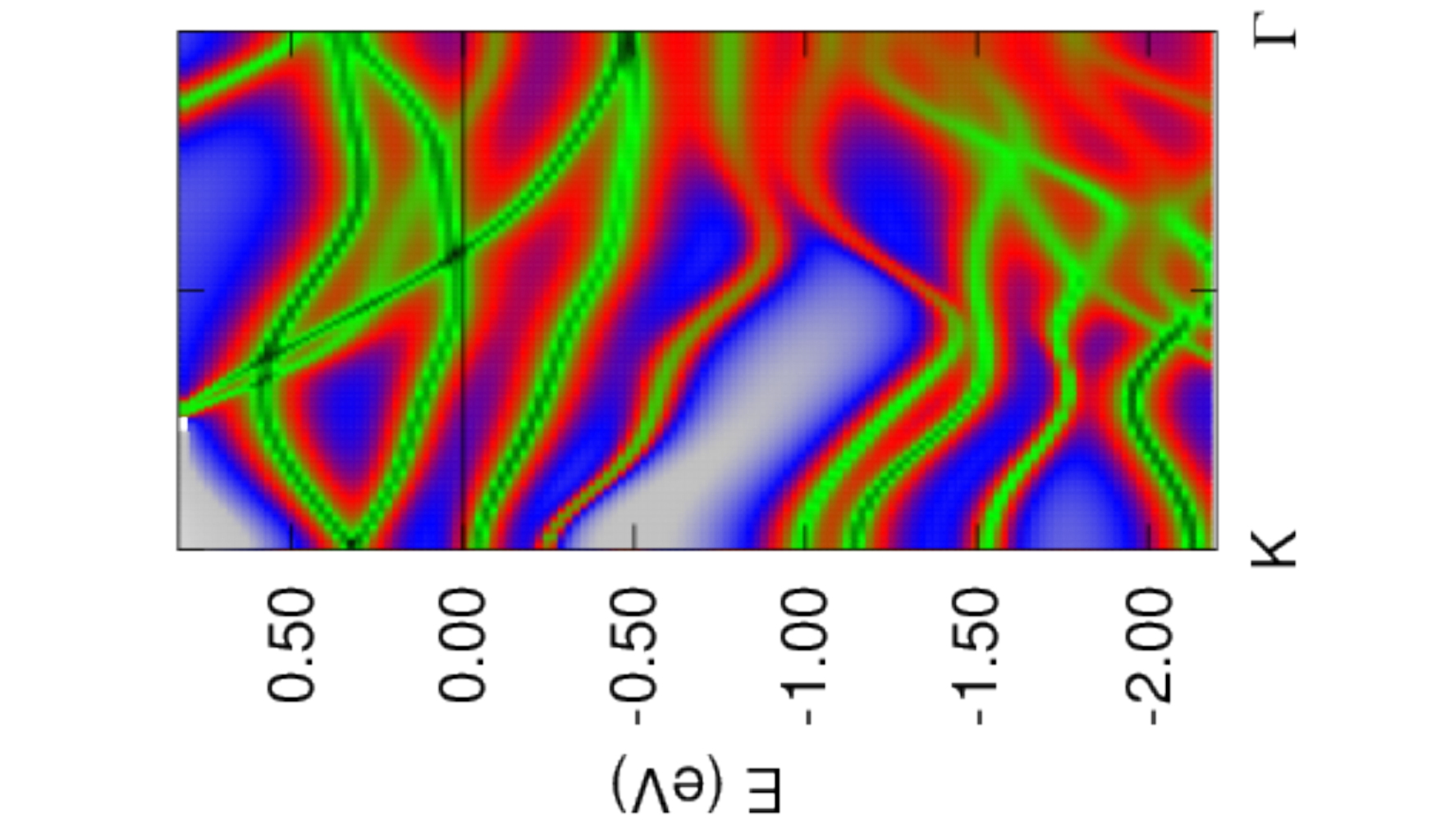}
\includegraphics[width=0.28\textwidth,angle=270,trim= 0 8.7cm 0 9.5cm,clip]{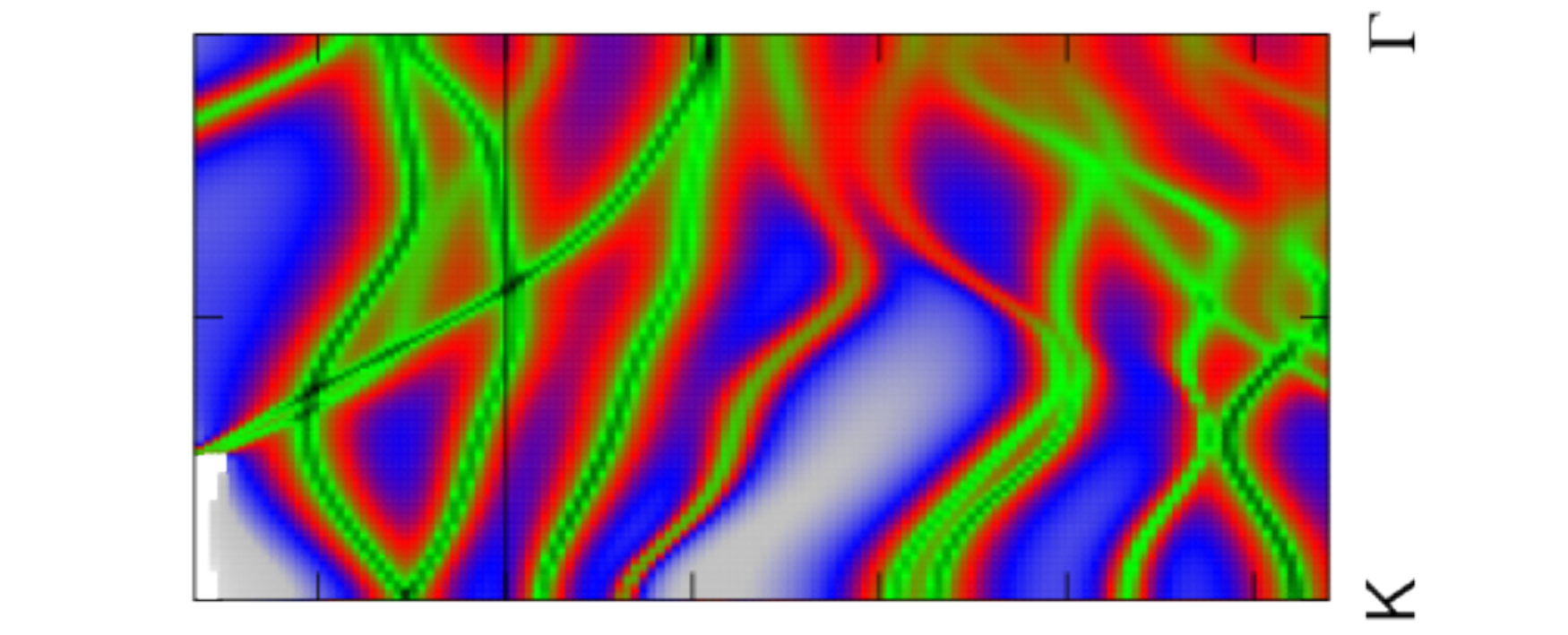}
\includegraphics[width=0.28\textwidth,angle=270,trim= 0 8.7cm 0 6cm,clip]{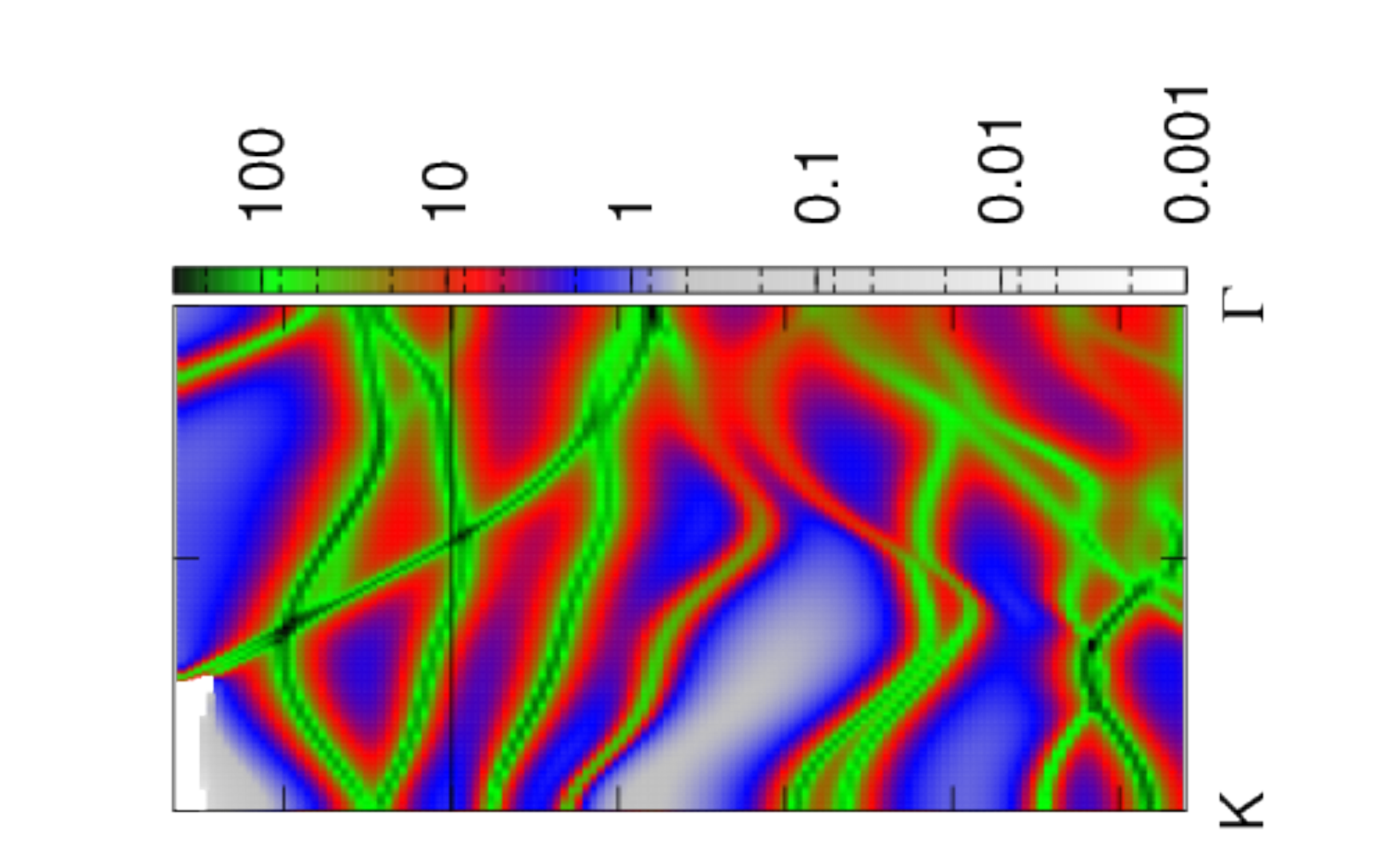}
\caption{\label{fig:Fe-BSF_norm_1MLFe_TMDC} Calculated Bloch spectral
  function $A({\cal E},\vec{k},0)$ (a),  $A({\cal E},\vec{k},E)$, $E
  = -10 \frac{V}{nm}$ (b) and $A({\cal E},\vec{k},E)$, $E = 10
  \frac{V}{nm}$ representing the states localized in Fe layer in 1 ML Fe on WS$_2$. }     
\end{figure}

Note also that  in contrast to Fe/Pt(111), the whole Fe/1H-WS$_2$ system
experiences the effect of the applied electric field due to its finite thickness.
As a result, an impact of the electric field on the electronic structure
is much stronger, as it can be seen in the BSF $A({\cal
  E},\vec{k},E)$ representing the Fe projected energy bands given in
Fig.\ \ref{fig:Fe-BSF_norm_1MLFe_TMDC} for three different cases,  $E
=-10$, $0$ and $10 \frac{V}{nm}$.  

\begin{figure}[t!]
\centering
\includegraphics[width=0.8\columnwidth]{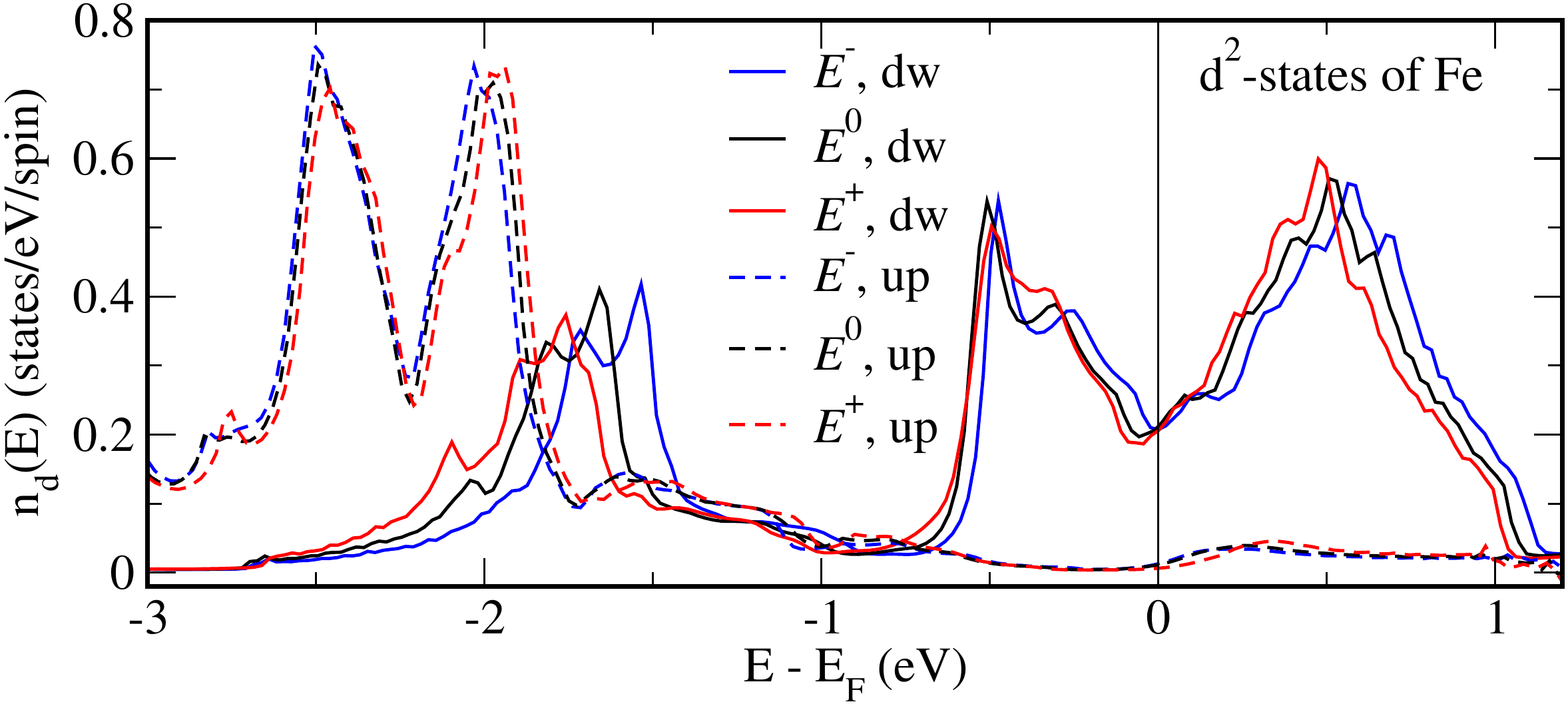}(a)
\includegraphics[width=0.8\columnwidth]{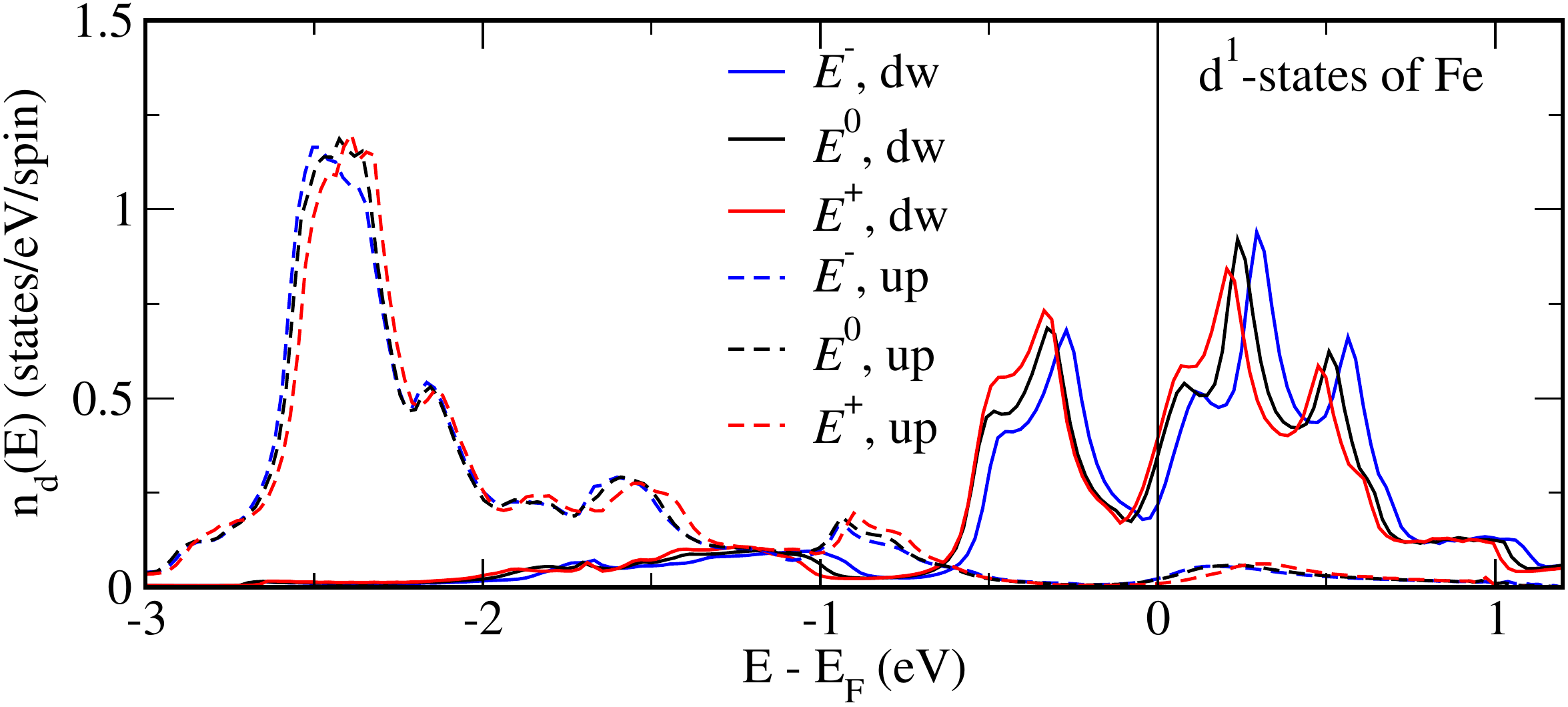}(b)
\includegraphics[width=0.8\columnwidth]{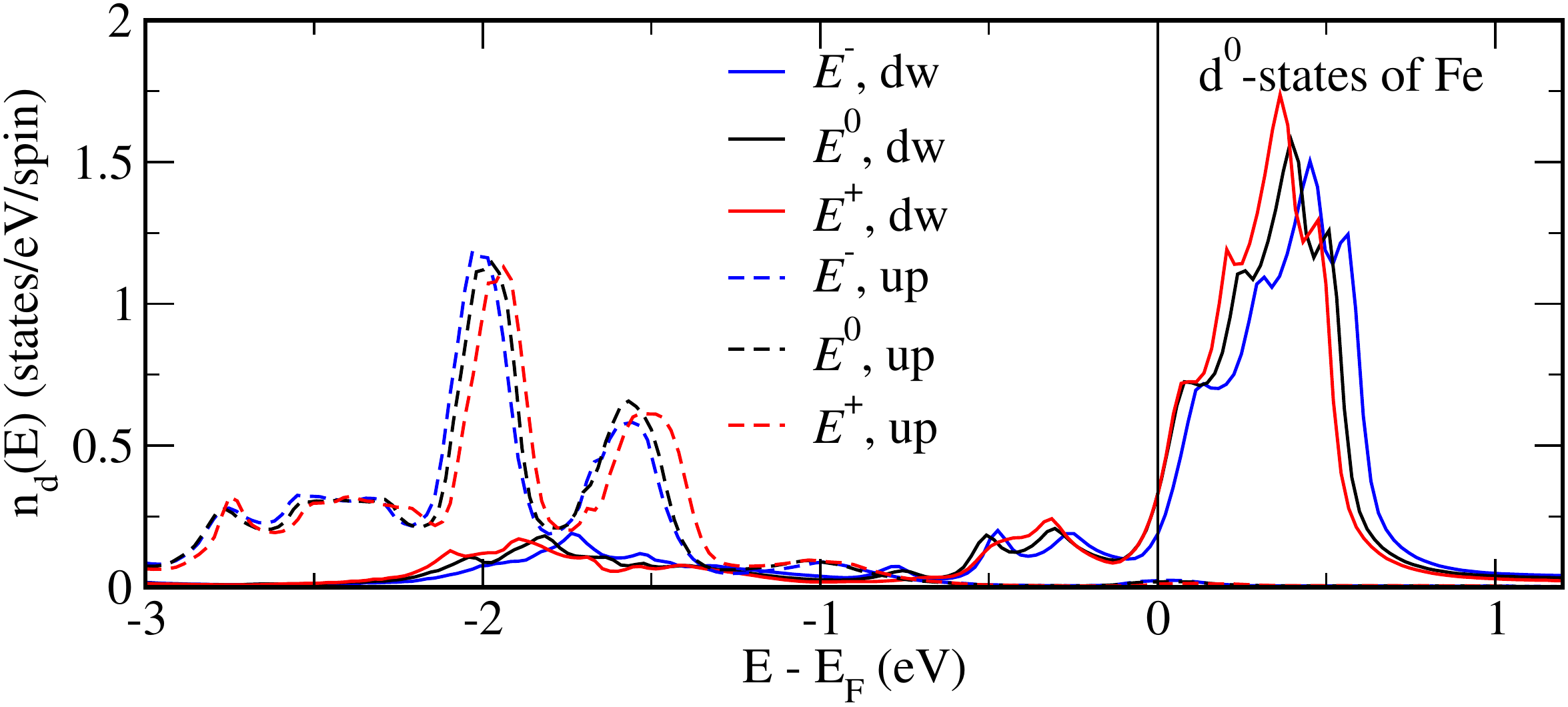}(c)
\caption{Electric-field-induced change of the $(l,m,s)$ resolved density
  of states in the Fe layer on top of WS$_2$. The applied electric fied $E^0 =
  0.0$ V/nm,  $E^+ = +10$ V/nm and $E^- = -10$ V/nm. $d^0$, $d^1$, $d^2$
  denote the $d_{z^2}$, ($d_{xz}$, $d_{yz}$), and ($d_{xy}$,
  $d_{x^2-y^2}$) states, respectively.
   }
\label{fig_DOS_RLM_Fe-WS2}
\end{figure}

In the $(l,m,s)$-resolved DOS of Fe plotted in Fig.\
\ref{fig_DOS_RLM_Fe-WS2} one can see that the bandwidth of the $d_{xy}$
and $d_{x^2-y^2}$ states is largest, indicating their 
strong hybridization with the states of the substrate, i.e. first of all
$p$ states of S. An applied electric field leads obviously to a rather complicated
modification of the electronic structure. One notes a
field-induced up- and down- shift of the minority-spin $d$-states of Fe
arranged around the Fermi level, depending on the direction of 
electric field (Fig.\ \ref{fig_DOS_RLM_Fe-WS2} (a)-(c)).

The DOS peaks at $E \approx -1.5$ eV  in Fig.\ \ref{fig_DOS_RLM_Fe-WS2} (a)
appear due to a hybridization of the Fe minority-spin $d_{xy}$ 
and $d_{x^2-y^2}$ states with the $p$-states of S and $d$-states of W. 
They show more pronounced field-induced shifts when compared to the
states around the Fermi energy. This can be attributed to a
field-induced change of the hybridization of these states caused by the
shifts of the Fe $d_{xy}$ and $d_{x^2-y^2}$ states.
As a common trend one notes for the majority-spin
states a shift in the opposite direction in comparison with the minority-spin
states, implying a field-induced change of the exchange splitting,
leading in turn to a corresponding change of the spin magnetic moment of Fe.

\begin{figure}
\includegraphics[width=0.19\textwidth,angle=0,clip]{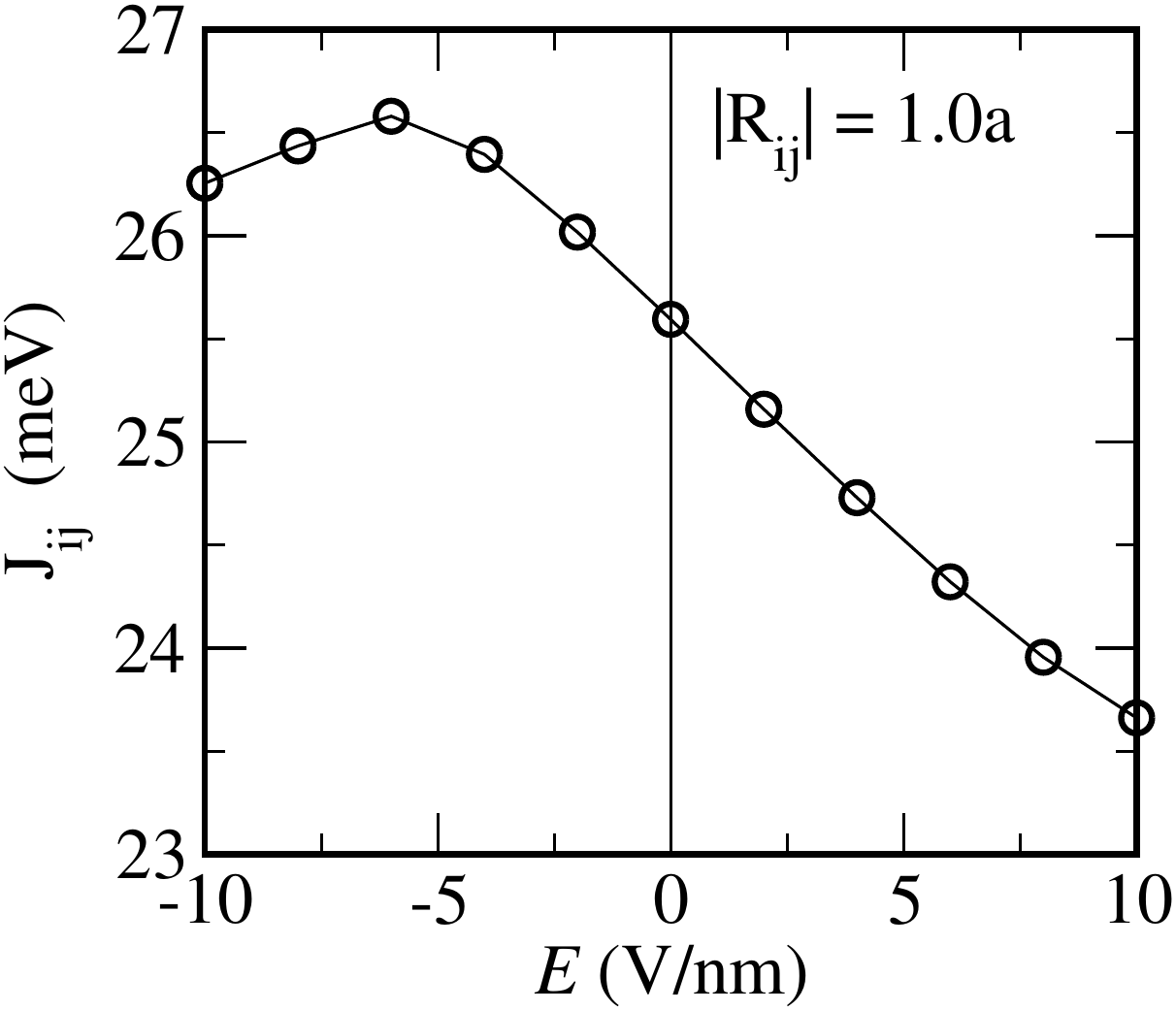}
\includegraphics[width=0.19\textwidth,angle=0,clip]{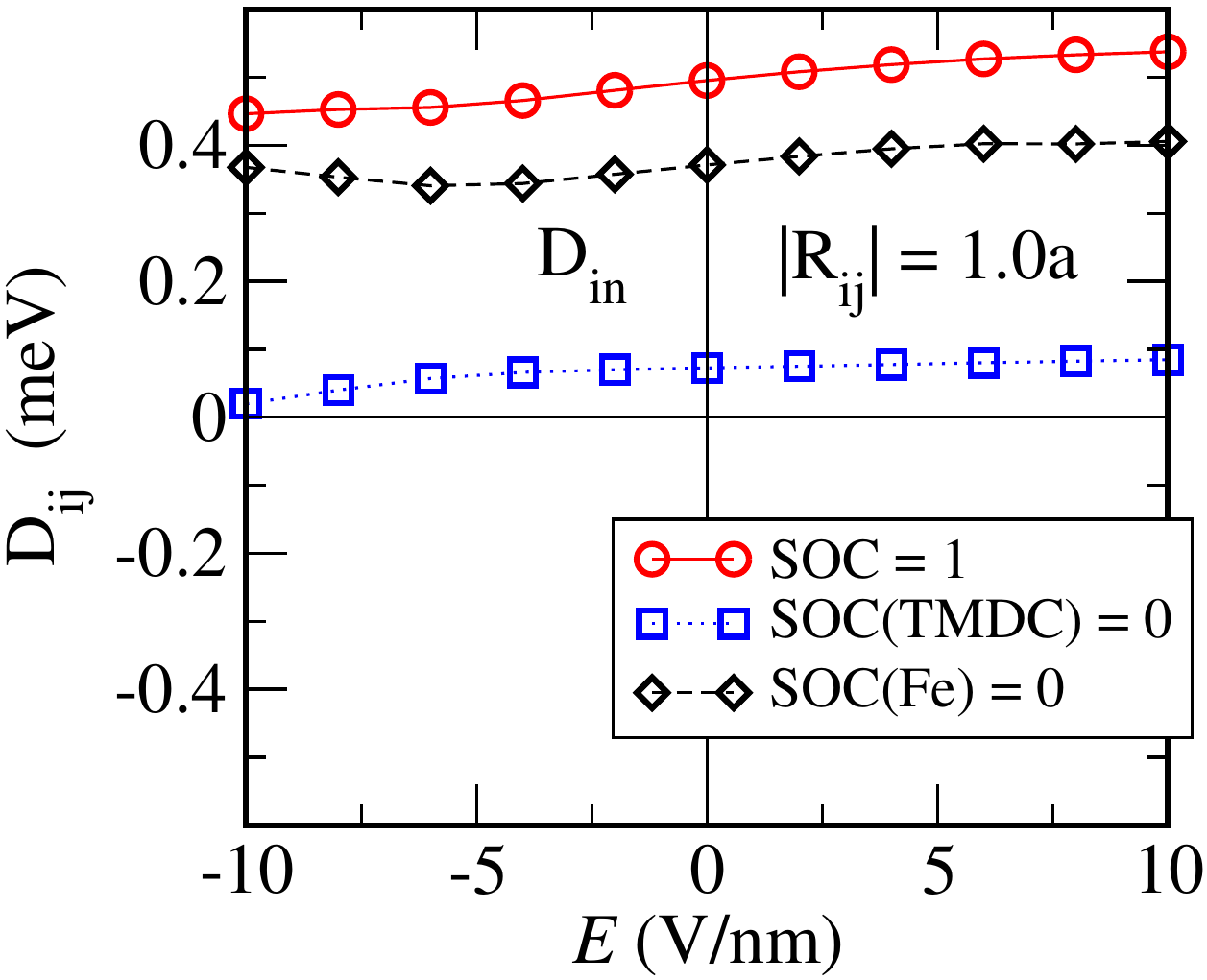}\;(a)
\includegraphics[width=0.19\textwidth,angle=0,clip]{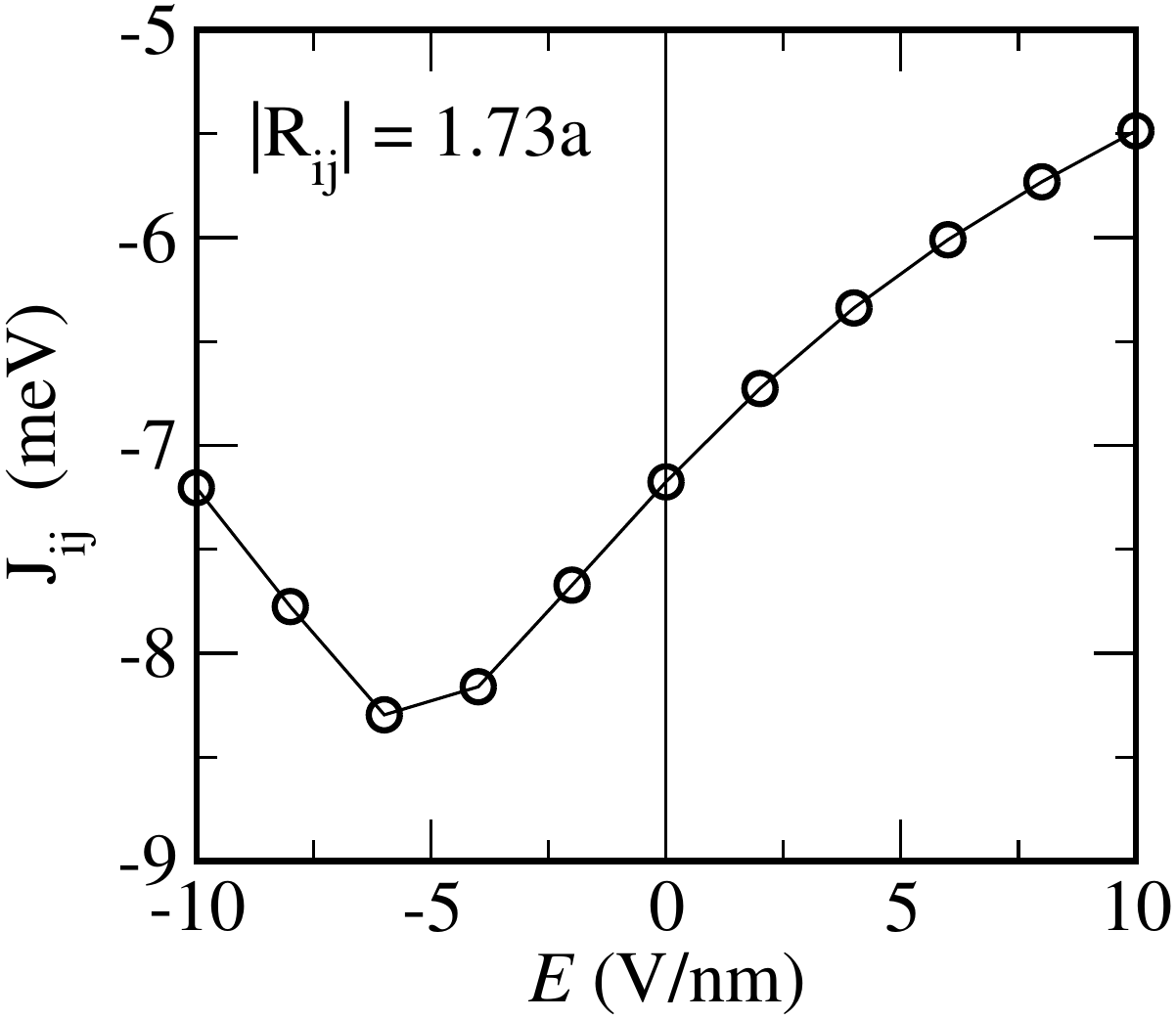}
\includegraphics[width=0.19\textwidth,angle=0,clip]{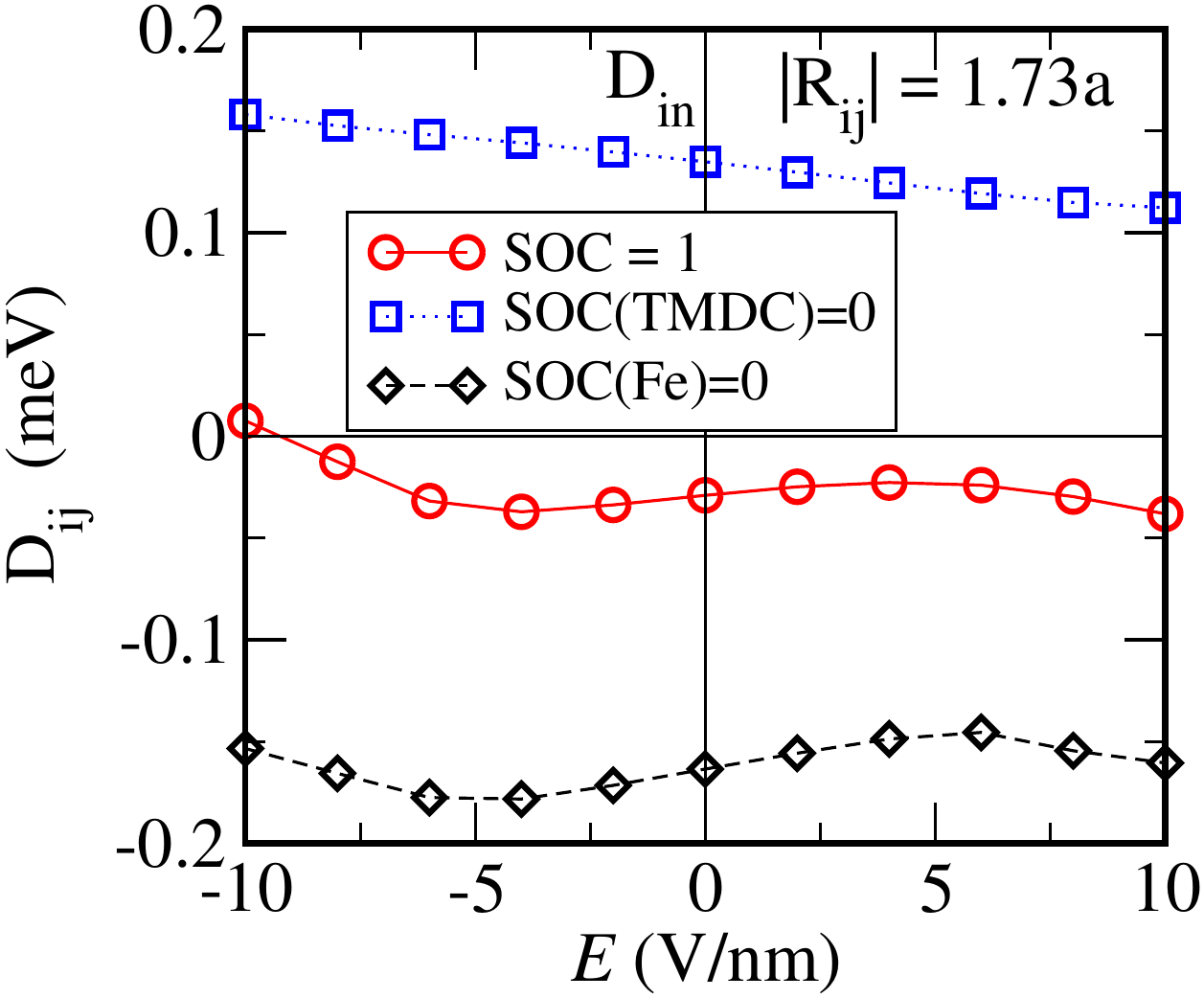}\;(b)
\includegraphics[width=0.19\textwidth,angle=0,clip]{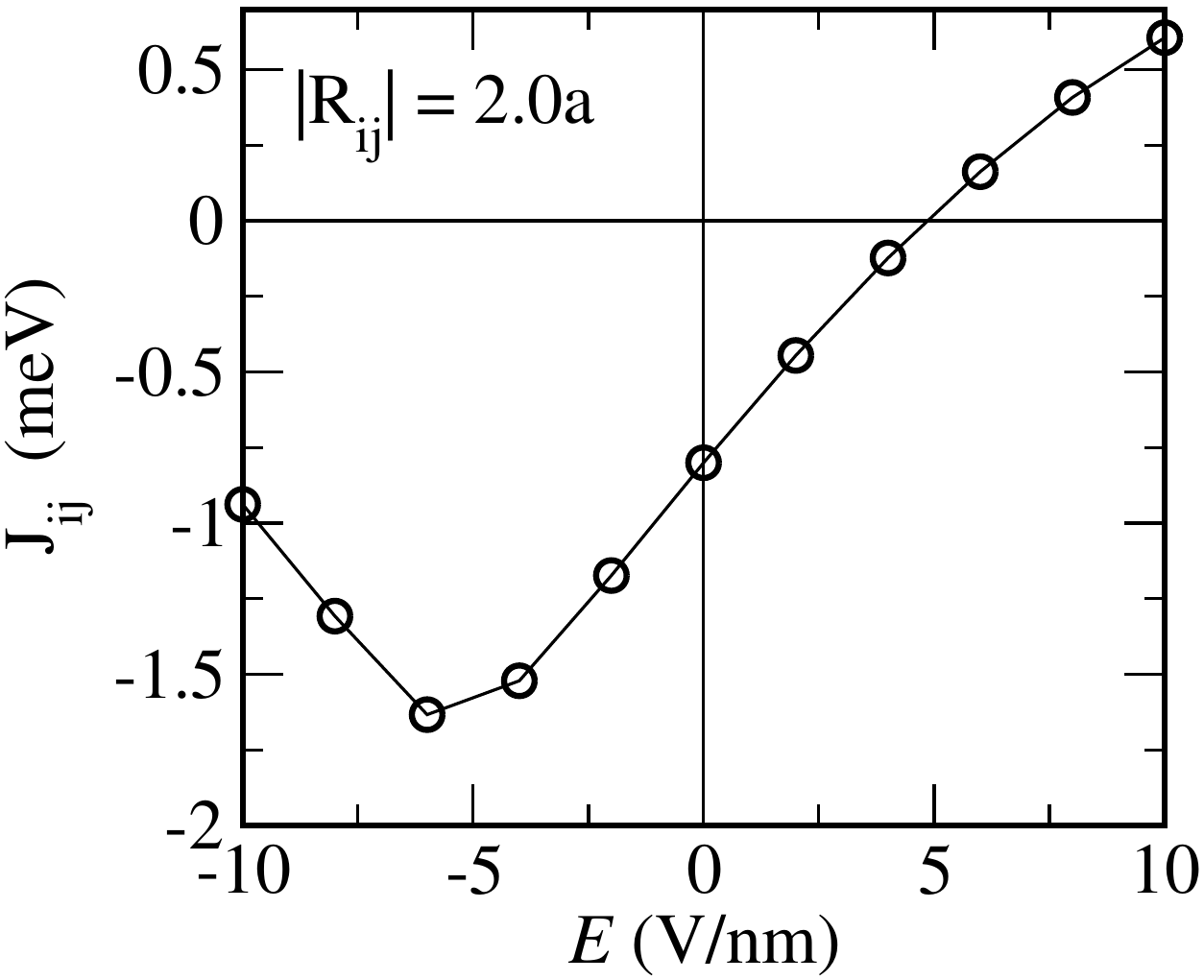}
\includegraphics[width=0.19\textwidth,angle=0,clip]{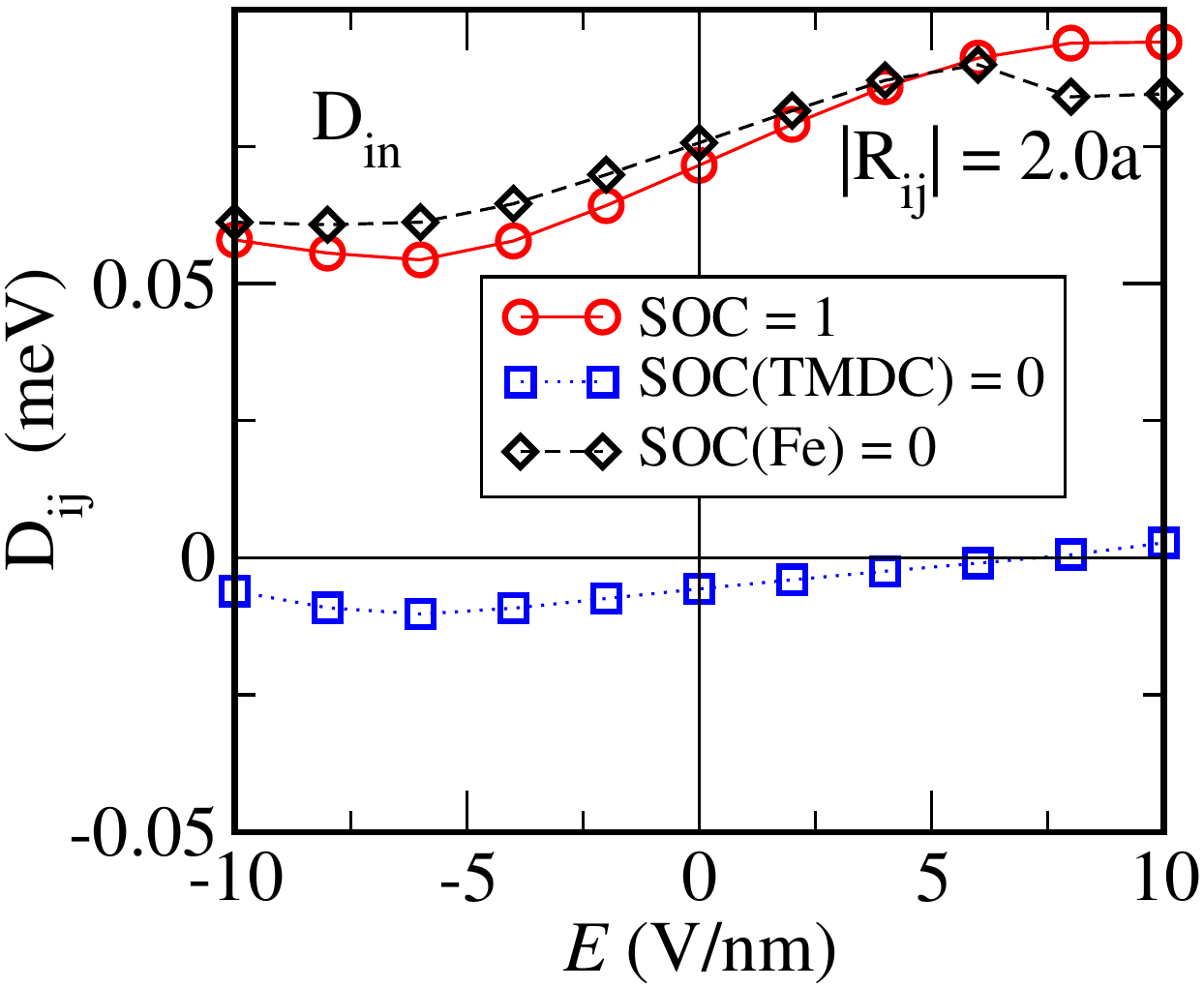}\;(c)
\includegraphics[width=0.4\textwidth,angle=0,clip]{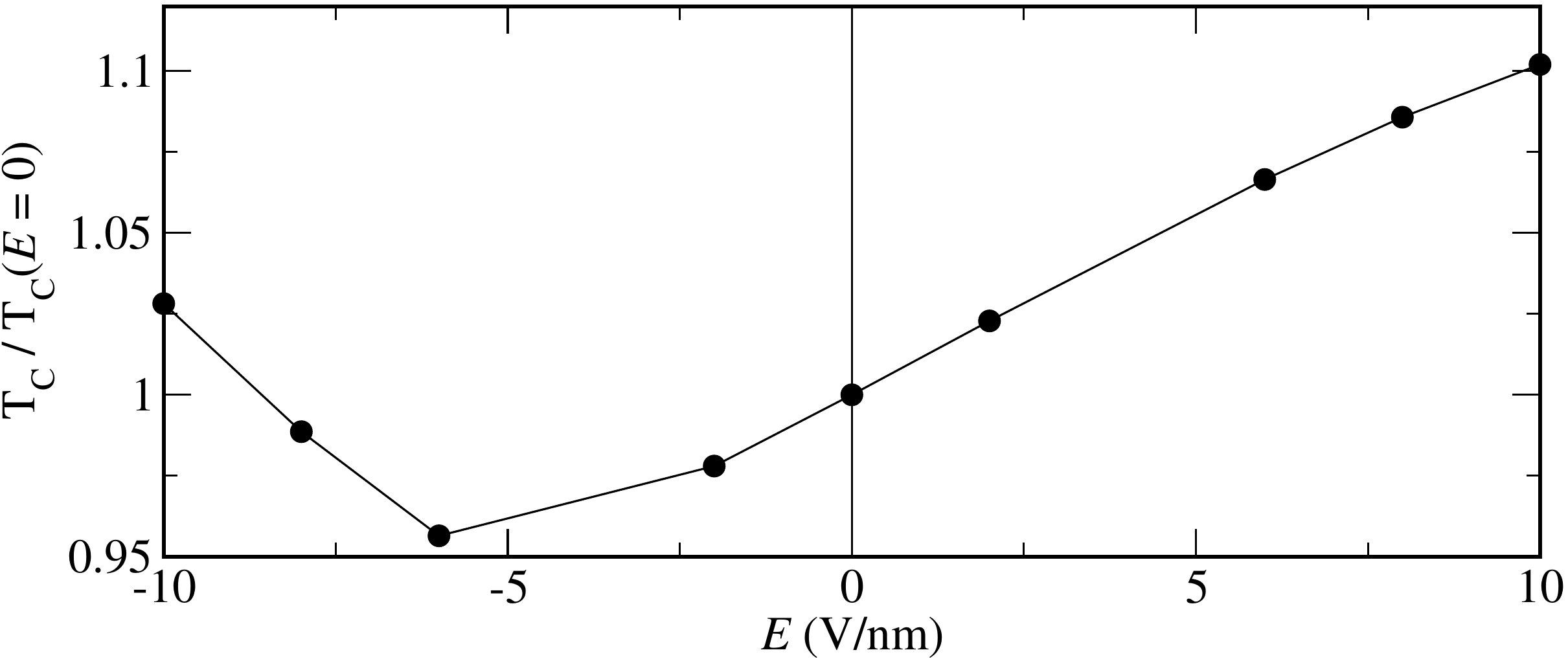}\;(d)
\caption{\label{fig:JXC_WS2} Isotropic Fe-Fe exchange coupling parameter
  $J_{ij}$ (left panel) and in-plane (parallel to the surface plane) components of the Fe-Fe
  Dzyaloshinskii-Moriya interactions $\vec{D}_{ij}$ (right panel) with the
  first-neighbor, $R_{01} = a$ (a),  second-neighbor, $R_{02} =
  1.73a$ (b) and third-neighbor, $R_{03} = 2.0a$ (c), for 1 ML Fe on
  WS$_2$. The parameters are presented 
  as a function of the applied electric fied. (d) the reduced mean-field
  Curie temperature $T_C/T_C(E=0)$, with $T_C(E=0) = 799$ K. }    
\end{figure}

The isotropic Fe-Fe exchange coupling parameters, $J_{ij}$, in 1 ML Fe
on WS$_2$ are plotted in Fig. \ref{fig:JXC_WS2}, left panel, as a function of applied
electric field. Figs. \ref{fig:JXC_WS2} (a), (b) and (c) represent the
exchange parameters for the distances $R_{01} = a$,  $R_{02} = 1.73a$ and
 $R_{03} = 2.0a$, respectively. One can see in all cases an almost linear
 variation of $J_{ij}$ for small electric fields.
 At larger fields,
 $J_{ij}$ changes almost linearly with the field strength for 'positive'
 field, and reach some extremum in the case of the 'negative'
 field. Fig. \ref{fig:JXC_WS2} (d) represents the reduced mean-field
 $T_C/T_C(E=0)$ 
 evaluated assuming FM ordering in the system, demonstrating rather
 pronounced impact of the electric field on the critical temperature.

\begin{figure}[h]
\includegraphics[width=0.4\textwidth,angle=0,clip]{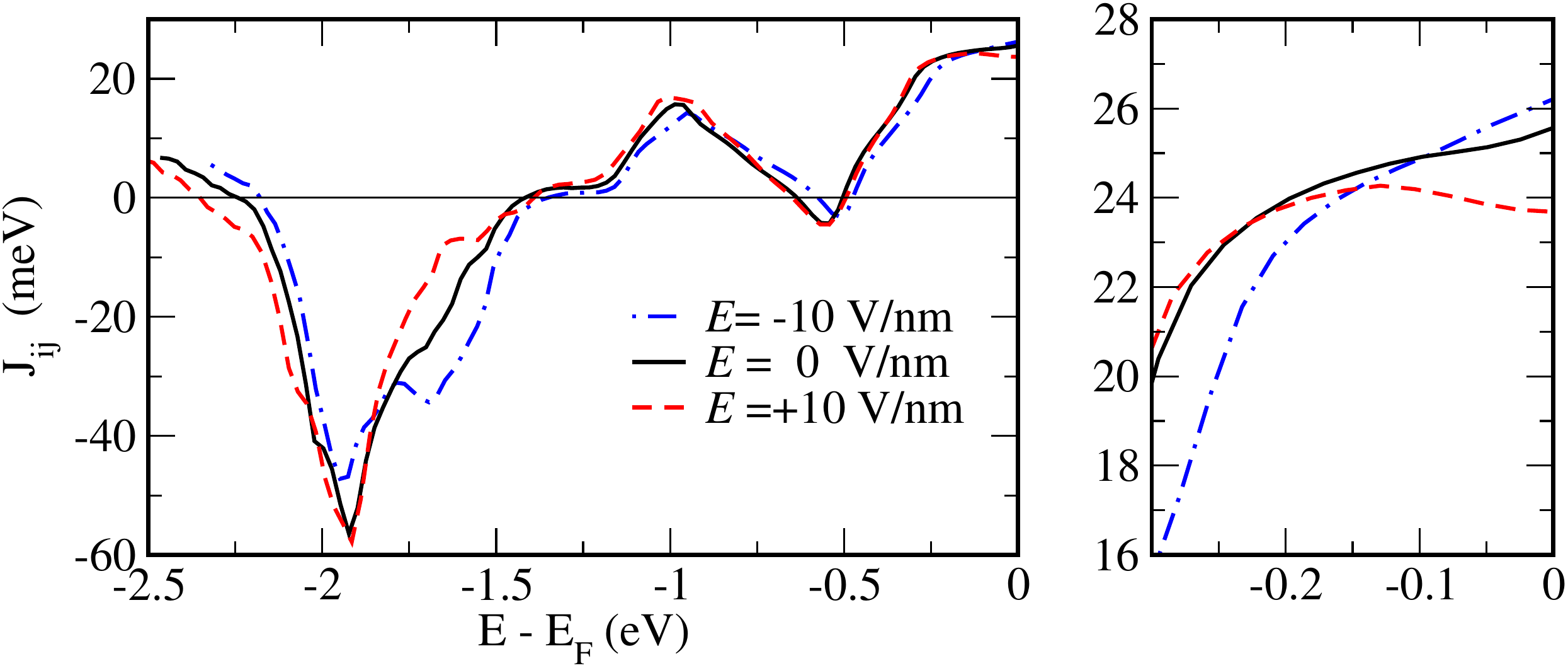}\;(a)
\includegraphics[width=0.4\textwidth,angle=0,clip]{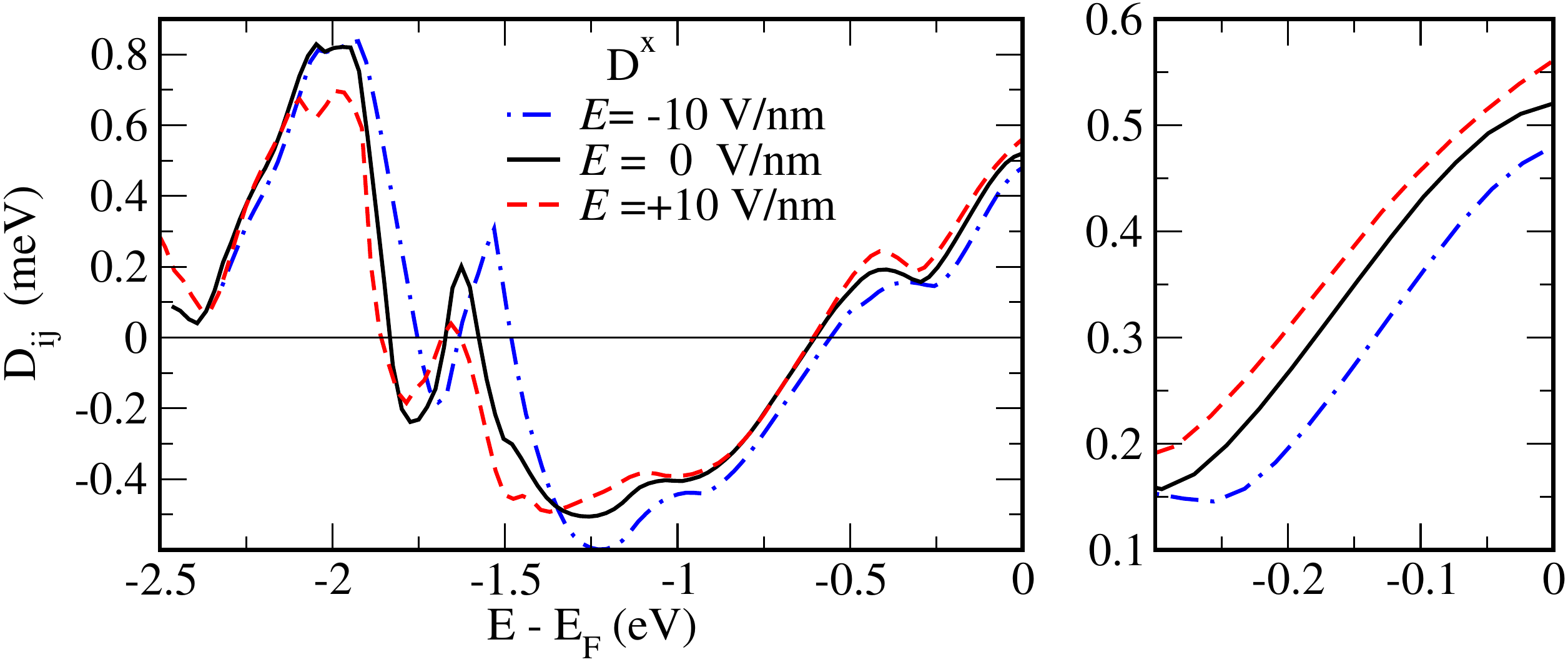}\;(b)
\caption{\label{fig:JXC-EF_WS2} Isotropic Fe-Fe exchange coupling parameter
  $J^1$ (a) and the in-plane $D^{1,in}$ (b)
  represented as a
  function of occupation of energy bands for 1 ML Fe on WS$_2$. The
  applied electric fied $E^0 = 0.0$ V/nm,  $E^+ = +10$ V/nm and $E^- =
  -10$ V/nm.   }    
\end{figure}

The in-plane components Dzyaloshinskii-Moriya interactions in 1 ML Fe on
WS$_2$ calculated for the distances $R_{01} = a$,  $R_{02} = 1.73a$ and
$R_{03} = 2.0a$ are plotted in Fig. \ref{fig:JXC_WS2}, right panel, as a
function of applied electric field. In contrast to the Fe/Pt(111)
system, the DMI exhibits a weaker dependence on the electric field,
implying a crucial role of the states localized at the
Fe/substrate interface, being rather sensitive to the influence of an
electric field. Such states occur in the vicinity of the Fermi energy for the 
Fe/Pt(111) interface, but not for the Fe/WS$_2$ system. 
A strong field induced modification of the Fe $d$-states in Fe/WS$_2$
occurs due to a change of their hybridization with the electronic states
of substrate. This however corresponds mainly to the states below the
Fermi energy. As one can see in 
Fig. \ref{fig:JXC-EF_WS2}(b) representing the in-plane DMI as a function
of the occupation of the electronic states, the field induced change of the
hybridization could result in a much stronger field-dependence of DMI in
the case of Fermi level shifted down by about 1.5 eV.

Note that according to the results shown in Fig.\ \ref{fig:JXC_WS2}, the
SOC of the substrate plays a leading role for the $\vec{D}_{01}$ and  $\vec{D}_{03}$
parameters, similar to the case of Fe/Pt(111), while for $\vec{D}_{03}$ the
SOC for Fe and the TMDC substrate compete with each other.

\section{Summary}

In summary, considering an Fe monolayer, free-standing and deposited on
two different substrates, we demonstrated the impact of an applied
electric field on the exchange parameters, both, isotropic $J_{ij}$ and
DMI, $\vec{D}_{ij}$. In the case of the free-standing Fe monolayer the
electric field has a key role creating the DMI by breaking the inversion
symmetry in the system. In the case of deposited Fe films rather
prominent changes of the exchange parameters occur for the Fe/Pt(111) 
system due to the localized electronic states at the Fe/Pt
interface, that are strongly affected by the electric field.
In the case of an TMDC substrate the dependence of DMI on the electric
field is much weaker, although the isotropic interactions still exhibit a
rather strong modification.

\appendix


%


\end{document}